\documentclass[
    reprint,
    superscriptaddress,
    aps,
    prl,
    amsmath,
    amssymb,
    floatfix,
        longbibliography,
]{revtex4-2}

\usepackage{graphicx}
\usepackage{bm}
\usepackage{xcolor}
\usepackage{microtype}

\usepackage{booktabs,makecell,tabularx,multirow,array}
\usepackage{mathtools,mathrsfs,amsthm}

\usepackage{tikz}
\usepackage{pgfplots}
\pgfplotsset{compat=1.18}
\usepgfplotslibrary{groupplots}

\usepackage{hyperref}
\hypersetup{
    colorlinks=true,
    linkcolor=blue,
    citecolor=blue,
    urlcolor=blue
}
\allowdisplaybreaks

\newtheorem{theorem}{Theorem}

\theoremstyle{definition}

\newcommand{\ket}[1]{\lvert #1\rangle}
\newcommand{\bra}[1]{\langle #1\rvert}

\colorlet{revred}{red!65!black}

\begin{document}

\preprint{APS/123-QED}

%\title{Near-optimal synthesis of non-Gaussian phase gates \\via qubit-oscillator Rabi control}
\title{Near-optimal synthesis of non-Gaussian phase gates via qubit-oscillator Rabi control}

\begin{abstract}

%Non-Gaussian gates remain a key bottleneck for universal continuous-variable (CV) quantum computation due to the difficulty of engineering sufficiently strong nonlinearities. To address this challenge, we develop an efficient qubit-oscillator Rabi synthesis scheme for polynomial phase gates, with a total interaction time that scales polylogarithmically with the inverse target error. Specifically, for a class of readily preparable initial states, we show that a degree-\(R\) phase gate can be approximated by an analytically constructed Rabi sequence with total time \(O(\log^{(R-1)/2+o(1)}(1/\varepsilon))\) for error \(\varepsilon\). This construction requires no numerical optimization and therefore extends naturally to arbitrarily large multimode systems. We further establish the total-time lower bound \(\Omega(\log^{(R-1)/2}(1/\varepsilon))\), showing that the synthesis is near optimal in its error dependence. As applications, we use the resulting sequences to simulate representative CV quantum dynamics and implement a CV quantum algorithm for solving linear partial differential equations. These results establish qubit-oscillator Rabi control as an analytically compilable, efficient, and near-optimal primitive for CV quantum computation. 

Non-Gaussian gates remain a key bottleneck for universal continuous-variable (CV) quantum computation because the nonlinearities they require are difficult to engineer. To address this challenge, we develop an efficient qubit-oscillator Rabi synthesis scheme for polynomial phase gates, with a total interaction time that scales polylogarithmically with the inverse target error \(\varepsilon\). Specifically, for a class of readily preparable initial states, we show that a degree-\(R\) phase gate can be approximated by an analytically constructed Rabi sequence with total time \(O(\log^{(R-1)/2+o(1)}(1/\varepsilon))\). This construction requires no numerical optimization and therefore extends naturally to arbitrarily large multimode systems. We further establish a total-time lower bound of \(\Omega(\log^{(R-1)/2}(1/\varepsilon))\), showing that the synthesis is near optimal. As applications, we use this scheme to simulate representative CV quantum dynamics and implement a CV quantum algorithm for solving linear partial differential equations. These results establish qubit-oscillator Rabi control as an efficient, analytically compilable, and near-optimal primitive for CV quantum information processing.

\end{abstract}

\author{Zhen Yang}
\affiliation{
    Institute of Fundamental and Frontier Sciences,
    University of Electronic Science and Technology of China,
    Chengdu 610051, China
}

\author{Shan Jin}
\email{jinshan@tgqs.net}
\affiliation{
    Institute of Fundamental and Frontier Sciences,
    University of Electronic Science and Technology of China,
    Chengdu 610051, China
}

\affiliation{
   Yangtze Delta Industrial Innovation Center of Quantum Science and Technology, Suzhou, China
}

\author{Zi-Wen Liu}
\email{zwliu0@tsinghua.edu.cn}
\affiliation{Yau Mathematical Sciences Center, Tsinghua University, Beijing, 100084, China
}

\author{Xiaoting Wang}
\email{xiaoting@uestc.edu.cn}
\affiliation{
    Institute of Fundamental and Frontier Sciences,
    University of Electronic Science and Technology of China,
    Chengdu 610051, China
}

\date{\today}

\maketitle

\paragraph{Introduction.---}
Continuous-variable (CV) quantum computation provides a natural framework for processing quantum information encoded in bosonic modes, whose continuous degrees of freedom naturally match the structure of many problems in quantum simulation~\cite{LloydBraunstein1999,BraunsteinLoock2005,Weedbrook2012}. This representation supports applications ranging from bosonic simulation and CV quantum search to algorithms for solving differential equations, with potential computational advantages under suitable assumptions~\cite{PatiBraunsteinLloyd2000,JinHuangWuEtAl2025ContinuousSearch,LauPooserSiopsisWeedbrook2017,ArrazolaKalajdzievskiWeedbrookLloyd2019}. In optical and other bosonic platforms, Gaussian states and operations can be generated and controlled with high precision and at large scale~\cite{YokoyamaEtAl2013,AsavanantEtAl2019,LarsenEtAl2019,AsavanantEtAl2021,SuEtAl2013GateSequence,EnomotoEtAl2021GaussianGates,LarsenEtAl2021MultimodeGates,JiaEtAl2026MonolithicCV}. Universal CV quantum computation, however, requires supplementing Gaussian operations with a suitable non-Gaussian element~\cite{LloydBraunstein1999,KalajdzievskiQuesada2021,BartlettSandersBraunsteinNemoto2002,MariEisert2012,Walschaers2021,ChabaudWalschaers2023,CalcluthReichelFerraroFerrini2024}. A canonical choice is the cubic phase gate, yet many bosonic platforms lack a native and independently tunable cubic interaction~\cite{GottesmanKitaevPreskill2001,LloydBraunstein1999,SefiVanLoock2011,KalajdzievskiQuesada2021}. Existing approaches based on non-Gaussian resource states, measurement-induced nonlinearities~\cite{MarekFilipFurusawa2011,YukawaEtAl2013,MarshallPooserSiopsisWeedbrook2015,Miyata2016,TakedaFurusawa2017,ArzaniTrepsFerrini2017,MarekEtAl2018General,SabapathyWeedbrook2018,ZhengEtAl2021,KudraEtAl2022CubicPhaseState,KonnoEtAl2021,SakaguchiEtAl2023}, or native higher-order interactions~\cite{Yanagimoto2020,HillmannEtAl2020,ErikssonEtAl2024NativeCubic} can realize these gates but remain experimentally demanding.

%Instead of directly generating the non-Gaussian element from within the bosonic system, we can alternatively use hybrid qubit-oscillator systems, such as trapped ions and superconducting systems, to indirectly generate it~\cite{KrastanovEtAl2015,HeeresEtAl2015,EickbuschEtAl2022,DiringerEtAl2024,LiuEtAl2026Hybrid,CraneEtAl2024Hybrid}. 
%In these hybrid platforms, readily implemented Rabi pulses couple a qubit linearly to an oscillator quadrature~\cite{Leibfried2003,FluhmannEtAl2019,SutherlandSrinivas2021,Blais2021}.
%Sequences combining these pulses with single-qubit rotations can induce effective nonlinear operations on the oscillator via final qubit postselection. 
%Numerical studies of Rabi-based protocols have demonstrated high-fidelity approximations to nonlinear phase gates~\cite{ParkMarekFilip2018,ParkFilip2024}.
%These numerical results, however, do not establish how the total Rabi interaction time scales as the target error $\varepsilon$ approaches zero.
%This motivates asking whether exact realization by a finite sequence could make the total time independent of $\varepsilon$.
%We show, however, that no finite postselected Rabi sequence restricted to a single oscillator quadrature can exactly implement a nonlinear polynomial phase gate.
%The challenge is therefore to improve accuracy while limiting the interaction time to reduce exposure to decoherence.
%A polylogarithmic rather than polynomial dependence on $1/\varepsilon$ would offer an exponential improvement in this time cost.

Rather than generating non-Gaussian operations directly from native bosonic nonlinearities, hybrid qubit-oscillator platforms offer an alternative route in which an ancillary qubit mediates effective nonlinear dynamics of the oscillator~\cite{KrastanovEtAl2015,HeeresEtAl2015,EickbuschEtAl2022,DiringerEtAl2024,LiuEtAl2026Hybrid,CraneEtAl2024Hybrid}. Such architectures arise naturally in trapped-ion and superconducting systems, where experimentally accessible Rabi interactions couple the qubit linearly to an oscillator quadrature~\cite{Leibfried2003,FluhmannEtAl2019,SutherlandSrinivas2021,Blais2021}. By interleaving these interactions with single-qubit rotations and postselecting the qubit at the end of the sequence, one can induce effective nonlinear transformations on the oscillator. Numerical Rabi-based protocols have demonstrated high-fidelity approximations to nonlinear phase gates~\cite{ParkMarekFilip2018,ParkFilip2024}, but do not reveal how the required interaction time scales as the target error $\varepsilon$ decreases. An exact finite synthesis would eliminate this precision dependence altogether; however, we will show that no finite postselected Rabi sequence restricted to a single oscillator quadrature can exactly realize a nonlinear polynomial phase gate. The relevant question is therefore how slowly the required interaction time can grow with increasing precision. In particular, a polylogarithmic dependence on $1/\varepsilon$, rather than a polynomial one, would provide an exponential improvement in the asymptotic time cost.

%Generalized quantum signal processing (GQSP) provides an analytic construction of control sequences for bounded Laurent polynomials $P(U)$ of a signal unitary $U$~\cite{Motlagh2024}.
%The sequence length scales with the polynomial degree needed to approximate the target operation~\cite{LowChuang2017,Haah2019ProductDecomposition,DongMengWhaleyLin2021,Ying2022StableFactorization}.
%QSP-Control uses this polynomial approach to suppress cross-Kerr errors under dispersive coupling, with a sequence length of $O(\log(1/\varepsilon))$~\cite{Majumdar2026QSPControl}.
%Related QSP constructions have also been developed to engineer non-Gaussian bosonic gates under dispersive qubit--oscillator coupling~\cite{FongLau2025}.
%With linear quadrature coupling, oscillator-qubit GQSP uses Fourier approximation to achieve $O(\log(1/\varepsilon))$ sequence length for fixed periodic analytic phase functions~\cite{Hong2025OQGQSP}.
%Fourier-Trotter methods also generate phase gates with linear couplings, using sequences of length $O(1/\varepsilon)$ to reduce the first-order Trotter error to $\varepsilon$~\cite{Chalermpusitarak2025Programmable,McGarry2026AnharmonicDynamics}.
%Unlike the periodic phase functions considered above, nonlinear polynomial phases vary increasingly rapidly as the quadrature grows.
%The fundamental challenge is therefore to implement nonlinear polynomial phase gates with linear Rabi sequences while retaining polylogarithmic resource scaling.

Achieving such polylogarithmic precision scaling calls for a systematic way to compile linear qubit-oscillator interactions into increasingly accurate nonlinear transformations. Generalized quantum signal processing (GQSP) provides a natural analytic framework for this purpose, constructing control sequences whose matrix elements realize bounded Laurent polynomials \(P(U)\) of a signal unitary \(U\)~\cite{Motlagh2024}, with sequence length governed by the polynomial degree required to approximate the target operation~\cite{LowChuang2017,Haah2019ProductDecomposition,DongMengWhaleyLin2021,Ying2022StableFactorization}. This framework has recently been adapted to bosonic control: QSP-control suppresses cross-Kerr errors under dispersive coupling with \(O(\log(1/\varepsilon))\) sequence length~\cite{Majumdar2026QSPControl}. For linear quadrature coupling, oscillator-qubit GQSP combines QSP with Fourier approximation to achieve \(O(\log(1/\varepsilon))\) sequence length for fixed periodic analytic phase functions~\cite{Hong2025OQGQSP}. Fourier-Trotter approaches can likewise generate nonlinear phase gates, but reducing first-order Trotter error to \(\varepsilon\) requires sequence length \(O(1/\varepsilon)\)~\cite{Chalermpusitarak2025Programmable,McGarry2026AnharmonicDynamics}. Polynomial phase gates pose a distinct challenge: their phases are nonperiodic and oscillate increasingly rapidly with quadrature amplitude, so the approximation region and the required Fourier bandwidth must both grow as higher precision is demanded. Whether such gates can nevertheless be synthesized by linear Rabi control with polylogarithmic total interaction time, and whether this scaling can approach the fundamental limit, is therefore the central question we address.

Here we answer these questions by developing an efficient Rabi synthesis scheme for the polynomial phase gate
\(U_p(T)=e^{iT p(\hat X)}\),
where
\(p(x)=\sum_{r=0}^{R}a_rx^r\)
is a fixed real polynomial of degree \(R\geq2\) and \(T>0\), with \(p(x)=x^3\) corresponding to the cubic phase gate.
We consider inputs from the fixed finite Fock subspace
\(\mathcal H_n=\operatorname{span}\{\ket{0},\ldots,\ket{n}\}\),
which contains a class of experimentally accessible low-energy states, and construct Rabi sequences that approximate the target gate to error at most \(\varepsilon\) uniformly over all normalized inputs in \(\mathcal H_n\).
For fixed \(n\) and \(T\), the required pulse number scales as
\(N=O(\log^{R/2+o(1)}(1/\varepsilon))\),
while the total Rabi interaction time scales as
\(T_{\rm tot}=O(\log^{(R-1)/2+o(1)}(1/\varepsilon))\).
The pulse parameters are obtained analytically through GQSP, without variational optimization, and the same approximation guarantee yields a postselection success probability of at least \((1-\varepsilon)^2\).
We further prove the lower bound
\(T_{\rm tot}=\Omega(\log^{(R-1)/2}(1/\varepsilon))\)
for arbitrary same-quadrature Rabi sequences, showing that the synthesis is near optimal.
The construction extends to fixed compositions of polynomial phase gates and thereby enables efficient simulation of a standard universal CV gate set~\cite{KalajdzievskiQuesada2021} while preserving polylogarithmic dependence on \(1/\varepsilon\).
As applications, we use the resulting scheme to simulate representative CV dynamics and implement a CV quantum algorithm for solving a linear elliptic equation.
These results establish qubit-oscillator Rabi control as an efficient, analytically compilable, and near-optimal primitive for CV quantum information processing.

\paragraph{Polynomial phase gates with Rabi control.---}
In many quantum algorithms, one is primarily interested in preparing a target output state from a readily preparable input state. For the infinite-dimensional bosonic system considered here, we take the fixed finite Fock subspace
$\mathcal H_n=\operatorname{span}\{\ket{0},\ldots,\ket{n}\}$
as the computational input space. This space contains a class of experimentally accessible low-energy states. The corresponding target output states are not restricted to $\mathcal H_n$.

Our goal is to approximate the polynomial phase $e^{iT p(\hat X)}$ using linear qubit-oscillator coupling, with $p(x)=\sum_{r=0}^{R}a_rx^r$ and $a_R\neq0$.
Since this target depends only on $\hat X$, the elementary controls are
$X$-type Rabi pulses $\mathcal X_{\boldsymbol\omega}(t)=\exp[i t\sigma_{\boldsymbol\omega}\otimes\hat X/2]$, 
where $t>0$ is the pulse duration for the unit-strength Rabi Hamiltonian
$H_{\boldsymbol\omega}=-\sigma_{\boldsymbol\omega}\otimes\hat X/2$, $\boldsymbol\sigma=(\sigma_x,\sigma_y,\sigma_z)$, $\sigma_{\boldsymbol\omega}=\boldsymbol\omega\cdot\boldsymbol\sigma$, and $\boldsymbol\omega$ is a unit Bloch vector.
An $N$-pulse sequence is then written as $\mathcal U_N=\mathcal X_{\boldsymbol\omega_N}(t_N)\cdots \mathcal X_{\boldsymbol\omega_1}(t_1)$.
With normalized ancilla input $\ket{\phi_{\mathrm{in}}}_a$ and postselection state $\ket{\phi_{\mathrm{out}}}_a$, the induced postselected block on the oscillator is $P_N=(\bra{\phi_{\mathrm{out}}}_a\otimes I)\mathcal U_N
(\ket{\phi_{\mathrm{in}}}_a\otimes I)$. We require, for every normalized $\ket{\psi}\in\mathcal H_n$,
\begin{equation}
\left\|\left[P_N-e^{iT p(\hat X)}\right]\ket{\psi}\right\|
\leq\varepsilon .
\label{eq:statewise-error-main}
\end{equation}

Before turning to approximation, we first show that no finite sequence of these pulses can realize the target exactly. In the $\hat X$ representation, let $\mathcal U_N(x)$ denote the single-qubit matrix obtained by replacing $\hat X$ with $x$ in $\mathcal U_N$, and define
$P_N(x)=\bra{\phi_{\mathrm{out}}}_a\mathcal U_N(x)
\ket{\phi_{\mathrm{in}}}_a$.
Each pulse contributes only the two phase factors $e^{\pm i t_jx/2}$, so the resulting $P_N(x)$ is a finite sum of exponentials. Consequently, $P_N(x)$ satisfies a nontrivial finite-order linear differential equation with constant coefficients. The target function $e^{iT p(x)}$, however, does not satisfy any such equation when $p$ is nonlinear.
If a finite sequence realized the target exactly on $\mathcal H_n$, applying it to the vacuum, whose wave function is nonzero for every real $x$, would imply $P_N(x)=e^{iT p(x)}$ for all $x\in\mathbb R$.
This equality is impossible because the two functions have incompatible differential-equation properties. A rigorous proof is given in Sec.~A of the Supplemental Material~\cite{SupplementalMaterial}.

\paragraph{Polylogarithmic Rabi synthesis and near-optimal interaction time.---}
We use an even number of Rabi pulses, all with duration $\tau$. Every pulse introduces only the frequencies $\pm\tau/2$, so after $N$ pulses the postselected function is a Laurent polynomial in $e^{i\tau x}$. 
This polynomial also obeys $\sup_{x\in\mathbb R}|P_N(x)|\leq1$ by the unitarity of $\mathcal U_N(x)$, precisely as required by GQSP~\cite{Motlagh2024}.
In the $\hat X$ representation, normalized inputs in $\mathcal H_n$ obey a uniform Gaussian tail bound~\cite{SupplementalMaterial}. 
The error contribution from $|x|>L$ is then exponentially small in $L^2$.
It therefore suffices to approximate $e^{iTp(x)}$ by $P_N(x)$ on the finite window $[-L,L]$.

\begin{theorem}[Polylogarithmic Rabi synthesis of polynomial phase gates]
\label{thm:polylog-rabi-main}
For all sufficiently small $\varepsilon>0$, one can choose a window $L=\Theta\!\left(\sqrt{n+1+\log(1/\varepsilon)}\right)$ and a common pulse duration $\tau=\pi/(4L)$. 
There exist an even integer $N$, Pauli axes, and ancilla input and postselection states such that the resulting postselected block $P_N$ satisfies the error bound~\eqref{eq:statewise-error-main} for every normalized input $\ket{\psi}\in\mathcal H_n$. The required pulse count obeys 
\begin{equation}
    N=O\!\left(\left(T\sum_{r=0}^{R}|a_r|L^r+\log(1/\varepsilon)\right)^{1+o(1)}\right).
\end{equation}
For fixed $p$, $n$, and $T$, this bound simplifies to $N=O\!\left(\log^{R/2+o(1)}(1/\varepsilon)\right)$, where $o(1)\to0$ as $\varepsilon\to0$.
\end{theorem}

For every such input, the error bound~\eqref{eq:statewise-error-main} also guarantees a postselection success probability $\|P_N\ket{\psi}\|^2\geq(1-\varepsilon)^2$.
The normalized output $P_N\ket{\psi}/\|P_N\ket{\psi}\|$ differs from the ideal state $e^{iTp(\hat X)}\ket{\psi}$ by at most $2\varepsilon$ in norm.

Once the Laurent polynomial $P_N$ and its complement $Q_N$ are known, the GQSP recursion determines the Rabi pulse axes and ancilla input and postselection states analytically, without variational optimization~\cite{Motlagh2024}.
Proofs and construction details are provided in Secs.~B-D of the Supplemental Material~\cite{SupplementalMaterial}.
The same construction and bounds apply to any rotated single-mode quadrature $\hat Q=\hat X\cos\theta+\hat P\sin\theta$, with $\hat X$ replaced by $\hat Q$ in both the target phase and every Rabi pulse.

In standard finite-dimensional gate models, elementary gates are treated as $O(1)$-time operations, so gate count is often used to characterize time complexity. However, CV gate parameters, can grow without bound. Squeezing, for example, counts as a single gate, but its interaction time grows with the squeezing parameter at fixed coupling strength.
We therefore also consider the total Rabi interaction time at unit coupling strength, $T_{\mathrm{tot}}=\sum_{j=1}^{N}t_j$.
For the equal-duration construction above, $T_{\mathrm{tot}}=N\tau$. At fixed $p$, $n$, and $T$, 
\begin{equation}
    T_{\mathrm{tot}}=O\!\left(\log^{(R-1)/2+o(1)}({1}/{\varepsilon})\right).
\end{equation}
The exponent $R/2$ follows from the leading $L^R$ dependence of the pulse-count bound and the window scaling $L=\Theta\!\left(\sqrt{\log(1/\varepsilon)}\right)$.
The pulse duration $\tau=\Theta(L^{-1})$ then reduces the logarithmic exponent of $T_{\mathrm{tot}}=N\tau$ by $1/2$.
Linear Rabi control can therefore approximate polynomial phase gates without directly engineering higher-order oscillator interactions.

We now show that this scaling is a fundamental limit for Rabi sequences in which every pulse couples to the same quadrature $\hat X$, rather than a consequence of the equal-duration choice. 
Consider any same-quadrature Rabi sequence $\mathcal U_{N(\varepsilon)}$ of the form defined above.
If the corresponding postselected block $P_N$ satisfies Eq.~\eqref{eq:statewise-error-main} for every normalized $\ket{\psi}\in\mathcal H_n$, then the following lower bound holds.

\begin{theorem}[Lower bound on total Rabi interaction time]
\label{thm:total-rabi-time-lower-bound-main}
There exists a constant $c_p>0$ depending only on $p$ such that, for all sufficiently small $\varepsilon$, $T_{\mathrm{tot}}\geq c_pT\log^{(R-1)/2}(1/\varepsilon)$. Hence, for fixed $T$, $T_{\mathrm{tot}}=\Omega\left(\log^{(R-1)/2}(1/\varepsilon)\right)$.
\end{theorem}
This lower bound shows that the leading logarithmic exponent achieved by the equal-duration construction cannot be improved within the same-quadrature setting. A detailed proof is given in Supplemental Material~\cite{SupplementalMaterial}, Sec.~F. Combining this lower bound with the constructive upper bound yields
\begin{equation}
    \Omega\left(\log^{(R-1)/2}(1/\varepsilon)\right)
    \leq T_{\mathrm{tot}}
    \leq O\left(\log^{(R-1)/2+o(1)}(1/\varepsilon)\right).
\end{equation}
Thus, the equal-duration construction is near-optimal in total Rabi interaction time, up to a factor of $\log^{o(1)}(1/\varepsilon)$. 
In particular, any scaling $o\!\left(\log^{(R-1)/2}(1/\varepsilon)\right)$,
including $O(\log\log(1/\varepsilon))$, is ruled out.

\paragraph{Polylogarithmic Rabi synthesis of a universal CV gate set.---}
Quantum evolutions are generally built from sequences of elementary unitaries, motivating the use of a finite set of gates sufficient for general quantum processing. To effectively simulate a universal CV gate set within our Rabi framework, we decompose each of its gates into a finite product of polynomial phase gates. The key question is then whether the polylogarithmic resource scaling is preserved under composition. 

Consider a finite product of phase blocks $U_{\mathrm{tot}}=U_M\cdots U_1$, where $U_j=e^{iT_jp_j(\hat Q_j)}$, $p_j$ is a fixed real polynomial of degree $R_j$, and $\hat Q_j=\sum_{\ell=1}^{m}(\alpha_{j\ell}\hat X_\ell+\beta_{j\ell}\hat P_\ell)$ is a real linear multimode quadrature. Let $P_{\mathrm{tot}}=P_M\cdots P_1$ be the product of the corresponding postselected blocks. 
The polylogarithmic scaling is preserved under composition provided that, for each block, the preceding ideal evolution $U_{j-1}\cdots U_1$ transforms $\hat Q_j$ in the Heisenberg picture into an $\varepsilon$-independent self-adjoint polynomial in the canonical quadrature operators of total degree $\mu_j$.
Under this condition, the postselected blocks can be chosen so that $|(P_{\mathrm{tot}}-U_{\mathrm{tot}})\ket{\psi}|\leq\varepsilon$ for every normalized $\ket{\psi}\in\mathcal H_n^{(m)}$, where $\mathcal H_n^{(m)}$ is the $m$-mode input subspace with a fixed total Fock cutoff. If $R_j\mu_j/2\geq1$ for all $j$, the total Rabi interaction time obeys
\begin{equation}
    T_{\mathrm{tot}}=O\left(\sum_{j=1}^{M}\log^{(R_j-1)\mu_j/2+o(1)}(M/\varepsilon)\right).
\end{equation}
The corresponding pulse count is likewise polylogarithmic in $1/\varepsilon$, while the joint postselection success probability remains at least $(1-\varepsilon)^2$. See Sec.~G.1 of the Supplemental Material~\cite{SupplementalMaterial} for the precise conditions and proof.

This result applies to the standard universal CV gate set $\mathcal G=\{R(\varphi),Z(\xi),\Phi(\kappa),V(\gamma),CZ_{12}(\zeta)\}$, where $R(\varphi)=e^{i\varphi(\hat X^2+\hat P^2)}$, $Z(\xi)=e^{i\xi\hat X}$, $\Phi(\kappa)=e^{i\kappa\hat X^2}$, $V(\gamma)=e^{i\gamma\hat X^3}$, and $CZ_{12}(\zeta)=e^{i\zeta\hat X_1\hat X_2}$~\cite{SefiVanLoock2011,KalajdzievskiWeedbrookRebentrost2018,KalajdzievskiArrazola2019,KalajdzievskiQuesada2021}. Within this set, $Z(\xi)$ can be implemented exactly by a single linear Rabi pulse with the ancilla prepared in an eigenstate of the coupled Pauli operator, requiring no postselection and $T_{\mathrm{tot}}=O(1)$ for fixed $\xi$. 
The gates $\Phi(\kappa)$ and $V(\gamma)$ follow directly from the $R=2$ and $R=3$ constructions above, respectively. The gates $R(\varphi)$ and $CZ_{12}(\zeta)$ can each be decomposed exactly into finitely many quadratic phase blocks. 
For these decompositions, the preceding ideal evolution is Gaussian, so each quadrature remains a linear combination of the canonical quadratures in the Heisenberg picture, giving $\mu_j=1$ and thereby retaining polylogarithmic resource scaling. Specifically, for fixed gate parameters and mode number, $\Phi(\kappa)$, $R(\varphi)$, and $CZ_{12}(\zeta)$ satisfy $T_{\mathrm{tot}}=O\!\left(\log^{1/2+o(1)}(1/\varepsilon)\right)$, whereas $V(\gamma)$ satisfies $T_{\mathrm{tot}}=O\!\left(\log^{1+o(1)}(1/\varepsilon)\right)$.

Among the gates in $\mathcal G$, $CZ_{12}(\zeta)$ is the only two-mode operation. Defining $\hat Q_\pm=(\hat X_1\pm\hat X_2)/\sqrt2$, it admits the exact decomposition
$CZ_{12}(\zeta)=e^{i\zeta \hat Q_+^2/2}e^{-i\zeta \hat Q_-^2/2}$~\cite{SefiVanLoock2011,KalajdzievskiQuesada2021}.
Since $\hat X_1$ and $\hat X_2$ commute, each collective-quadrature Rabi pulse factorizes exactly as
$\exp[i\tau\sigma_{\boldsymbol\omega}\otimes\hat Q_\pm/2]=\mathcal X_{\boldsymbol\omega,1}(\tau/\sqrt2)\mathcal X_{\pm\boldsymbol\omega,2}(\tau/\sqrt2)$, 
where $\mathcal X_{\boldsymbol\omega,\ell}(t)=\exp[i t\sigma_{\boldsymbol\omega}\otimes\hat X_\ell/2]$ is a mode-selective Rabi pulse on mode $\ell$. This implementation requires only that the same ancilla remain coherent throughout the two mode-selective interactions, with its coupling switched from one mode to the other. It therefore avoids both a native two-mode interaction and a simultaneous multimode Rabi coupling. Such mode-selective linear couplings are available in trapped-ion and circuit-QED platforms~\cite{Leibfried2003,Blais2021,Hastrup2022PRL}.

On platforms where Gaussian operations are natively supported, these gates can be implemented directly. The results above provide a Rabi synthesis of all gates in $\mathcal G$, thereby enabling efficient simulation of the standard universal CV gate set for a class of readily preparable input states. Explicit gate decompositions are given in Sec.~G.2 of the Supplemental Material~\cite{SupplementalMaterial}.

\paragraph{Applications to CV dynamics and elliptic-equation solving.---}
\begin{figure}[t]
    \centering
    \includegraphics[width=0.7\linewidth]{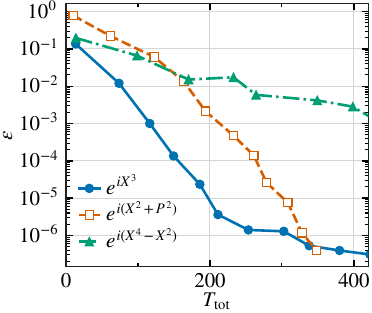}
    \caption{Finite-resource convergence for representative CV evolutions. The curves show the numerical implementation errors as functions of the total Rabi interaction time $T_{\mathrm{tot}}$ for the cubic phase gate $e^{i\hat X^3}$, the rotation $e^{i(\hat X^2+\hat P^2)}$, and the double-well evolution $e^{i(\hat X^4-\hat X^2)}$.}
    \label{fig:applications-main}
\end{figure}
The constructions above cover both single-quadrature polynomial evolutions and finite compositions of phase gates involving different quadratures.
We numerically evaluate the state-dependent implementation errors for the cubic phase gate $e^{i\hat X^3}$ and the double-well evolution $e^{i(\hat X^4-\hat X^2)}$, both acting on $\ket{0}$, as well as the rotation $e^{i(\hat X^2+\hat P^2)}$ acting on $(\ket{0}+\ket{1})/\sqrt2$, with evolution time $T=1$ in all cases. As shown in Fig.~\ref{fig:applications-main}, the errors decrease with increasing total Rabi interaction time for all three evolutions. These finite-resource results illustrate the efficient convergence underlying the asymptotic scaling established above.

The polynomial phase gates can also serve as a computational primitive in CV quantum algorithms, rather than only for direct dynamical simulation. 
Linear differential equations offer a direct example: CV formulations represent functions as bosonic wave functions and spatial derivatives as momentum operators, reducing the solution of a linear PDE to preparation of the corresponding solution state~\cite{HarrowHassidimLloyd2009,LauPooserSiopsisWeedbrook2017,ArrazolaKalajdzievskiWeedbrookLloyd2019}.
We consider the regularized biharmonic equation $(\mu-\alpha\Delta_D+\beta\Delta_D^2)u(\mathbf x)=f(\mathbf x)$ and write $\hat A_D=\mu-\alpha\Delta_D+\beta\Delta_D^2$. The goal is to prepare $\ket u\propto\hat A_D^{-1}\ket f$. The required core unitary is $U_A=\exp[-i\hat X_1\hat X_2\hat A_D]$. We therefore focus on the Rabi implementation of this core unitary.

With $-\Delta_D=\sum_{j=1}^{D}\hat P_{j+2}^{\,2}$, repeated use of the polarization identity gives an exact decomposition of $U_A$ into $M=O(D^2)$ phase blocks $e^{i\theta_\ell \hat Q_\ell^{r_\ell}}$, where $r_\ell\in\{2,4,6\}$ and $\hat Q_\ell$ is a real linear multimode quadrature. 
Since every $\hat Q_\ell$ is a linear combination of the mutually commuting operators $\{\hat X_1,\hat X_2,\hat P_3,\ldots,\hat P_{D+2}\}$, every preceding phase block commutes with each subsequent $\hat Q_\ell$.
Its Heisenberg transform therefore remains $\hat Q_\ell$ itself, giving $\mu_\ell=1$. With $\max_\ell r_\ell=6$, the core unitary can be implemented with total Rabi interaction time $T_{\mathrm{tot}}=O\!\left(D^2\log^{5/2+o(1)}(D/\varepsilon)\right)$.

For a $D=2$ example, we take the source term $f$ to have an antisymmetric double-Gaussian spatial profile. After combining identical contributions, the decomposition of the core unitary contains $28$ polynomial phase blocks. The normalized solution reproduces the expected antisymmetric profile with state fidelity $F_{\mathrm{sol}}=0.9998$ relative to a Fourier-spectral reference. The full decomposition and numerical results are given in Sec.~H of the Supplemental Material~\cite{SupplementalMaterial}.

\paragraph{Discussion.---}

We have developed an efficient Rabi synthesis of polynomial phase gates using linear qubit-oscillator interactions. For a class of readily preparable initial states, the approximation error can be made arbitrarily small with polylogarithmic total Rabi interaction time, while the postselection success probability approaches unity. The upper and lower bounds match up to subpolynomial factors, showing that this scaling is near-optimal. Finite compositions of phase blocks further enable efficient simulation of a standard universal CV gate set. We apply these constructions to representative CV dynamics and to the solution of a linear fourth-order elliptic equation.

A natural direction for future work is to close the residual $\log^{o(1)}(1/\varepsilon)$ gap between the upper and lower bounds. Beyond this complexity question, extending the present results to more general bosonic Hamiltonians involving noncommuting functions of $\hat X$ and $\hat P$ would broaden the scope of Rabi synthesis. At the practical level, incorporating finite control precision and platform-specific noise would help assess its performance in realistic qubit-oscillator systems.

\begin{acknowledgments}
\paragraph{Acknowledgments.---}
This work was supported by the National Natural Science Foundation of China (Grants No.~92265208;12475023) and the Sichuan Science and Technology Program (Grant No.~2025YFHZ0336). Z.-W.~L. also acknowledges support from the Dushi Program and startup funding from YMSC.
\end{acknowledgments}
\textit{Note added.—}While completing this work, we became aware of an independent related work~\cite{2609.02405v1}.

\bibliography{ref}

\end{document}

\title{Supplemental Material for
``Near-Optimal Synthesis of Non-Gaussian Phase Gates via Qubit--Oscillator Rabi Control''}

\author{Zhen Yang}
\affiliation{
    Institute of Fundamental and Frontier Sciences,
    University of Electronic Science and Technology of China,
    Chengdu 610051, China
}

\author{Shan Jin}
\email{jinshan@tgqs.net}
\affiliation{
    Institute of Fundamental and Frontier Sciences,
    University of Electronic Science and Technology of China,
    Chengdu 610051, China
}

\affiliation{
   Yangtze Delta Industrial Innovation Center of Quantum Science and Technology, Suzhou, China
}

\author{Zi-Wen Liu}
\email{zwliu0@tsinghua.edu.cn}
\affiliation{Yau Mathematical Sciences Center, Tsinghua University, Beijing, 100084, China   
}

\author{Xiaoting Wang}
\email{xiaoting@uestc.edu.cn}
\affiliation{
    Institute of Fundamental and Frontier Sciences,
    University of Electronic Science and Technology of China,
    Chengdu 610051, China
}

\date{\today}

\maketitle

This Supplemental Material provides the proofs and technical details supporting the results in the main text. Section I establishes that no finite same-\(\hat X\) Rabi sequence can exactly realize a nonlinear polynomial phase gate. Section II develops the constructive synthesis based on windowed Laurent approximation and generalized quantum signal processing, and derives the corresponding upper bounds on the sequence length and total Rabi interaction time. Section III proves the matching lower bound on the total interaction time for arbitrary same-quadrature Rabi sequences. Section IV extends the construction to finite products of polynomial phase gates and gives the Gaussian-gate decompositions used in the main text. Section V presents the finite-resource numerical implementations of representative CV evolutions and the regularized biharmonic-equation example.

\section{Finite \(X\)-Rabi pulses cannot exactly generate polynomial phase gates}
\label{app:finite-rabi-function-class}

This section gives the rigorous proof that no finite \(X\)-Rabi sequence can realize the polynomial phase \(e^{iTp(\hat X)}\) exactly, where \(p(x)=\sum_{r=0}^{R}a_r x^r\), \(R\ge2\), and \(a_R\ne0\). Since \(T>0\) is fixed, it can be absorbed into the coefficients of \(p\), so it is sufficient to consider the target \(e^{ip(\hat X)}\).
Using the notation of the main text, consider an arbitrary finite sequence \(\mathcal U_N := \prod_{j=1}^{N} \exp\left(i \frac{t_j}{2}\sigma_{\omega_j}\otimes \hat X\right)\), where \(t_j>0\) and \(\|\omega_j\|=1\).

We first determine the functional form of the corresponding postselected block in the \(X\)-representation. Replacing \(\hat X\) by the real variable \(x\), define
\begin{equation}
F_N(x) := {}_a\!\langle\phi_{\rm out}| \exp\left(i \frac{t_Nx}{2}\sigma_{\omega_N}\right)
\cdots \exp\left(i \frac{t_{1}x}{2}\sigma_{\omega_{1}}\right) |\phi_{\rm in}\rangle_a .
\end{equation}
Since \(\|\omega_j\|=1\), one has \(\sigma_{\omega_j}^2=I\). Therefore,
\begin{equation}
\begin{aligned}
\exp\left(i \frac{t_jx}{2}\sigma_{\omega_j}\right)
&= \cos\left(\frac{t_jx}{2}\right)I + i\sin\left(\frac{t_jx}{2}\right)\sigma_{\omega_j}\\
&= \frac{e^{it_jx/2}+e^{-it_jx/2}}{2}I + \frac{e^{it_jx/2}-e^{-it_jx/2}}{2}\sigma_{\omega_j} \\
&= \frac{I+\sigma_{\omega_j}}{2}e^{it_jx/2} + \frac{I-\sigma_{\omega_j}}{2}e^{-it_jx/2}\\
:&= \sum_{s_j=\pm1}\Pi_{\omega_j}^{s_j}e^{is_jt_jx/2},
\end{aligned}
\end{equation}
where \(\Pi_{\omega_j}^{s_j} := \frac{I+s_j\sigma_{\omega_j}}{2}\), \(s_j=\pm1\). Expanding every Rabi factor gives
\begin{equation}
\begin{aligned}
F_N(x)=\sum_{s_1=\pm1}\cdots\sum_{s_N=\pm1}c_{s_1,\ldots,s_N}\exp\left(\frac{i}{2}\sum_{j=1}^{N}s_jt_jx\right),
\end{aligned}
\end{equation}
where \(c_{s_1,\ldots,s_N} := {}_a\!\langle\phi_{\rm out}|\prod_{j=1}^{N}\Pi_{\omega_j}^{s_j}|\phi_{\rm in}\rangle_a\). The coefficients \(c_{s_1,\ldots,s_N}\) are independent of \(x\), while the frequencies \(\frac{1}{2}\sum_{j=1}^{N}s_jt_j\) can take only finitely many real values. Let the distinct frequencies with nonzero merged coefficients be \(\lambda_1,\ldots,\lambda_M\). For each \(\lambda_\ell\), define
\begin{equation}
\begin{aligned}
c_\ell := \sum_{\substack{s_1,\ldots,s_N=\pm1\\ \frac{1}{2}\sum_{j=1}^{N}s_jt_j=\lambda_\ell}}c_{s_1,\ldots,s_N}
= \sum_{\substack{s_1,\ldots,s_N=\pm1\\ \frac{1}{2}\sum_{j=1}^{N}s_jt_j=\lambda_\ell}}{}_a\!\langle\phi_{\rm out}|\prod_{j=1}^{N}\Pi_{\omega_j}^{s_j}|\phi_{\rm in}\rangle_a .
\end{aligned}
\end{equation}
The postselected function can therefore be written as \(F_N(x)=\sum_{\ell=1}^{M}c_\ell e^{i\lambda_\ell x}\). Hence any finite-length postselected \(X\)-Rabi sequence produces a finite exponential polynomial in the \(X\)-representation.

We next use this finite-frequency structure to compare \(F_N(x)\) with the nonlinear target \(e^{ip(x)}\). If all merged coefficients vanish, then \(F_N(x)\equiv0\), which clearly cannot implement the target gate with a nonzero success probability. We therefore restrict attention to \(F_N\not\equiv0\).

Let \(D:=\frac{d}{dx}\). For each frequency \(\lambda_s\), one has \((D-i\lambda_s)e^{i\lambda_sx}=0\). Since the factors \(D-i\lambda_\ell\) commute, applying their product to \(F_N\) gives
\begin{equation}
\left[\prod_{\ell=1}^{M}(D-i\lambda_\ell)\right]F_N(x)
= \sum_{s=1}^{M}c_s\left[\prod_{\ell=1}^{M}(D-i\lambda_\ell)\right]e^{i\lambda_sx}=0.
\end{equation}
Thus every finite-sequence function \(F_N\) satisfies a finite-order linear differential equation with constant coefficients.

We now evaluate the same differential operator on the target function \(e^{ip(x)}\). Define \(Q_0(x)=1\) and recursively set
\begin{equation}\label{Eq:Q_k}
Q_k(x) = Q_{k-1}'(x) + i\bigl(p'(x)-\lambda_k\bigr)Q_{k-1}(x),
\end{equation}
where \(k=1,\ldots,M\). Repeated application of the product rule gives
\begin{equation}\label{eq:diff_D}
\left[\prod_{\ell=1}^{k}(D-i\lambda_\ell)\right]e^{ip(x)}=e^{ip(x)}Q_k(x).
\end{equation}
Indeed, assuming this identity holds for \(k-1\), we obtain
\begin{equation}
\begin{aligned}
\left[\prod_{\ell=1}^{k}(D-i\lambda_\ell)\right]e^{ip(x)}
&= (D-i\lambda_k)\left(e^{ip(x)}Q_{k-1}(x)\right) \\
&= \left(De^{ip(x)}\right)Q_{k-1}(x) + e^{ip(x)}Q_{k-1}'(x) - i\lambda_k e^{ip(x)}Q_{k-1}(x)\\
&= e^{ip(x)}\left[ip'(x)Q_{k-1}(x) + Q_{k-1}'(x) - i\lambda_kQ_{k-1}(x)\right] \\
&= e^{ip(x)}Q_k(x).
\end{aligned}
\end{equation}
This proves Eq.~\eqref{eq:diff_D} by induction.

It remains to show that \(Q_M(x)\) is not the zero polynomial.
Since
\(p'(x)=Ra_Rx^{R-1}+\text{lower-order terms}\)
with \(a_R\ne0\), we show by induction that
\begin{equation}
    Q_k(x)
    =
    (iRa_R)^k x^{k(R-1)}
    +
    \text{lower-order terms}.
    \label{eq:Qk-leading-term}
\end{equation}
The claim holds for \(k=0\) because \(Q_0(x)=1\).
Suppose it holds for \(Q_{k-1}\).
In the recurrence relation~\eqref{Eq:Q_k}, the product
\(ip'(x)Q_{k-1}(x)\) contributes the leading term
\((iRa_R)^k x^{k(R-1)}\).
The terms \(Q_{k-1}'(x)\) and
\(-i\lambda_kQ_{k-1}(x)\) have strictly lower degree because \(R-1\ge1\).
Equation~\eqref{eq:Qk-leading-term} therefore also holds for \(Q_k\).
By induction, \(Q_M(x)\) has a nonzero leading coefficient and hence \(Q_M(x)\not\equiv0\).

Taking \(k=M\) in Eq.~\eqref{eq:diff_D} gives
\begin{equation}
\left[\prod_{\ell=1}^{M}(D-i\lambda_\ell)\right]e^{ip(x)}=e^{ip(x)}Q_M(x)\not\equiv0.
\end{equation}
Here \(\not\equiv0\) means that the function is not identically zero and does not exclude isolated zeros of \(Q_M(x)\). The target \(e^{ip(x)}\) therefore does not satisfy the differential equation obeyed by every finite-sequence function \(F_N\).

We now return to the exact implementation condition on \(\mathcal H_n\). Suppose, for contradiction, that a finite-length postselected \(X\)-Rabi sequence exactly implements the target gate on \(\mathcal H_n\). Since postselection may contribute an input-independent overall amplitude, the most general exact implementation allows a constant \(\alpha\ne0\) such that
\begin{equation}
({}_a\!\langle\phi_{\rm out}|\otimes I)\mathcal U_N(|\phi_{\rm in}\rangle_a\otimes|\psi\rangle)
= \alpha e^{ip(\hat X)}|\psi\rangle
\end{equation}
for all \(|\psi\rangle\in\mathcal H_n\). If the unnormalized postselected block itself equals the target gate, this corresponds to \(\alpha=1\).

Because the vacuum state \(|0\rangle\) belongs to \(\mathcal H_n\), the above condition holds in particular for \(|\psi\rangle=|0\rangle\). In the \(X\)-representation, \(\langle x|0\rangle=\pi^{-1/4}e^{-x^2/2}\), which is nonzero for every \(x\in\mathbb R\). The exact implementation condition therefore implies \(F_N(x)\langle x|0\rangle=\alpha e^{ip(x)}\langle x|0\rangle\) almost everywhere. Dividing by the nonzero vacuum wave function gives
\begin{equation}\label{eq:F_N}
F_N(x)=\alpha e^{ip(x)}
\end{equation}
almost everywhere. Both sides are continuous, so Eq.~\eqref{eq:F_N} holds for all \(x\in\mathbb R\).

Applying \(\prod_{\ell=1}^{M}(D-i\lambda_\ell)\) to both sides of Eq.~\eqref{eq:F_N}, the left-hand side gives
\begin{equation}
\left[\prod_{\ell=1}^{M}(D-i\lambda_\ell)\right]F_N(x)=0,
\end{equation}
while Eq.~\eqref{eq:diff_D} gives
\begin{equation}
\left[\prod_{\ell=1}^{M}(D-i\lambda_\ell)\right]\alpha e^{ip(x)}=\alpha e^{ip(x)}Q_M(x)\not\equiv0.
\end{equation}
The two results are incompatible, contradicting Eq.~\eqref{eq:F_N}. Therefore, no finite-length postselected \(X\)-Rabi sequence can exactly implement \(e^{ip(\hat X)}\) on the fixed Fock subspace \(\mathcal H_n\) with a nonzero success amplitude. At any finite sequence length, the Rabi construction can only approximate the target phase rather than realize it exactly.

\section{Constructive Rabi synthesis of polynomial phase gates}

This section establishes the constructive upper bound stated in Theorem~1 of the main text.
Section~II.B constructs the bounded Laurent polynomial needed for the synthesis, using the Gevrey estimates from Sec.~II.A, and Sec.~II.C extends the resulting approximation to the finite Fock input space.
Section~II.D then realizes this polynomial by equal-duration \(X\)-Rabi pulses through GQSP.
Finally, Sec.~II.E combines the approximation and realization results to derive the complete resource and postselection bounds.

\subsection{Auxiliary Gevrey estimates}
\label{app:gevrey}

The Laurent approximation in Sec.~II.B relies on quantitative control of the Fourier decay of a smooth periodic extension of the target phase.
We therefore establish here the Gevrey derivative bounds for the switching and cutoff functions used in that construction.

\subsubsection{Definition of Gevrey classes}

\begin{definition}[Gevrey class]
Let \(s\ge 1\), and let \(I\subset\mathbb R\) be an interval. Suppose that \(f\in C^\infty(I)\). If there exist constants \(A,B>0\) such that, for every integer \(n\ge0\),
\begin{equation}
    \sup_{x\in I}\abs{f^{(n)}(x)} \le A B^{\,n}(n!)^s,
\end{equation}
then \(f\) is said to satisfy an order-\(s\) Gevrey derivative bound on \(I\). We also write \(f\in G^s(I)\) for such a function.
\end{definition}
The constants \(A,B\) may depend on the function \(f\), the order \(s\), and the interval \(I\), but they are independent of the derivative order \(n\).

\subsubsection{Gevrey derivative bound for \(\xi_\delta(t)\)}
\begin{lemma}[Gevrey derivative bound for \(\xi_\delta(t)\)]\label{lem:xi-gevrey}
Fix \(\delta>0\), set \(\alpha:=1/\delta\), and define
\(
\xi_\delta(t)=
\begin{cases}
e^{-t^{-\alpha}}, & t>0,\\
0, & t\le 0.
\end{cases}
\)
Then \(\xi_\delta\in C^\infty(\mathbb R)\), and \(\xi_\delta\) belongs to the Gevrey class of order \(1+\delta\). More precisely, for every integer \(n\ge0\),
\begin{equation}\label{d_xi_up}
    \sup_{t\in\mathbb R}\abs{\xi_\delta^{(n)}(t)}
\le A_{\delta,\xi}B_{\delta,\xi}^{\,n}(n!)^{1+\delta}.
\end{equation}
Here \(A_{\delta,\xi}=1,\; B_{\delta,\xi}=(1+2\alpha)e^{1+2\delta}(1+\delta)^\delta.\)

\end{lemma}

\begin{proof}
The proof has two parts. We first show that \(\xi_\delta\in C^\infty(\mathbb R)\). We then estimate its derivatives and derive Eq.~\eqref{d_xi_up}.

We begin with the case \(t>0\), where \(\xi_\delta(t)=e^{-t^{-\alpha}}\). We first prove by induction that, for each integer \(n\ge1\), there exist real coefficients \(c_{n,1},\ldots,c_{n,n}\) such that
\begin{equation}\label{d_xi_n}
    \xi_\delta^{(n)}(t)=e^{-t^{-\alpha}}\sum_{\ell=1}^{n}c_{n,\ell}t^{-n-\alpha\ell},
\end{equation}
and these coefficients satisfy the recurrence relation
\(c_{1,1}=\alpha,\; c_{n+1,\ell}=-(n+\alpha\ell)c_{n,\ell}+\alpha c_{n,\ell-1},\)
with the convention \(c_{n,0}=c_{n,n+1}=0\).

For the initial case \(n=1\), since \(\frac{\dd}{\dd t}(-t^{-\alpha})=\alpha t^{-\alpha-1},\) we have
\(\xi_\delta'(t)=e^{-t^{-\alpha}}\alpha t^{-\alpha-1}.\)
This is precisely the form of Eq.~\eqref{d_xi_n} for \(n=1\), with \(c_{1,1}=\alpha\). Suppose now that the claim holds for some \(n\ge1\). Differentiating both sides of Eq.~\eqref{d_xi_n} and using the product rule gives
\begin{equation}
    \begin{split}
    \xi_\delta^{(n+1)}(t)
    &=\frac{\dd}{\dd t}\left(e^{-t^{-\alpha}}    \sum_{\ell=1}^{n}c_{n,\ell}t^{-n-\alpha\ell}\right)\\
    &=e^{-t^{-\alpha}}\sum_{\ell=1}^{n}\alpha c_{n,\ell}t^{-n-1-\alpha(\ell+1)}
    +e^{-t^{-\alpha}}\sum_{\ell=1}^{n}-(n+\alpha\ell)c_{n,\ell}t^{-n-1-\alpha\ell}.
    \end{split}
\end{equation}
The exponent in the first sum is \(-n-1-\alpha(\ell+1)\). Reindexing that sum by replacing \(\ell+1\) with the new summation index, and using the convention \(c_{n,0}=c_{n,n+1}=0\), the preceding expression can be written uniformly as
\begin{equation}
    \xi_\delta^{(n+1)}(t)
    =e^{-t^{-\alpha}}\sum_{\ell=1}^{n+1}\bigl(-(n+\alpha\ell)c_{n,\ell}+\alpha c_{n,\ell-1}\bigr)t^{-(n+1)-\alpha\ell}.
\end{equation}
Thus \(\xi_\delta^{(n+1)}(t)\) has the same structural form, with the updated coefficients satisfying
\(c_{n+1,\ell}=-(n+\alpha\ell)c_{n,\ell}+\alpha c_{n,\ell-1}.\)
The induction is complete.

We next use Eq.~\eqref{d_xi_n} to prove that \(\xi_\delta\in C^\infty(\mathbb R)\). For \(t<0\), \(\xi_\delta(t)\equiv0\), so all derivatives exist and vanish on the negative half-line. For \(t>0\), the function \(\xi_\delta(t)=e^{-t^{-\alpha}}\) is clearly \(C^\infty\). It remains only to check smoothness at \(t=0\).
For \(n=0\), as \(t\to0^+\), one has \(t^{-\alpha}\to+\infty\), and hence \(\xi_\delta(t)=e^{-t^{-\alpha}}\to0\). Since \(\xi_\delta(t)=0\) for \(t\le0\), the function is continuous at \(0\), with \(\xi_\delta(0)=0\).
We now argue inductively. Suppose that for some \(n\ge0\), \(\xi_\delta^{(n)}(0)=0\) and \(\xi_\delta^{(n)}\) is continuous at \(0\). We prove that \(\xi_\delta^{(n+1)}(0)\) exists, is equal to \(0\), and is continuous at \(0\). By the definition of the derivative,
\begin{equation}
    \xi_\delta^{(n+1)}(0)=\lim_{h\to0}\frac{\xi_\delta^{(n)}(h)-\xi_\delta^{(n)}(0)}{h}=\lim_{h\to0} \frac{\xi_\delta^{(n)}(h)}{h}.
\end{equation}
For \(h<0\), \(\xi_\delta^{(n)}(h)=0\), so the left limit is \(0\). For \(h>0\), if \(n=0\), then \(\xi_\delta(h)=e^{-h^{-\alpha}}\). If \(n\ge1\), Eq.~\eqref{d_xi_n} shows that \(\xi_\delta^{(n)}(h)\) is a finite linear combination of terms of the form \(e^{-h^{-\alpha}}h^{-M}\). In either case, \(\xi_\delta^{(n)}(h)/h\) is a finite linear combination of terms of the form \(e^{-h^{-\alpha}}h^{-M'}\). Setting \(y=h^{-\alpha}\), we have \(h^{-M'}=y^{M'/\alpha}\). As \(h\to0^+\), \(y\to+\infty\), and \(y^{M'/\alpha}e^{-y}\to0\). Therefore the right limit is also \(0\). This proves that \(\xi_\delta^{(n+1)}(0)\) exists and equals \(0\).

It remains to verify continuity. On the left side, \(t<0\), one has \(\xi_\delta^{(n+1)}(t)=0\). On the right side, \(t>0\), applying Eq.~\eqref{d_xi_n} to the \((n+1)\)-st derivative shows that \(\xi_\delta^{(n+1)}(t)\) is a finite linear combination of terms of the form \(e^{-t^{-\alpha}}t^{-M}\). Again setting \(y=t^{-\alpha}\), each such term becomes \(e^{-y}y^{M/\alpha}\), which tends to \(0\) as \(t\to0^+\). Hence
\(\lim_{t\to0^+}\xi_\delta^{(n+1)}(t)=0=\xi_\delta^{(n+1)}(0)\), so \(\xi_\delta^{(n+1)}\) is continuous at \(0\). The induction is complete.
Thus, for every integer \(n\ge0\), the derivative \(\xi_\delta^{(n)}\) exists, is continuous, and satisfies \(\xi_\delta^{(n)}(0)=0\). Combining this with smoothness on the positive and negative half-lines gives \(\xi_\delta\in C^\infty(\mathbb R)\).

We now prove the Gevrey derivative bound. We first estimate the coefficients in Eq.~\eqref{d_xi_n}. Define
\(D_n:=\max_{1\le\ell\le n}\frac{\ell!\abs{c_{n,\ell}}}{n!}.\)
Using the recurrence relation, for any \(1\le\ell\le n+1\) we have
\begin{equation}
    \begin{aligned}
    \frac{\ell!\abs{c_{n+1,\ell}}}{(n+1)!}
    &\le\frac{\ell!(n+\alpha\ell)\abs{c_{n,\ell}}}{(n+1)!}
    +\frac{\alpha\ell!\abs{c_{n,\ell-1}}}{(n+1)!}\\
    &=\frac{n+\alpha\ell}{n+1}\frac{\ell!\abs{c_{n,\ell}}}{n!}
    +\frac{\alpha\ell}{n+1}\frac{(\ell-1)!\abs{c_{n,\ell-1}}}{n!}.
    \end{aligned}
\end{equation}
If \(\ell=1\) or \(\ell=n+1\), the corresponding boundary coefficient is interpreted as zero according to the convention above. Since \(1\le\ell\le n+1\), we have
\(\frac{n+\alpha\ell}{n+1}\le 1+\alpha\) and \(\frac{\alpha\ell}{n+1}\le \alpha.\)
Therefore,
\begin{equation}
    \frac{\ell!\abs{c_{n+1,\ell}}}{(n+1)!}\le(1+\alpha)D_n+\alpha D_n=(1+2\alpha)D_n.
\end{equation}
Taking the maximum over \(\ell\) gives \(D_{n+1}\le (1+2\alpha)D_n.\) Since \(D_1=\alpha\), it follows that
\(D_n\le \alpha(1+2\alpha)^{n-1}.\)
Consequently, for all \(1\le \ell\le n\),
\begin{equation}
    \abs{c_{n,\ell}}\le\alpha(1+2\alpha)^{n-1}\frac{n!}{\ell!}.
\end{equation}
Substituting this estimate into the structural formula for the higher derivatives, for \(t>0\) we obtain
\begin{equation}
\begin{split}
    \abs{\xi_\delta^{(n)}(t)} 
    &\le\alpha(1+2\alpha)^{n-1}n! e^{-t^{-\alpha}} \sum_{\ell=1}^{n}\frac{t^{-n-\alpha\ell}}{\ell!}\\
    &=\alpha(1+2\alpha)^{n-1}n! e^{-y} \sum_{\ell=1}^{n}\frac{y^{n\delta+\ell}}{\ell!}.
\end{split}
\end{equation}
In the second line we used the change of variables \(y:=t^{-\alpha}\) and the identity \(\alpha=1/\delta\). We now estimate the supremum of the right-hand side with respect to \(y\). For fixed \(\ell\), the function \(y\longmapsto y^{n\delta+\ell}e^{-y}\) attains its maximum at \(y=n\delta+\ell\). Hence
\(\sup_{y>0}y^{n\delta+\ell}e^{-y} = (n\delta+\ell)^{n\delta+\ell}e^{-(n\delta+\ell)}.\)
Since \(1\le \ell\le n\), we have \(n\delta+\ell\le (1+\delta)n\). Using also \(e^{-(n\delta+\ell)}\le1\), we get
\((n\delta+\ell)^{n\delta+\ell}e^{-(n\delta+\ell)} \le \bigl((1+\delta)n\bigr)^{n\delta} \bigl((1+\delta)n\bigr)^\ell.\)
Therefore,
\begin{equation}
    \sup_{t>0}\abs{\xi_\delta^{(n)}(t)} \le
    \alpha(1+2\alpha)^{n-1}n! \bigl((1+\delta)n\bigr)^{n\delta} \sum_{\ell=1}^{n}\frac{\bigl((1+\delta)n\bigr)^\ell}{\ell!}.
\end{equation}
We bound the finite sum by the exponential series:
\(\sum_{\ell=1}^{n}\frac{\bigl((1+\delta)n\bigr)^\ell}{\ell!} \le \sum_{\ell=0}^{\infty}\frac{\bigl((1+\delta)n\bigr)^\ell}{\ell!} = e^{(1+\delta)n}.\)
Moreover, the lower bound \(n!\ge (n/e)^n\) implies
\(n^{n\delta}\le e^{\delta n}(n!)^\delta. \)
Thus
\begin{equation}
\begin{split}
    \sup_{t>0}\abs{\xi_\delta^{(n)}(t)}
    &\le \alpha(1+2\alpha)^{n-1} e^{(1+2\delta)n}(1+\delta)^{\delta n} (n!)^{1+\delta}\\
    &= \frac{\alpha}{1+2\alpha} \left((1+2\alpha)e^{1+2\delta}(1+\delta)^\delta\right)^n (n!)^{1+\delta}\\
    &\le A_{\delta,\xi}B_{\delta,\xi}^{\,n}(n!)^{1+\delta}.
\end{split}
\end{equation}
Here
\(A_{\delta,\xi}:=1\ge \frac{\alpha}{1+2\alpha},\; B_{\delta,\xi}:=(1+2\alpha)e^{1+2\delta}(1+\delta)^\delta.\)
Therefore, for \(n\ge1\),
\(\sup_{t>0}\abs{\xi_\delta^{(n)}(t)} \le A_{\delta,\xi}B_{\delta,\xi}^{\,n}(n!)^{1+\delta}. \)
For \(n=0\), the definition gives \(0\le\xi_\delta(t)\le1\), and hence
\(\sup_{t\in\mathbb R}\abs{\xi_\delta(t)}
\le1= A_{\delta,\xi}.\)

Finally, all derivatives vanish for \(t<0\), and all derivatives at \(t=0\) also vanish. Since the derivative estimate has been established on \(t>0\), Eq.~\eqref{d_xi_up} holds on the entire real line. This proves the lemma.
\end{proof}

\subsubsection{Gevrey derivative bounds for \(q_\delta(\theta)\)}
\begin{lemma}\label{lem:q-gevrey-AB}
Fix a real polynomial \(p(x)=\sum_{r=0}^{R}a_rx^r\) of degree \(R\ge2\). Let \(L>0\), set \(\tau:=\frac{\pi}{4L}\) and \(S_L:=\sum_{r=0}^{R}|a_r|L^r\), and define \(\widetilde q_\delta(\theta):=\chi_\delta(\theta)p(\theta/\tau)\) on \([-\pi,\pi]\). For any fixed \(\delta>0\), let \(q_\delta\) be the \(2\pi\)-periodic extension of \(\widetilde q_\delta\). Then \(q_\delta\) belongs to the Gevrey class of order \(1+\delta\). Equivalently, there exist constants \(A_\delta,B_\delta>0\), depending only on \(\delta\) and the fixed degree \(R\), such that for every integer \(n\ge0\),
\begin{equation}\label{q_delta_gevrey_appendix}
    \sup_{\theta\in\mathbb R}\abs{q_\delta^{(n)}(\theta)}
    \le S_LA_\delta B_\delta^{\,n}(n!)^{1+\delta}.
\end{equation}
Moreover, there exists a constant \(C_0>0\), which may depend on the fixed degree \(R\) but is independent of \(L\), \(\varepsilon\), \(\delta\), and the coefficients \(a_r\), such that, for all sufficiently small \(\delta>0\),
\begin{equation}
    A_\delta B_\delta^2 \le \exp\!\left(C_0\exp\!\left(\frac{\log\frac{32}{3\pi^2}}{\delta}\right)\right).
\end{equation}
In particular, for sufficiently large \(X\), if \(\delta=\frac{2\log\frac{32}{3\pi^2}}{\log\log X}\), where \(X:=e^e+S_L+\log\frac1\varepsilon\), with \(0<\varepsilon<1\) and \(L>0\), then \(A_\delta B_\delta^2 \le \exp\!\left(C_0\sqrt{\log X}\right)\).
\end{lemma}

\begin{proof}
We first prove the Gevrey derivative bound for \(q_\delta\). By definition, on the interval \([-\pi,\pi]\) we have \(\widetilde q_\delta(\theta)=\chi_\delta(\theta)p(\theta/\tau)\). Here \(\chi_\delta(\theta)=\frac{a_\delta(\theta)}{a_\delta(\theta)+b_\delta(\theta)}\), where \(a_\delta(\theta):=\xi_\delta\!\left(\frac{\pi^2}{4}-\theta^2\right)\) and \(b_\delta(\theta):=\xi_\delta\!\left(\theta^2-\frac{\pi^2}{16}\right)\). Set \(w_\delta(\theta):=a_\delta(\theta)+b_\delta(\theta)\) and \(m_\delta:=\min_{\theta\in[-\pi,\pi]}w_\delta(\theta)\). From the definitions of \(a_\delta\) and \(b_\delta\), at least one of them is strictly positive at every point of \([-\pi,\pi]\). Hence \(m_\delta>0\).

By Lemma~\ref{lem:xi-gevrey}, the function \(\xi_\delta\) satisfies
\begin{equation}
    \sup_{t\in\mathbb R}\abs{\xi_\delta^{(n)}(t)}
    \le A_{\delta,\xi}B_{\delta,\xi}^{\,n}(n!)^{1+\delta},
    \qquad
    A_{\delta,\xi}=1,
    \qquad
    B_{\delta,\xi}=(1+2\alpha)e^{1+2\delta}(1+\delta)^\delta.
\end{equation}
We now propagate this derivative bound to \(q_\delta\).

We first treat the quadratic compositions \(a_\delta\) and \(b_\delta\). The two estimates are identical, so it suffices to consider \(a_\delta\). Since the inner function \(\frac{\pi^2}{4}-\theta^2\) is a quadratic polynomial, all derivatives of order three and higher vanish. Thus the Faà di Bruno formula reduces here to a finite sum:
\begin{equation}
    \begin{split}
    a_\delta^{(n)}(\theta)
    &= \sum_{k=\lceil n/2\rceil}^{n}\frac{n!}{(2k-n)!(n-k)!}(-2\theta)^{2k-n}(-1)^{n-k}\xi_\delta^{(k)}\!\left(\frac{\pi^2}{4}-\theta^2\right)\\
    \abs{a_\delta^{(n)}(\theta)}
    &\le \sum_{k=\lceil n/2\rceil}^{n}\frac{n!}{(2k-n)!(n-k)!}(2\pi)^nA_{\delta,\xi}B_{\delta,\xi}^{\,k}(k!)^{1+\delta}.
    \end{split}
\end{equation}
In the inequality we used \(|-2\theta|\le2\pi\) for \(\theta\in[-\pi,\pi]\). Let \(u:=2k-n\) and \(v:=n-k\). Then \(u,v\ge0\), and \(u+v=k\). Hence \(\frac{n!}{(2k-n)!(n-k)!}(k!)^{1+\delta}=\frac{n!}{u!v!}(k!)^{1+\delta}=n!\frac{k!}{u!v!}(k!)^\delta\). Since \(\frac{k!}{u!v!}=\binom{k}{v}\le2^k\le2^n\), and \(k\le n\) implies \((k!)^\delta\le(n!)^\delta\), we have \(\frac{n!}{(2k-n)!(n-k)!}(k!)^{1+\delta}\le2^n(n!)^{1+\delta}\). Moreover, after increasing \(B_{\delta,\xi}\) if necessary, we may assume \(B_{\delta,\xi}\ge1\), so that \(B_{\delta,\xi}^{\,k}\le B_{\delta,\xi}^{\,n}\). The number of summands is at most \(n+1\), and \(n+1\le2^n\) for all \(n\ge0\). Therefore,
\begin{equation}
    \sup_{\theta\in[-\pi,\pi]}\abs{a_\delta^{(n)}(\theta)} \le A_{\delta,\xi}(8\pi B_{\delta,\xi})^n(n!)^{1+\delta}:=A_{\delta,a}(B_{\delta,a})^n(n!)^{1+\delta}.
\end{equation}
Here \(A_{\delta,a}=1\) and \(B_{\delta,a}=8\pi B_{\delta,\xi}\). The same argument gives \(\sup_{\theta\in[-\pi,\pi]}\abs{b_\delta^{(n)}(\theta)}\le A_{\delta,a}(B_{\delta,a})^n(n!)^{1+\delta}\). Thus \(a_\delta\) and \(b_\delta\) satisfy the same Gevrey derivative bound. For \(w_\delta=a_\delta+b_\delta\), the triangle inequality gives
\[
\sup_{\theta\in[-\pi,\pi]}\abs{w_\delta^{(n)}(\theta)}
\le
A_{\delta,w}B_{\delta,w}^{\,n}(n!)^{1+\delta},
\qquad
A_{\delta,w}:=2,
\qquad
B_{\delta,w}:=8\pi B_{\delta,\xi}.
\]

We next estimate the reciprocal. Let \(v_\delta(\theta):=\frac1{w_\delta(\theta)}\). Then \(\abs{v_\delta(\theta)}\le m_\delta^{-1}\). Differentiating the identity \(w_\delta(\theta)v_\delta(\theta)\equiv1\) \(n\) times, for \(n\ge1\), the Leibniz rule gives \(\sum_{k=0}^{n}\binom{n}{k}w_\delta^{(k)}(\theta)v_\delta^{(n-k)}(\theta)=0\). Moving the \(k=0\) term to the left-hand side and dividing by \(w_\delta(\theta)\), we obtain the recurrence
\begin{equation}
    v_\delta^{(n)}(\theta) = -\frac1{w_\delta(\theta)} \sum_{k=1}^{n}\binom{n}{k}w_\delta^{(k)}(\theta)v_\delta^{(n-k)}(\theta).
\end{equation}
Set \(A_{\delta,v}:=m_\delta^{-1}\) and \(B_{\delta,v}:=B_{\delta,w}\left(1+\frac{A_{\delta,w}}{m_\delta}\right)\). We show that these constants control the derivatives of \(v_\delta\). For \(n=0\), the bound follows immediately from \(\abs{v_\delta(\theta)}\le m_\delta^{-1}\). Suppose that the bound has been proved for all \(j<n\). The recurrence above yields
\begin{equation}
    \sup_\theta\abs{v_\delta^{(n)}(\theta)} \le \frac{A_{\delta,w}A_{\delta,v}}{m_\delta} \sum_{k=1}^{n} \binom{n}{k} B_{\delta,w}^{\,k}B_{\delta,v}^{\,n-k} (k!)^{1+\delta}((n-k)!)^{1+\delta}.
\end{equation}
Since \(\binom{n}{k}(k!)^{1+\delta}((n-k)!)^{1+\delta}=n!(k!)^\delta((n-k)!)^\delta\le(n!)^{1+\delta}\), we obtain
\begin{equation}
    \sup_\theta\abs{v_\delta^{(n)}(\theta)}
    \le \frac{A_{\delta,w}A_{\delta,v}}{m_\delta}(n!)^{1+\delta}\sum_{k=1}^{n}B_{\delta,w}^{\,k}B_{\delta,v}^{\,n-k}.
\end{equation}
Since \(B_{\delta,v}=B_{\delta,w}(1+A_{\delta,w}/m_\delta)>B_{\delta,w}\), we have \(\sum_{k=1}^{n}B_{\delta,w}^{\,k}B_{\delta,v}^{\,n-k}\le B_{\delta,v}^{\,n}\frac{B_{\delta,w}}{B_{\delta,v}-B_{\delta,w}}\). Moreover, \(B_{\delta,v}-B_{\delta,w}=B_{\delta,w}A_{\delta,w}/m_\delta\), and hence \(\frac{A_{\delta,w}A_{\delta,v}}{m_\delta}\frac{B_{\delta,w}}{B_{\delta,v}-B_{\delta,w}}=A_{\delta,v}\). Therefore,
\begin{equation}
    \sup_{\theta\in[-\pi,\pi]}\abs{v_\delta^{(n)}(\theta)} \le A_{\delta,v}B_{\delta,v}^{\,n}(n!)^{1+\delta}.
\end{equation}
The induction is complete.

We now estimate \(\chi_\delta=a_\delta v_\delta\). By the Leibniz rule, \(\chi_\delta^{(n)}(\theta)=\sum_{k=0}^{n}\binom{n}{k}a_\delta^{(k)}(\theta)v_\delta^{(n-k)}(\theta)\). Using the derivative bounds for \(a_\delta\) and \(v_\delta\), we obtain
\begin{equation}
\begin{split}
    \sup_{\theta\in[-\pi,\pi]}\abs{\chi_\delta^{(n)}(\theta)}
    &\le A_{\delta,a}A_{\delta,v} \sum_{k=0}^{n} \binom{n}{k} B_{\delta,a}^{\,k}B_{\delta,v}^{\,n-k} (k!)^{1+\delta}((n-k)!)^{1+\delta}\\
    &\le A_{\delta,a}A_{\delta,v}(n!)^{1+\delta} \sum_{k=0}^{n}B_{\delta,a}^{\,k}B_{\delta,v}^{\,n-k}.
\end{split}
\end{equation}
Here the second inequality uses \(\binom{n}{k}(k!)^{1+\delta}((n-k)!)^{1+\delta}\le(n!)^{1+\delta}\). Since \(B_{\delta,v}\ge B_{\delta,a}\), each term in the last sum is bounded by \(B_{\delta,v}^n\). The number of terms is \(n+1\le2^n\), and therefore
\begin{equation}
    \sup_{\theta\in[-\pi,\pi]}\abs{\chi_\delta^{(n)}(\theta)} \le A_{\delta,a}A_{\delta,v}(2B_{\delta,v})^n(n!)^{1+\delta}.
\end{equation}
Thus we may take \(A_{\delta,\chi}=A_{\delta,a}A_{\delta,v}=m_\delta^{-1}\) and \(B_{\delta,\chi}=2B_{\delta,v}=16\pi B_{\delta,\xi}\left(1+\frac2{m_\delta}\right)\).

It remains to estimate \(\widetilde q_\delta(\theta)=\chi_\delta(\theta)r_L(\theta)\), where \(r_L(\theta):=p(\theta/\tau)=\sum_{r=0}^{R}a_rL^r\left(\frac{4\theta}{\pi}\right)^r\). For \(0\le k\le R\), one has \(r_L^{(k)}(\theta)=\sum_{r=k}^{R}a_rL^r\left(\frac4\pi\right)^r\frac{r!}{(r-k)!}\theta^{r-k}\). Hence \(\sup_{\theta\in[-\pi,\pi]}\abs{r_L^{(k)}(\theta)}\le4^RR!S_L\), while \(r_L^{(k)}=0\) for \(k>R\). By the Leibniz rule, \(\widetilde q_\delta^{(n)}(\theta)=\sum_{k=0}^{\min\{n,R\}}\binom{n}{k}\chi_\delta^{(n-k)}(\theta)r_L^{(k)}(\theta)\). Hence
\begin{equation}
    \sup_{\theta\in[-\pi,\pi]}\abs{\widetilde q_\delta^{(n)}(\theta)}
    \le S_LA_{\delta,\chi}4^RR! \sum_{k=0}^{\min\{n,R\}} \binom{n}{k} B_{\delta,\chi}^{\,n-k}((n-k)!)^{1+\delta}.
\end{equation}
For every \(0\le k\le\min\{n,R\}\), we have \(\binom{n}{k}((n-k)!)^{1+\delta}=\frac{n!}{k!}((n-k)!)^\delta\le\frac{(n!)^{1+\delta}}{k!}\). Since \(B_{\delta,\chi}>1\), one also has \(B_{\delta,\chi}^{\,n-k}\le B_{\delta,\chi}^{\,n}\). Summing over \(0\le k\le\min\{n,R\}\) and extending the resulting finite positive sum to \(k=R\), we obtain
\begin{equation}
    \sup_{\theta\in[-\pi,\pi]}\abs{\widetilde q_\delta^{(n)}(\theta)}
    \le S_LA_{\delta,\chi}4^RR!\left(\sum_{k=0}^{R}\frac1{k!}\right)B_{\delta,\chi}^{\,n}(n!)^{1+\delta}.
\end{equation}
Since \(\widetilde q_\delta\) is identically zero for \(|\theta|\ge\pi/2\), all of its derivatives vanish in neighborhoods of \(\theta=\pm\pi\). Its \(2\pi\)-periodic extension therefore introduces no loss of smoothness at the seams. On each period, the extended function \(q_\delta\) is a translate of \(\widetilde q_\delta\), so the same derivative bound holds on the entire real line. Thus we may take \(A_\delta=A_{\delta,\chi}4^RR!\sum_{k=0}^{R}\frac1{k!}\) and \(B_\delta=B_{\delta,\chi}\). Substituting \(A_{\delta,\chi}\) and \(B_{\delta,\chi}\), we obtain
\begin{equation}\label{AB_delta}
    A_\delta = 4^RR!\left(\sum_{k=0}^{R}\frac1{k!}\right)m_\delta^{-1}, \qquad
    B_\delta = 16\pi B_{\delta,\xi}\left(1+\frac2{m_\delta}\right).
\end{equation}
This proves the Gevrey derivative bound for \(q_\delta\).

We now derive an upper bound on \(A_\delta B_\delta^2\). From Eq.~\eqref{AB_delta}, we have
\begin{equation}
    A_\delta B_\delta^2 = 4^RR!\left(\sum_{k=0}^{R}\frac1{k!}\right)(16\pi)^2B_{\delta,\xi}^2m_\delta^{-1}\left(1+\frac2{m_\delta}\right)^2.
\end{equation}
Since \(0\le\xi_\delta\le1\), we have \(0<w_\delta(\theta)\le2\), and hence \(m_\delta\le2\). Thus \(2/m_\delta\ge1\), which implies \(1+\frac2{m_\delta}\le\frac4{m_\delta}\). Consequently, there exists a constant \(C_0>0\), which may depend on the fixed degree \(R\) but is independent of \(L\), \(\varepsilon\), \(\delta\), and the coefficients \(a_r\), such that
\begin{equation}
    A_\delta B_\delta^2 \le C_0B_{\delta,\xi}^2m_\delta^{-3}.
\end{equation}
We next estimate \(B_{\delta,\xi}\). For sufficiently small \(\delta>0\), we have \(1+2\alpha=1+\frac2\delta\le\frac3\delta\), \(e^{1+2\delta}\le e^2\), and \((1+\delta)^\delta\le2\). Therefore, \(B_{\delta,\xi}=(1+2\alpha)e^{1+2\delta}(1+\delta)^\delta\le\frac6\delta e^2\). After enlarging the constant \(C_0\), still independently of \(L\), \(\varepsilon\), \(\delta\), and the coefficients \(a_r\), we may write \(B_{\delta,\xi}\le C_0\delta^{-1}\). Substituting this into the previous estimate gives
\begin{equation}
    A_\delta B_\delta^2 \le C_0\delta^{-2}m_\delta^{-3}.
\end{equation}
It remains to bound \(m_\delta^{-1}\). For every \(\theta\in[-\pi,\pi]\), \(\left(\frac{\pi^2}{4}-\theta^2\right)+\left(\theta^2-\frac{\pi^2}{16}\right)=\frac{3\pi^2}{16}\). Hence at least one of the two terms is not smaller than \(\frac{3\pi^2}{32}\). Since \(\xi_\delta(t)=e^{-t^{-1/\delta}}\) is increasing for \(t>0\), it follows that \(w_\delta(\theta)\ge\exp\!\left[-\left(\frac{32}{3\pi^2}\right)^{1/\delta}\right]\). Thus \(m_\delta^{-1}\le\exp\!\left[\left(\frac{32}{3\pi^2}\right)^{1/\delta}\right]=\exp\!\left[\exp\!\left(\frac{\log\frac{32}{3\pi^2}}{\delta}\right)\right]\). Substituting this bound, we obtain
\begin{equation}
    A_\delta B_\delta^2
    \le C_0\delta^{-2}\exp\!\left[3\exp\!\left(\frac{\log\frac{32}{3\pi^2}}{\delta}\right)\right].
\end{equation}
The factor \(\delta^{-2}\) grows much more slowly than the double-exponential factor on the right-hand side. Hence, by enlarging the constant if necessary, it can be absorbed into the exponent. Therefore, for all sufficiently small \(\delta>0\),
\begin{equation}
    A_\delta B_\delta^2
    \le \exp\!\left(C_0\exp\!\left(\frac{\log\frac{32}{3\pi^2}}{\delta}\right)\right).
\end{equation}

Finally, substitute the special choice \(\delta=\frac{2\log\frac{32}{3\pi^2}}{\log\log X}\). Then \(\exp\!\left(\frac{\log\frac{32}{3\pi^2}}{\delta}\right)=\exp\!\left(\frac12\log\log X\right)=\sqrt{\log X}\). Hence, for sufficiently large \(X\),
\begin{equation}
    A_\delta B_\delta^2 \le \exp\!\left(C_0\sqrt{\log X}\right).
\end{equation}
This proves the lemma.
\end{proof}

\subsection{Windowed Laurent approximation}
\label{app:bounded-laurent-approximation}

This subsection establishes a bounded Laurent approximation to \(e^{ip(x)}\) on a finite quadrature window, with the unit-circle constraint needed for the GQSP realization in Sec.~II.D.

\begin{theorem}
\label{thm:polynomial-laurent-approx}
Let \(p(x)=\sum_{r=0}^{R}a_rx^r\) be an arbitrary real polynomial of finite degree \(R\ge2\), with \(a_R\ne0\), and define \(S_L:=\sum_{r=0}^{R}|a_r|L^r\). For any \(0<\varepsilon<1\) and \(L>0\), set \(\tau=\frac{\pi}{4L}\). Then there exists a Laurent polynomial \(P_d(z)=\sum_{|m|\le d}p_mz^m\) such that
\begin{equation}
\sup_{|x|\le L}\left|P_d(e^{i\tau x})-\e^{ip(x)}\right|<\varepsilon,
\qquad
\sup_{|z|=1}|P_d(z)|\le 1,
\end{equation}
and the bandwidth can be chosen as \(d=O\!\left(\left(S_L+\log\frac1\varepsilon\right)^{1+o(1)}\right)\). Here \(o(1)\) denotes a quantity that tends to \(0\) as \(S_L+\log\frac1\varepsilon\to\infty\), with the degree \(R\) fixed. 
\end{theorem}

\begin{proof}
The proof is organized into three steps. First, for a fixed \(\delta>0\), we prove that the Fourier coefficients of a periodic extension of the target phase decay at a stretched-exponential rate. Second, we estimate the Laurent truncation error \(\eta_d\), and show that a normalized Laurent polynomial satisfies both the unit-circle constraint and the desired approximation bound on \([-L,L]\). Third, we choose \(\delta\) as a function of \(S_L\) and \(\varepsilon\), which gives the stated upper bound on the bandwidth \(d\).

Let \(q_\delta\) be the \(2\pi\)-periodic function constructed in Lemma~\ref{lem:q-gevrey-AB} from \(\widetilde q_\delta(\theta)=\chi_\delta(\theta)p(\theta/\tau)\) on \([-\pi,\pi]\), and define \(F_\delta(\theta):=\e^{iq_\delta(\theta)}\). Since \(\chi_\delta(\theta)=1\) for \(|\theta|\le\pi/4\), whenever \(|x|\le L\) one has \(|\tau x|\le\pi/4\) and therefore \(q_\delta(\tau x)=p(x)\). Consequently, \(F_\delta(\tau x)=\e^{ip(x)}\) on the target window.
The windowing construction is illustrated in
Fig.~\ref{fig:smooth-window-example} for the cubic phase as a representative example.
\begin{figure}[t]
\centering
\begin{tikzpicture}
\begin{groupplot}[
group style={group size=2 by 1, horizontal sep=2.4cm},
width=0.40\columnwidth,
height=5.0cm,
xmin=-3.1416,
xmax=3.1416,
samples=300,
domain=-3.1415:3.1415,
axis lines=left,
clip=false,
every axis plot/.append style={very thick},
xtick={-3.14159265,-1.57079633,0,1.57079633,3.14159265},
xticklabels={$-\pi$,$-\pi/2$,$0$,$\pi/2$,$\pi$},
ticklabel style={font=\scriptsize},
label style={font=\small},
legend style={font=\scriptsize,draw=none,fill=none},
legend cell align={left}
]

% ---------- (a) ----------
\nextgroupplot[
ymin=-0.08,
ymax=1.08,
xlabel={$\theta$},
legend style={
at={(0.68,0.93)},
anchor=north west,
font=\scriptsize,
draw=none,
fill=none
}
]

\addplot[black,solid]
{((abs(x)>=pi/2) ? 0 : exp(-1/(pi*pi/4 - x^2)))
/
(((abs(x)>=pi/2) ? 0 : exp(-1/(pi*pi/4 - x^2)))
+
((abs(x)<=pi/4) ? 0 : exp(-1/(x^2 - pi*pi/16))))};
\addlegendentry{$\chi_\delta$}

\addplot[blue,densely dashed]
{(abs(x)>=pi/2) ? 0 : exp(-1/(pi*pi/4 - x^2))};
\addlegendentry{$\xi_\delta\!\left(\frac{\pi^2}{4}-\theta^2\right)$}

\addplot[red,dotted]
{(abs(x)<=pi/4) ? 0 : exp(-1/(x^2 - pi*pi/16))};
\addlegendentry{$\xi_\delta\!\left(\theta^2-\frac{\pi^2}{16}\right)$}

\draw[densely dashed] (axis cs:-0.78539816,-0.08) -- (axis cs:-0.78539816,1.05);
\draw[densely dashed] (axis cs:0.78539816,-0.08) -- (axis cs:0.78539816,1.05);
\draw[densely dashed] (axis cs:-1.57079633,-0.08) -- (axis cs:-1.57079633,1.05);
\draw[densely dashed] (axis cs:1.57079633,-0.08) -- (axis cs:1.57079633,1.05);

\node[anchor=north west,font=\bfseries\small]
at (rel axis cs:0.05,0.95) {(a)};

% ---------- (b) ----------
\nextgroupplot[
ymin=-1.2,
ymax=1.2,
xlabel={$\theta$}
]

\addplot[teal!70!black,solid]
{(((abs(x)>=pi/2) ? 0 : exp(-1/(pi*pi/4 - x^2)))
/
(((abs(x)>=pi/2) ? 0 : exp(-1/(pi*pi/4 - x^2)))
+
((abs(x)<=pi/4) ? 0 : exp(-1/(x^2 - pi*pi/16)))))
*(4*x/pi)^3};

\draw[densely dashed] (axis cs:-0.78539816,-1.2) -- (axis cs:-0.78539816,1.2);
\draw[densely dashed] (axis cs:0.78539816,-1.2) -- (axis cs:0.78539816,1.2);
\draw[densely dashed] (axis cs:-1.57079633,-1.2) -- (axis cs:-1.57079633,1.2);
\draw[densely dashed] (axis cs:1.57079633,-1.2) -- (axis cs:1.57079633,1.2);

\node[anchor=north west,font=\bfseries\small]
at (rel axis cs:0.05,0.95) {(b)};

\end{groupplot}
\end{tikzpicture}

\caption{
Smooth windowing used in the Laurent approximation.
(a) The cutoff function \(\chi_\delta(\theta)\) and the two switching functions from which it is constructed.
The cutoff equals unity for \(|\theta|\le\pi/4\) and vanishes for \(|\theta|\ge\pi/2\).
(b) The windowed phase
\(\widetilde q_\delta(\theta)=\chi_\delta(\theta)p(\theta/\tau)\)
for the representative cubic case \(p(x)=x^3\), with
\(L=1\), \(\tau=\pi/4\), and \(\delta=1\).
The proof applies to arbitrary fixed-degree real polynomials.
}
\label{fig:smooth-window-example}
\end{figure}
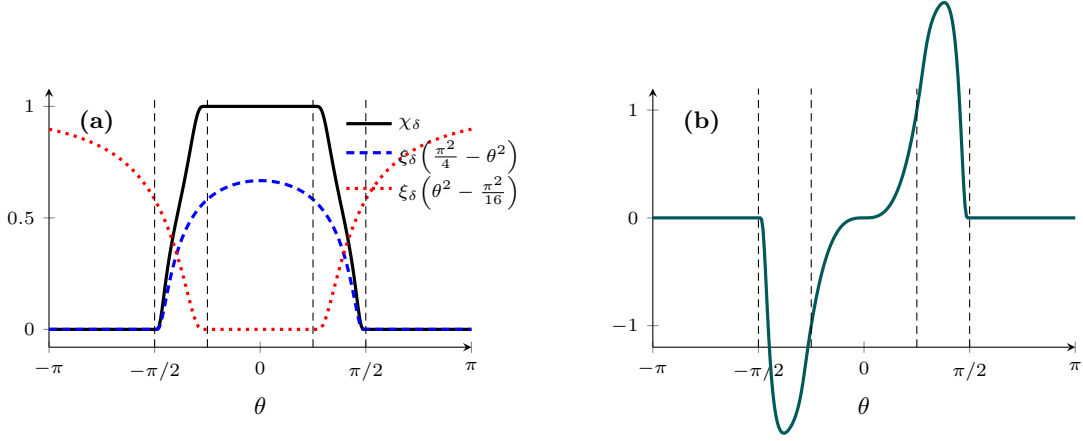

We begin by estimating the Fourier coefficients of \(q_\delta\), from which the corresponding estimate for the Fourier coefficients of \(F_\delta\) will be derived. Write \(q_\delta(\theta)=\sum_{m\in\mathbf Z}q_me^{im\theta}\), where \(q_m:=\frac{1}{2\pi}\int_{-\pi}^{\pi}q_\delta(\theta)\e^{-\ii m\theta}\,\dd\theta\). By Lemma~\ref{lem:q-gevrey-AB}, there exist constants \(A_\delta,B_\delta>0\), depending only on \(\delta\) and the fixed degree \(R\), such that for every integer \(N\ge0\),
\begin{equation}
\sup_{\theta\in\mathbb R}\abs{q_\delta^{(N)}(\theta)}
\le S_LA_\delta B_\delta^{\,N}(N!)^{1+\delta}.
\label{eq:q-gevrey-main}
\end{equation}
In particular, the case \(N=0\) gives \(\abs{q_0}\le\sup_{\theta\in\mathbb R}\abs{q_\delta(\theta)}\le S_LA_\delta\).

We now consider \(m\ne0\). Integrating the Fourier coefficient \(q_m\) by parts once gives
\begin{equation}
    q_m
    =\frac{1}{2\pi}\left[\frac{q_\delta(\theta)\e^{-\ii m\theta}}{-\ii m}\right]_{-\pi}^{\pi}
    -\frac{1}{2\pi}\int_{-\pi}^{\pi}\frac{q_\delta'(\theta)\e^{-\ii m\theta}}{-\ii m}\,\dd\theta.
\end{equation}
Since \(q_\delta\) is a \(C^\infty\), \(2\pi\)-periodic function, every derivative \(q_\delta^{(k)}\) is also \(2\pi\)-periodic. Moreover, \(\e^{-\ii m\pi}=\e^{\ii m\pi}=(-1)^m\). Hence every boundary term arising from repeated integration by parts vanishes. Repeating the argument \(N\) times, for any integer \(N\ge1\), gives
\begin{equation}
q_m=\frac{1}{2\pi(\ii m)^N}\int_{-\pi}^{\pi}q_\delta^{(N)}(\theta)\e^{-\ii m\theta}\,\dd\theta.
\label{eq:IBP}
\end{equation}
Taking absolute values and using Eq.~\eqref{eq:q-gevrey-main}, we obtain
\begin{equation}\label{upper_qm}
    \abs{q_m}\le\frac{\sup_{\theta\in\mathbb R}\abs{q_\delta^{(N)}(\theta)}}{\abs{m}^N}
    \le S_LA_\delta\frac{B_\delta^{\,N}(N!)^{1+\delta}}{\abs{m}^N}.
\end{equation}
Using \(N!\le N^N\), for every integer \(N\ge1\) this implies
\begin{equation}\label{eq:fourier-start2}
    \abs{q_m}\le S_LA_\delta\left(\frac{B_\delta N^{1+\delta}}{\abs{m}}\right)^N.
\end{equation}

Assume first that \(\abs{m}\ge2B_\delta\), and choose \(N:=\left\lfloor\left(\frac{\abs{m}}{2B_\delta}\right)^{1/(1+\delta)}\right\rfloor\). Then \(N\ge1\), \(N\le(\abs{m}/(2B_\delta))^{1/(1+\delta)}\), and hence \(\frac{B_\delta N^{1+\delta}}{\abs{m}}\le\frac12\). Substituting this choice into Eq.~\eqref{eq:fourier-start2} gives
\begin{equation}\label{qm_up2}
\begin{split}
    \abs{q_m}
    &\le S_LA_\delta2^{-N}
    \le2S_LA_\delta\cdot2^{-(\abs{m}/(2B_\delta))^{1/(1+\delta)}}\\
    &=2S_LA_\delta\exp\!\left(-\frac{\log2}{(2B_\delta)^{1/(1+\delta)}}\abs{m}^{1/(1+\delta)}\right)
    :=2S_LA_\delta\exp\!\left(-C_\delta\abs{m}^{1/(1+\delta)}\right).
\end{split}
\end{equation}
Here \(C_\delta:=\frac{\log2}{(2B_\delta)^{1/(1+\delta)}}\), and the second inequality uses \(N\ge(\abs{m}/(2B_\delta))^{1/(1+\delta)}-1\).

If \(0<\abs{m}<2B_\delta\), then the zeroth-order estimate gives \(\abs{q_m}\le\sup_{\theta\in\mathbb R}\abs{q_\delta(\theta)}\le S_LA_\delta\). Since \(C_\delta\abs{m}^{1/(1+\delta)}<\log2\), one has \(2\exp(-C_\delta\abs{m}^{1/(1+\delta)})>1\), and therefore
\(\abs{q_m}\le2S_LA_\delta\exp(-C_\delta\abs{m}^{1/(1+\delta)})\).
Thus Eq.~\eqref{qm_up2} holds for every \(m\ne0\).

We next transfer this decay estimate to the Fourier coefficients \(f_m\) of \(F_\delta\), defined by \(F_\delta(\theta)=\sum_{m\in\mathbf Z}f_me^{im\theta}\). For a periodic function \(g\) with Fourier coefficients \(g_m\), define
\begin{equation}
\|g\|_{1+\delta}:=
\sum_{m\in\mathbf Z}\abs{g_m}\e^{\frac12C_\delta\abs{m}^{1/(1+\delta)}}.
\label{eq:weighted-norm}
\end{equation}
It follows from Eq.~\eqref{qm_up2}, together with \(\abs{q_0}\le S_LA_\delta\), that
\(\|q_\delta\|_{1+\delta}\le S_LQ_\delta\),
where
\(Q_\delta:=2A_\delta\sum_{m\in\mathbf Z}\e^{-\frac12C_\delta\abs{m}^{1/(1+\delta)}}\).
In particular, \(Q_\delta\) depends only on \(\delta\) and \(R\), and not on \(L\) or the coefficients \(a_r\).

For any two periodic functions \(u,v\), this norm is submultiplicative. Indeed, the convolution formula for Fourier coefficients and the inequality \(\abs{m}^{1/(1+\delta)}\le\abs{k}^{1/(1+\delta)}+\abs{m-k}^{1/(1+\delta)}\) imply \(\|uv\|_{1+\delta}\le\|u\|_{1+\delta}\|v\|_{1+\delta}\). The corresponding weighted Fourier space is therefore a Banach algebra. Hence the exponential series converges in this norm and represents the pointwise function \(F_\delta=\e^{iq_\delta}\). Consequently,
\begin{equation}
    \|F_\delta\|_{1+\delta}
    =\left\|\sum_{r=0}^{\infty}\frac{(\ii q_\delta)^r}{r!}\right\|_{1+\delta}
    \le\sum_{r=0}^{\infty}\frac{\|q_\delta\|_{1+\delta}^{\,r}}{r!}
    \le\sum_{r=0}^{\infty}\frac{(S_LQ_\delta)^r}{r!}
    =\e^{S_LQ_\delta}.
\end{equation}
By the definition of the weighted norm, each nonnegative summand is bounded by the complete sum. Therefore,
\begin{equation}\label{fm_up}
    \abs{f_m}
    \le\e^{S_LQ_\delta}\e^{-\frac12C_\delta\abs{m}^{1/(1+\delta)}}.
\end{equation}

For an integer \(d\ge0\), define \(\widetilde P_d(z):=\sum_{|m|\le d}f_mz^m\). The corresponding Laurent truncation error satisfies
\begin{equation}\label{eta_up}
    \eta_d:=\sup_{\theta\in[-\pi,\pi]}\left|F_\delta(\theta)-\widetilde P_d(e^{i\theta})\right|
    \le\sum_{\abs{m}>d}\abs{f_m}
    \le2\e^{S_LQ_\delta}\sum_{m>d}\e^{-\frac12C_\delta m^{1/(1+\delta)}}.
\end{equation}
It remains to estimate the final tail sum. Set \(s:=1+\delta\) and \(h(x):=\e^{-\frac12C_\delta x^{1/s}}\) for \(x\ge0\). Since \(h\) is monotonically decreasing,
\begin{equation}
\sum_{m>d}\e^{-\frac12C_\delta m^{1/s}}
\le\int_d^\infty\e^{-\frac12C_\delta x^{1/s}}\,\dd x
=s\int_{d^{1/s}}^\infty t^\delta\e^{-\frac12C_\delta t}\,\dd t.
\label{eq:tail-int}
\end{equation}
The second equality follows from the change of variables \(x=t^s\). Consider \(\Phi(t):=t^\delta\e^{-\frac14C_\delta t}\) for \(t\ge0\). This function is continuous on \([0,\infty)\), tends to zero as \(t\to\infty\), and, for \(t>0\), satisfies \(\Phi'(t)=t^{\delta-1}\e^{-\frac14C_\delta t}\left(\delta-\frac{C_\delta t}{4}\right)\). Its maximum is therefore attained at \(t=4\delta/C_\delta\), with value \(\left(\frac{4\delta}{C_\delta}\right)^\delta\e^{-\delta}\). Hence
\begin{equation}
t^\delta\e^{-\frac12C_\delta t}
=\bigl(t^\delta\e^{-\frac14C_\delta t}\bigr)\e^{-\frac14C_\delta t}
\le\left(\frac{4\delta}{C_\delta}\right)^\delta\e^{-\delta}\e^{-\frac14C_\delta t}.
\label{eq:poly-absorb}
\end{equation}
Substituting Eq.~\eqref{eq:poly-absorb} into Eq.~\eqref{eq:tail-int} gives
\begin{equation}
\begin{split}
    \sum_{m>d}\e^{-\frac12C_\delta m^{1/s}}
    &\le(1+\delta)\left(\frac{4\delta}{C_\delta}\right)^\delta\e^{-\delta}\int_{d^{1/s}}^\infty\e^{-\frac14C_\delta t}\,\dd t\\
    &\le(1+\delta)\left(\frac{4\delta}{C_\delta}\right)^\delta\e^{-\delta}\cdot\frac4{C_\delta}\e^{-\frac14C_\delta d^{1/s}}.
\end{split}
\label{eq:tail-step3}
\end{equation}
Combining this estimate with Eq.~\eqref{eta_up}, we obtain
\begin{equation}
\eta_d
\le\frac{8(1+\delta)}{C_\delta}\left(\frac{4\delta}{C_\delta}\right)^\delta\e^{-\delta}\e^{S_LQ_\delta}\e^{-\frac14C_\delta d^{1/(1+\delta)}}
:=K_\delta\e^{S_LQ_\delta}\e^{-\frac14C_\delta d^{1/(1+\delta)}}.
\label{eq:eta-main}
\end{equation}
Here \(K_\delta:=\frac{8(1+\delta)}{C_\delta}\left(\frac{4\delta}{C_\delta}\right)^\delta\e^{-\delta}\).

We next impose the unit-circle bound. Since \(\abs{F_\delta(\theta)}=1\), the definition of \(\eta_d\) implies
\begin{equation}
\sup_{\abs{z}=1}\abs{\widetilde P_d(z)}\le1+\eta_d.
\label{eq:Ptilde-bound}
\end{equation}
Define \(P_d(z):=\widetilde P_d(z)/(1+\eta_d)\). Then \(\sup_{|z|=1}|P_d(z)|\le1\). For any \(\abs{x}\le L\), using \(F_\delta(\tau x)=\e^{ip(x)}\), we have
\begin{equation}
\begin{aligned}
\abs{P_d(\e^{\ii\tau x})-\e^{ip(x)}}
&\le\abs{P_d(\e^{\ii\tau x})-\widetilde P_d(\e^{\ii\tau x})}
+\abs{\widetilde P_d(\e^{\ii\tau x})-\e^{ip(x)}}\\
&=\frac{\eta_d}{1+\eta_d}\abs{\widetilde P_d(z)}
+\abs{\widetilde P_d(\e^{\ii\tau x})-F_\delta(\tau x)}\\
&\le2\eta_d,
\end{aligned}
\label{eq:window-error-split}
\end{equation}
where \(z=\e^{\ii\tau x}\) in the second line. Thus it is sufficient to impose \(2\eta_d<\varepsilon\). By Eq.~\eqref{eq:eta-main}, a sufficient condition is
\begin{equation}\label{d_lower}
d>\left[\frac4{C_\delta}\left(S_LQ_\delta+\log\frac{2K_\delta}{\varepsilon}\right)\right]^{1+\delta}.
\end{equation}

It remains to choose \(\delta\) and estimate the right-hand side of Eq.~\eqref{d_lower}. Since the chosen value of \(\delta\) will tend to zero, it suffices to work in the regime \(0<\delta\le1/2\). By increasing \(A_\delta\) and \(B_\delta\) if necessary, assume \(A_\delta\ge1\) and \(B_\delta\ge1\). From \(C_\delta=\frac{\log2}{(2B_\delta)^{1/(1+\delta)}}\), one obtains
\begin{equation}\label{C_up}
    C_\delta^{-1}
    =\frac{(2B_\delta)^{1/(1+\delta)}}{\log2}
    \le C_0B_\delta.
\end{equation}
Here and below, \(C_0>0\) may change from line to line and may depend on the fixed degree \(R\), but is independent of \(L\), \(\varepsilon\), \(\delta\), and the coefficients \(a_r\).

We next estimate \(Q_\delta\). Separating the zero mode and using symmetry,
\(Q_\delta=2A_\delta\left(1+2\sum_{m=1}^\infty\e^{-\frac12C_\delta m^{1/(1+\delta)}}\right)\).
Letting \(s=1+\delta\), comparison with an integral gives
\begin{equation}
    \sum_{m=1}^{\infty}\e^{-\frac12C_\delta m^{1/s}}
    \le\int_0^\infty\e^{-\frac12C_\delta x^{1/s}}\,\dd x
    =s\int_0^\infty t^{s-1}\e^{-\frac12C_\delta t}\,\dd t
    =s\Gamma(s)\left(\frac2{C_\delta}\right)^s.
\end{equation}
For \(0<\delta\le1/2\), one has \(s\in(1,3/2]\), so \(s\Gamma(s)\) is uniformly bounded. Moreover,
\(C_\delta^{-(1+\delta)}=\frac{2B_\delta}{(\log2)^{1+\delta}}\le C_0B_\delta\).
It follows that
\begin{equation}\label{Q_up}
    Q_\delta\le C_0A_\delta B_\delta.
\end{equation}

From the definition of \(K_\delta\), the factors \(8(1+\delta)\e^{-\delta}\) and \((4\delta)^\delta\) are uniformly bounded for \(0<\delta\le1/2\). Hence \(K_\delta\le C_0C_\delta^{-(1+\delta)}\), and therefore
\begin{equation}\label{K_up}
    K_\delta\le C_0B_\delta.
\end{equation}

Using Eqs.~\eqref{C_up}, \eqref{Q_up}, and \eqref{K_up}, and increasing \(C_0\) so that \(C_0\ge1\), we obtain
\begin{equation}\label{d1}
\begin{split}
\frac4{C_\delta}\left(S_LQ_\delta+\log\frac{2K_\delta}{\varepsilon}\right)
&\le C_0B_\delta\left(S_LA_\delta B_\delta+\log\frac1\varepsilon+\log(2C_0B_\delta)\right)\\
&=C_0S_LA_\delta B_\delta^2+C_0B_\delta\log\frac1\varepsilon+C_0B_\delta\log(2C_0B_\delta)\\
&\le C_0\left(S_L+\log\frac1\varepsilon\right)A_\delta B_\delta^2+C_0A_\delta B_\delta^2\log(2C_0A_\delta B_\delta^2).
\end{split}
\end{equation}

Set \(X:=e^e+S_L+\log\frac1\varepsilon\), and choose \(\delta:=\frac{2\log\frac{32}{3\pi^2}}{\log\log X}\). The additive term \(e^e\) ensures that \(\log\log X\ge1\) and does not affect the asymptotic order. Since \(\delta\to0\) as \(X\to\infty\), there exists \(X_0>0\) such that for every \(X\ge X_0\), one has \(0<\delta\le1/2\) and \(\delta\) lies in the sufficiently small regime required by Lemma~\ref{lem:q-gevrey-AB}. Hence
\(A_\delta B_\delta^2\le\exp(C_0\sqrt{\log X})\)
for \(X\ge X_0\). For the bounded range \(e^e\le X<X_0\), one may instead fix any sufficiently small \(\delta\); the right-hand side of Eq.~\eqref{d_lower} is then uniformly bounded and can be absorbed into the constant implicit in the final \(O(\cdot)\) estimate. It therefore suffices to consider \(X\ge X_0\).

Using \(S_L+\log\frac1\varepsilon\le X\), Eq.~\eqref{d1} gives
\begin{equation}
    \frac4{C_\delta}\left(S_LQ_\delta+\log\frac{2K_\delta}{\varepsilon}\right)
    \le C_0X\exp\!\left(C_0\sqrt{\log X}\right)+C_0\exp\!\left(C_0\sqrt{\log X}\right)\log\!\left(2C_0\exp\!\left(C_0\sqrt{\log X}\right)\right).
\end{equation}
Since \(\log(2C_0\exp(C_0\sqrt{\log X}))\le C_0'\log X\le C_0'X\) for sufficiently large \(X\), the second term can be absorbed into the first after increasing \(C_0\). Thus
\begin{equation}
\frac4{C_\delta}\left(S_LQ_\delta+\log\frac{2K_\delta}{\varepsilon}\right)
\le C_0X\exp\!\left(C_0\sqrt{\log X}\right).
\end{equation}
Since \(\exp(C_0\sqrt{\log X})=X^{C_0/\sqrt{\log X}}=X^{o(1)}\), we obtain
\begin{equation}
\frac4{C_\delta}\left(S_LQ_\delta+\log\frac{2K_\delta}{\varepsilon}\right)
\le X^{1+o(1)}.
\end{equation}
Moreover, \(\delta=O(1/\log\log X)=o(1)\). Therefore,
\begin{equation}
    \left[\frac4{C_\delta}\left(S_LQ_\delta+\log\frac{2K_\delta}{\varepsilon}\right)\right]^{1+\delta}
    \le\left(X^{1+o(1)}\right)^{1+o(1)}
    =X^{1+o(1)}.
\end{equation}

Taking
\(d:=1+\left\lceil\left[\frac4{C_\delta}\left(S_LQ_\delta+\log\frac{2K_\delta}{\varepsilon}\right)\right]^{1+\delta}\right\rceil\)
ensures the strict inequality in Eq.~\eqref{d_lower}. The additive constant and the ceiling do not affect the asymptotic order. Hence \(d=O(X^{1+o(1)})\). Since \(X=e^e+S_L+\log\frac1\varepsilon\), this yields
\begin{equation}
    d=O\!\left(\left(S_L+\log\frac1\varepsilon\right)^{1+o(1)}\right).
\end{equation}
This proves the theorem.
\end{proof}

\subsection{Approximation on a finite Fock input space}
\label{app:finite-fock-approximation}

The windowed Laurent approximation controls the target only on a finite quadrature interval.
This subsection extends it to a uniform state-norm approximation on the fixed finite Fock subspace \(\mathcal H_n\) by controlling the contribution outside the window.

\begin{theorem}
\label{thm:polynomial-operator-approx}
Fix an integer \(n\ge0\), and let \(p(x)=\sum_{r=0}^{R}a_rx^r\) be a fixed real polynomial of degree \(R\ge2\), with \(a_R\ne0\). For any \(0<\varepsilon<1\), there exist a step size \(\tau>0\) and a Laurent polynomial \(P_d(z)=\sum_{|m|\le d}p_mz^m\) satisfying \(\sup_{|z|=1}|P_d(z)|\le1\), such that, for every normalized state \(|\psi\rangle\in\mathcal H_n\),
\begin{equation}
\left\|\left(P_d(U)-e^{ip(\hat X)}\right)|\psi\rangle\right\|<\varepsilon.
\end{equation}
Here \(U=e^{i\tau\hat X}\). The construction can be chosen with a quadrature window
\(L=\Theta\!\left(\sqrt{n+1+\log(1/\varepsilon)}\right)\)
and \(\tau=\pi/(4L)\).
Moreover, for every normalized state \(|\psi\rangle\in\mathcal H_n\), one has
\(\|P_d(U)|\psi\rangle\|\ge1-\varepsilon\).
As \(\varepsilon\to0\), one can choose
\(d=O\!\left(\left(\log\frac1\varepsilon\right)^{R/2+o(1)}\right)\).

\end{theorem}

\begin{proof}
Fix an integer \(n\ge0\), a fixed real polynomial \(p(x)=\sum_{r=0}^{R}a_rx^r\) of degree \(R\ge2\), and an accuracy \(0<\varepsilon<1\). Let \(L\ge1\) be a window parameter to be chosen later, and denote by \(\Pi_L:=\int_{-L}^L|x\rangle\langle x|\dd x\) the spectral projection of \(\hat X\) onto the interval \([-L,L]\). By Theorem~\ref{thm:polynomial-laurent-approx}, for window size \(L\) and approximation accuracy \(\varepsilon/2\), one can construct a Laurent polynomial \(P_d(z)=\sum_{|m|\le d}p_mz^m\) such that \(\sup_{|x|\le L}\left|P_d(e^{i\tau x})-e^{ip(x)}\right|<\frac{\varepsilon}{2}\) and \(\sup_{|z|=1}|P_d(z)|\le1\). Let \(|\psi\rangle\in\mathcal H_n\) be an arbitrary normalized state. Using the spectral decomposition with respect to the window, we write \(|\psi\rangle=\Pi_L|\psi\rangle+(I-\Pi_L)|\psi\rangle\). The triangle inequality then gives
\begin{equation}
    \left\|\left(P_d(U)-e^{ip(\hat X)}\right)|\psi\rangle\right\|
    \le \left\|\left(P_d(U)-e^{ip(\hat X)}\right)\Pi_L|\psi\rangle\right\|
    + \left\|\left(P_d(U)-e^{ip(\hat X)}\right)(I-\Pi_L)|\psi\rangle\right\|.
\end{equation}
where \(U:=e^{i\tau\hat X}\) and \(\tau:=\frac{\pi}{4L}\).

We estimate the two terms separately. For the part supported inside the window, the spectral theorem implies that \(P_d(U)=P_d(e^{i\tau\hat X})\) acts in the \(x\)-representation as multiplication by the scalar function \(x\mapsto P_d(e^{i\tau x})\), while \(e^{ip(\hat X)}\) acts as multiplication by \(x\mapsto e^{ip(x)}\). Therefore, on the window \([-L,L]\),
\begin{equation}
    \left\|\Pi_L\left(P_d(U)-e^{ip(\hat X)}\right)\Pi_L\right\|_{\mathrm{op}}
    = \sup_{|x|\le L}\left|P_d(e^{i\tau x})-e^{ip(x)}\right|<\frac{\varepsilon}{2}.
\end{equation}
Moreover, \(\Pi_L\), \(P_d(U)\), and \(e^{ip(\hat X)}\) are all functions of \(\hat X\), and hence commute pairwise. It follows that \(\left(P_d(U)-e^{ip(\hat X)}\right)\Pi_L=\Pi_L\left(P_d(U)-e^{ip(\hat X)}\right)\Pi_L\). Thus
\begin{equation}
    \begin{aligned}
    \left\|\left(P_d(U)-e^{ip(\hat X)}\right)\Pi_L|\psi\rangle\right\|
    &=\left\|\Pi_L\left(P_d(U)-e^{ip(\hat X)}\right)\Pi_L|\psi\rangle\right\|\\
    &\le\left\|\Pi_L\left(P_d(U)-e^{ip(\hat X)}\right)\Pi_L\right\|_{\mathrm{op}}\cdot\|\Pi_L|\psi\rangle\|<\frac{\varepsilon}{2}.
\end{aligned}
\end{equation}
Here we used \(\|\Pi_L|\psi\rangle\|\le\||\psi\rangle\|=1\).

It remains to estimate the contribution outside the window. Since \(\sup_{|z|=1}|P_d(z)|\le1\), the spectral theorem gives \(\|P_d(U)\|_{\mathrm{op}}\le1\). Since \(p\) is real, \(e^{ip(\hat X)}\) is unitary, so \(\|e^{ip(\hat X)}\|_{\mathrm{op}}=1\). Hence
\begin{equation}\label{eq:out}
    \begin{aligned}
    \left\|\left(P_d(U)-e^{ip(\hat X)}\right)(I-\Pi_L)|\psi\rangle\right\|
        &\le\left(\|P_d(U)\|_{\mathrm{op}}+\|e^{ip(\hat X)}\|_{\mathrm{op}}\right)\|(I-\Pi_L)|\psi\rangle\|\\
        &\le2\|(I-\Pi_L)|\psi\rangle\|=2\left(\int_{|x|>L}|\psi(x)|^2\dd x\right)^{1/2}.
    \end{aligned}
\end{equation}
Since \(|\psi\rangle\in\mathcal H_n\), it can be written as \(|\psi\rangle=\sum_{k=0}^{n}c_k|k\rangle,\; \sum_{k=0}^{n}|c_k|^2=1\). In the position representation, denote the wave function of the \(k\)-th Fock state by \(\varphi_k(x)=\frac{1}{\pi^{1/4}\sqrt{2^k k!}}H_k(x)e^{-x^2/2}\), where \(H_k(x)\) is the \(k\)-th Hermite polynomial, \(H_k(x)=k!\sum_{r=0}^{\lfloor k/2\rfloor}\frac{(-1)^r(2x)^{k-2r}}{r!(k-2r)!}\). Thus the position-space wave function of \(|\psi\rangle\) is \(\psi(x)=\sum_{k=0}^{n}c_k\varphi_k(x)\). By the Cauchy--Schwarz inequality,
\begin{equation}
    |\psi(x)|^2
    =\left|\sum_{k=0}^{n}c_k\varphi_k(x)\right|^2
    \le\left(\sum_{k=0}^{n}|c_k|^2\right)\left(\sum_{k=0}^{n}|\varphi_k(x)|^2\right)
    =\sum_{k=0}^{n}|\varphi_k(x)|^2.
\end{equation}
We now bound the right-hand side. Write \(H_k(x)=\sum_{j=0}^{k}h_{k,j}x^j\). The sum of the absolute values of its coefficients is \(A_k:=\sum_{j=0}^{k}|h_{k,j}|=\sum_{r=0}^{\lfloor k/2\rfloor}\frac{k!2^{k-2r}}{r!(k-2r)!}\). In particular, for \(|x|\ge1\) and \(0\le k\le n\), we have \(|H_k(x)|\le A_k|x|^k\le A_k|x|^n\). Therefore, for \(|x|\ge1\),
\begin{equation}
    |\varphi_k(x)|^2
    =
    \frac{1}{\sqrt{\pi}\,2^k k!}H_k(x)^2e^{-x^2}
    \le
    \frac{A_k^2}{\sqrt{\pi}\,2^k k!}|x|^{2n}e^{-x^2}.
\end{equation}
Summing over \(k=0,\ldots,n\), we obtain \(|\psi(x)|^2\le C_n|x|^{2n}e^{-x^2}\), where \(C_n:=\frac1{\sqrt\pi}\sum_{k=0}^{n}\frac{A_k^2}{2^k k!}\). Here \(C_n\) depends only on the fixed truncation level \(n\), and is independent of the normalized state \(|\psi\rangle\), \(L\), and \(\varepsilon\). Hence, for \(L\ge1\),
\begin{equation}\label{eq:psi_x}
    \begin{aligned}
    \int_{|x|>L}|\psi(x)|^2\,\dd x
        \le2C_n\int_L^\infty x^{2n}e^{-x^2}\,\dd x
    =C_n\Gamma\!\left(n+\frac12,L^2\right),
    \end{aligned}
\end{equation}
where \(\Gamma(a,y):=\int_y^\infty t^{a-1}e^{-t}\dd t\) is the upper incomplete gamma function.

We next rewrite this tail bound in a form convenient for choosing \(L\). When \(n=0\), the space \(\mathcal H_0\) contains only the vacuum state, and \(C_0=\frac1{\sqrt\pi}\), \(C_0\Gamma\!\left(\frac12,L^2\right)=\operatorname{erfc}(L)\). Thus the above estimate reduces to the exact vacuum tail probability. Moreover, for \(L\ge1\), one has \(\operatorname{erfc}(L)\le e^{-L^2}/(\sqrt\pi L)\le e^{-L^2}\). Therefore,
\begin{equation}
    \begin{aligned}
    \int_{|x|>L}|\psi(x)|^2\,\dd x
    \le\operatorname{erfc}(L)
    \le e^{-L^2}.
    \end{aligned}
\end{equation}
To make the outside-window contribution in Eq.~\eqref{eq:out} smaller than \(\frac{\varepsilon}{2}\), it suffices to choose \(L^2>\log\frac{16}{\varepsilon^2}\). Then \(e^{-L^2}<\frac{\varepsilon^2}{16}\), and hence \(\int_{|x|>L}|\psi(x)|^2\dd x<\frac{\varepsilon^2}{16}\).

We now consider the case \(n\ge1\). If \(L^2>2n-1\), then for every \(u\ge L^2\),
\(u^{n-\frac12}=L^{2n-1}\left(1+\frac{u-L^2}{L^2}\right)^{n-\frac12}
\le L^{2n-1}\exp\!\left(\frac{n-\frac12}{L^2}(u-L^2)\right)\).
Consequently,
\begin{equation}
    \begin{aligned}
    \Gamma\!\left(n+\frac12,L^2\right)
    &=\int_{L^2}^{\infty}u^{n-\frac12}e^{-u}\dd u\\
    &\le
    L^{2n-1}e^{-L^2}
    \int_{L^2}^{\infty}
    \exp\!\left[
        -\left(1-\frac{n-\frac12}{L^2}\right)(u-L^2)
    \right]\dd u\\
    &=
    \frac{L^{2n-1}e^{-L^2}}
    {1-\frac{n-\frac12}{L^2}}
    \le2L^{2n-1}e^{-L^2}.
    \end{aligned}
\end{equation}
Substituting this bound into Eq.~\eqref{eq:psi_x}, it is sufficient to choose \(L\) such that
\begin{equation}
    \int_{|x|>L}|\psi(x)|^2\,\dd x
    \le2C_nL^{2n-1}e^{-L^2}
    <\frac{\varepsilon^2}{16}.
\end{equation}

For sufficiently small \(\varepsilon\), choose
\(L=2\sqrt{n+1+\log(1/\varepsilon)}\).
This choice satisfies \(L^2>2n-1\). Moreover, the ratio of the upper bound
\(2C_nL^{2n-1}e^{-L^2}\) to \(\varepsilon^2/16\) is
\begin{equation}
    32C_ne^{-4(n+1)}\varepsilon^2
    \left[
        4\left(
        n+1+\log\frac1\varepsilon
        \right)
    \right]^{n-\frac12},
\end{equation}
which tends to zero as \(\varepsilon\to0\) for every fixed \(n\).
Hence the required tail bound holds for all sufficiently small \(\varepsilon\).
The same choice also satisfies the vacuum-tail condition derived above for \(n=0\).
Therefore, for every fixed \(n\), the window can be chosen with
\(L=\Theta\!\left(\sqrt{n+1+\log(1/\varepsilon)}\right)\).

Combining the inside-window and outside-window estimates gives
\begin{equation}
\begin{aligned}
    \left\|\left(P_d(U)-e^{ip(\hat X)}\right)|\psi\rangle\right\|
    &\le\left\|\left(P_d(U)-e^{ip(\hat X)}\right)\Pi_L|\psi\rangle\right\|
    +\left\|\left(P_d(U)-e^{ip(\hat X)}\right)(I-\Pi_L)|\psi\rangle\right\|\\
    &<\frac{\varepsilon}{2}+\frac{\varepsilon}{2}
    =\varepsilon.
\end{aligned}
\end{equation}
Since \(e^{ip(\hat X)}\) is unitary, the reverse triangle inequality also gives \(\|P_d(U)|\psi\rangle\|\ge\|e^{ip(\hat X)}|\psi\rangle\|-\|(P_d(U)-e^{ip(\hat X)})|\psi\rangle\|>1-\varepsilon\).

It remains to estimate the bandwidth \(d\). By Theorem~\ref{thm:polynomial-laurent-approx}, for window size \(L\) and window accuracy \(\varepsilon/2\), one can choose \(d=O\!\left(\left(S_L+\log\frac{2}{\varepsilon}\right)^{1+o(1)}\right)\), where \(S_L:=\sum_{r=0}^{R}|a_r|L^r\). Since \(p\) is fixed and \(L\ge1\), one has \(S_L\le\left(\sum_{r=0}^{R}|a_r|\right)L^R=O(L^R)\). 
For fixed \(n\), the window choice above also gives \(L=O\!\left(\sqrt{\log(1/\varepsilon)}\right)\) as \(\varepsilon\to0\).
Together with \(R\ge2\), this gives \(S_L+\log\frac{2}{\varepsilon}=O\!\left(\left(\log\frac1\varepsilon\right)^{R/2}\right)\).
Substituting this estimate into the bandwidth bound yields \(d=O\!\left(\left(\log\frac1\varepsilon\right)^{R/2+o(1)}\right)\). This proves the theorem.

\end{proof}

\subsection{GQSP realization by Rabi control}
\label{app:gqsp-rabi-realization}

This subsection gives a constructive GQSP realization of the bounded Laurent polynomial and rewrites the resulting sequence as equal-duration \(X\)-Rabi pulses with suitable ancilla boundary states.

\begin{lemma}[Laurent-polynomial GQSP realization]
\label{lem:laurent-gqsp}
Let \(U_\tau=e^{i\tau\hat X}\), and define, in the ancilla basis \(\{\ket0_a,\ket1_a\}\), \(A_\tau=\mathrm{diag}(U_\tau,I)_a\) and \(A_\tau'=\mathrm{diag}(I,U_\tau^\dagger)_a\). If \(P_d(z)=\sum_{m=-d}^{d}p_mz^m\) satisfies \(\sup_{|z|=1}|P_d(z)|\le1\), then there exist single-qubit unitaries \(G_j\), \(j=0,\ldots,2d\), such that
\begin{equation}
    \mathcal V_d=\left(\prod_{j=d+1}^{2d}G_jA_\tau'\right)\left(\prod_{j=1}^{d}G_jA_\tau\right)G_0
    \label{gqsp_sequence}
\end{equation}
satisfies \((\bra0_a\otimes I)\mathcal V_d(\ket0_a\otimes I)=P_d(U_\tau)\).
\end{lemma}

\begin{proof}
The case \(d=0\) is immediate, so assume \(d\ge1\). Define the ordinary polynomial \(\widetilde P_{2d}(z):=z^dP_d(z)\). Then \(\widetilde P_{2d}\) has degree at most \(2d\), and \(|\widetilde P_{2d}(z)|=|P_d(z)|\le1\) for every \(|z|=1\). By the ordinary-polynomial GQSP realization theorem~\cite{Motlagh2024}, there exist single-qubit unitaries \(G_0,\ldots,G_{2d}\) such that the GQSP sequence \(\widetilde{\mathcal V}_d:=\left(\prod_{j=1}^{2d}G_jA_\tau\right)G_0\) satisfies \((\bra0_a\otimes I)\widetilde{\mathcal V}_d(\ket0_a\otimes I)=\widetilde P_{2d}(U_\tau)\).

Since \(A_\tau'=(I_a\otimes U_\tau^\dagger)A_\tau\), and \(I_a\otimes U_\tau^\dagger\) commutes with every single-qubit unitary \(G_j\) and with \(A_\tau\), replacing \(d\) of these \(2d\) copies of \(A_\tau\) by \(A_\tau'\) gives \(\mathcal V_d=(I_a\otimes U_\tau^{-d})\widetilde{\mathcal V}_d\). Therefore, \((\bra0_a\otimes I)\mathcal V_d(\ket0_a\otimes I)=U_\tau^{-d}\widetilde P_{2d}(U_\tau)=P_d(U_\tau)\). The unitaries \(G_j\) can be obtained constructively by first constructing a complementary polynomial for \(\widetilde P_{2d}\) and then applying the recursive GQSP factorization.
\end{proof}

\begin{theorem}[Postselected Rabi-sequence realization of polynomial phase gates]
\label{thm:polynomial-rabi-realization}
Fix an integer \(n\ge0\), and let \(p(x)=\sum_{r=0}^{R}a_rx^r\) be a fixed real polynomial of degree \(R\ge2\), with \(a_R\ne0\). For any \(0<\varepsilon<1\), there exist a step size \(\tau>0\), an integer \(d\), \(2d\) unit vectors \(\omega_j=(\omega_{j,x},\omega_{j,y},\omega_{j,z})\), and normalized ancilla states \(\ket{\phi_{\rm in}}_a\) and \(\ket{\phi_{\rm out}}_a\), such that the \(X\)-Rabi sequence with \(2d\) pulses,
\begin{equation}\label{Seq:rabi-sequence}
    \mathcal U_d:=\prod_{j=1}^{2d}\mathcal X_{\omega_j}(\tau)=\prod_{j=1}^{2d}\exp\!\left(i\frac{\tau}{2}\sigma_{\omega_j}\otimes\hat X\right)
\end{equation}
satisfies \((\bra{\phi_{\rm out}}_a\otimes I)\mathcal U_d(\ket{\phi_{\rm in}}_a\otimes I)=P_d(e^{i\tau\hat X})\), where \(P_d(z)=\sum_{|m|\le d}p_mz^m\) is the Laurent polynomial constructed above. Moreover, for every normalized state \(\ket\psi\in\mathcal H_n\),
\begin{equation}
    \left\|(\bra{\phi_{\rm out}}_a\otimes I)\mathcal U_d(\ket{\phi_{\rm in}}_a\otimes\ket\psi)-e^{ip(\hat X)}\ket\psi\right\|<\varepsilon.
\end{equation}
The success probability of postselecting the ancilla qubit onto \(\ket{\phi_{\rm out}}_a\) satisfies \(p_{\rm succ}(\psi)>(1-\varepsilon)^2\). As \(\varepsilon\to0\), the number \(2d\) of \(X\)-Rabi pulses obeys \(2d=O\!\left(\left(\log\frac1\varepsilon\right)^{R/2+o(1)}\right)\). The pulse directions $\omega_j$ and the ancilla boundary states $\ket{\phi_{\rm in}}_a,\ket{\phi_{\rm out}}_a$ can be obtained constructively from the Laurent coefficients \(p_m\) through complementary-polynomial construction and recursive GQSP factorization.
\end{theorem}

\begin{proof}
Fix \(n\ge0\), the polynomial \(p\), and \(0<\varepsilon<1\). By Theorem~\ref{thm:polynomial-operator-approx}, there exist a step size \(\tau>0\), an integer \(d\), and a Laurent polynomial \(P_d(z)=\sum_{|m|\le d}p_mz^m\) such that, for every normalized state \(\ket\psi\in\mathcal H_n\), \(\left\|\left(P_d(e^{i\tau\hat X})-e^{ip(\hat X)}\right)\ket\psi\right\|<\varepsilon\), and \(\sup_{|z|=1}|P_d(z)|\le1\). Let \(U_\tau:=e^{i\tau\hat X}\). Since \(P_d\) satisfies the unit-circle constraint, Lemma~\ref{lem:laurent-gqsp} yields single-qubit unitaries \(G_0,\ldots,G_{2d}\) such that the GQSP sequence
\begin{equation}\label{gqsp_sequence_2}
    \mathcal V_d=\left(\prod_{j=d+1}^{2d}G_jA_\tau'\right)\left(\prod_{j=1}^{d}G_jA_\tau\right)G_0
\end{equation}
satisfies \((\bra0_a\otimes I)\mathcal V_d(\ket0_a\otimes I)=P_d(U_\tau)\). We now rewrite \(A_\tau\) and \(A_\tau'\) as products of an \(X\)-Rabi pulse and an unconditional oscillator phase.
\begin{equation}
    \begin{aligned}
    A_\tau
    &=\begin{pmatrix}
    e^{i\tau\hat X}&0\\
    0&I
    \end{pmatrix}
    =\begin{pmatrix}
    e^{i\tau\hat X/2}&0\\
    0&e^{i\tau\hat X/2}
    \end{pmatrix}
    \begin{pmatrix}
    e^{i\tau\hat X/2}&0\\
    0&e^{-i\tau\hat X/2}
    \end{pmatrix}
    =(I_a\otimes e^{i\tau\hat X/2})\mathcal X_{\mathbf e_z}(\tau).
    \end{aligned}
\end{equation}
Similarly, \(A_\tau'=(I_a\otimes e^{-i\tau\hat X/2})\mathcal X_{\mathbf e_z}(\tau)\), where \(\mathbf e_z=(0,0,1)\). The factors \(I_a\otimes e^{\pm i\tau\hat X/2}\) commute with every single-qubit unitary \(G_j\) because they act on different subsystems. They also commute with \(\mathcal X_{\mathbf e_z}(\tau)\), since both are functions of \(\hat X\). As \(A_\tau\) and \(A_\tau'\) each occur \(d\) times, the unconditional oscillator phases cancel exactly. Therefore,
\begin{equation}\label{gqsp_rabi}
    \begin{aligned}
    \mathcal V_d
    &=\left(\prod_{j=d+1}^{2d}G_j(I_a\otimes e^{-i\tau\hat X/2})\mathcal X_{\mathbf e_z}(\tau)\right)\left(\prod_{j=1}^{d}G_j(I_a\otimes e^{i\tau\hat X/2})\mathcal X_{\mathbf e_z}(\tau)\right)G_0\\
    &=(I_a\otimes e^{-i\tau\hat X/2})^d(I_a\otimes e^{i\tau\hat X/2})^d\left(\prod_{j=d+1}^{2d}G_j\mathcal X_{\mathbf e_z}(\tau)\right)\left(\prod_{j=1}^{d}G_j\mathcal X_{\mathbf e_z}(\tau)\right)G_0\\
    &=\left(\prod_{j=d+1}^{2d}G_j\mathcal X_{\mathbf e_z}(\tau)\right)\left(\prod_{j=1}^{d}G_j\mathcal X_{\mathbf e_z}(\tau)\right)G_0\\
    &=G_{2d}\mathcal X_{\mathbf e_z}(\tau)G_{2d-1}\mathcal X_{\mathbf e_z}(\tau)\cdots G_1\mathcal X_{\mathbf e_z}(\tau)G_0\\
    &=(C_{2d}\otimes I)\prod_{j=1}^{2d}\left[(C_{j-1}^{\dagger}\otimes I)\mathcal X_{\mathbf e_z}(\tau)(C_{j-1}\otimes I)\right].
    \end{aligned}
\end{equation}
Here \(C_0:=G_0\) and \(C_j:=G_jG_{j-1}\cdots G_0\) for \(j=1,\ldots,2d\). Each \(G_j\) and \(C_j\) acts on the ancilla, with the identity on the oscillator suppressed. For each \(j\), define \(\sigma_{\omega_j}:=C_{j-1}^{\dagger}\sigma_zC_{j-1}\). Since \(C_{j-1}\) is a single-qubit unitary, \(\sigma_{\omega_j}\) is a traceless Hermitian unitary and hence a Pauli operator along a unit Bloch-sphere direction \(\omega_j\). It follows that \((C_{j-1}^{\dagger}\otimes I)\mathcal X_{\mathbf e_z}(\tau)(C_{j-1}\otimes I)=\mathcal X_{\omega_j}(\tau)\). Substituting this identity into Eq.~\eqref{gqsp_rabi}, we obtain
\begin{equation}
    \mathcal V_d=(C_{2d}\otimes I)\left(\prod_{j=1}^{2d}\mathcal X_{\omega_j}(\tau)\right)=(C_{2d}\otimes I)\mathcal U_d.
\end{equation}
Choose \(\ket{\phi_{\rm in}}_a:=\ket0_a\) and \(\ket{\phi_{\rm out}}_a:=C_{2d}^{\dagger}\ket0_a\). Since \(C_{2d}\) is unitary, both boundary states are normalized. Moreover, \(\bra{\phi_{\rm out}}_a=\bra0_aC_{2d}\), and therefore \((\bra{\phi_{\rm out}}_a\otimes I)\mathcal U_d(\ket{\phi_{\rm in}}_a\otimes I)=(\bra0_a\otimes I)\mathcal V_d(\ket0_a\otimes I)=P_d(e^{i\tau\hat X})\).

For every normalized state \(\ket\psi\in\mathcal H_n\), it follows that
\begin{equation}
    \left\|(\bra{\phi_{\rm out}}_a\otimes I)\mathcal U_d(\ket{\phi_{\rm in}}_a\otimes\ket\psi)-e^{ip(\hat X)}\ket\psi\right\|
    =\left\|\left(P_d(e^{i\tau\hat X})-e^{ip(\hat X)}\right)\ket\psi\right\|
    <\varepsilon.
\end{equation}
The success probability is \(p_{\rm succ}(\psi)=\|P_d(e^{i\tau\hat X})\ket\psi\|^2\). Since \(e^{ip(\hat X)}\) is unitary, the reverse triangle inequality gives
\begin{equation}
    p_{\rm succ}(\psi)=\|P_d(e^{i\tau\hat X})\ket\psi\|^2
    \ge\left(\|e^{ip(\hat X)}\ket\psi\|-\left\|\left(P_d(e^{i\tau\hat X})-e^{ip(\hat X)}\right)\ket\psi\right\|\right)^2
    >(1-\varepsilon)^2.
\end{equation}
Finally, Theorem~\ref{thm:polynomial-operator-approx} gives \(2d=O\!\left(\left(\log\frac1\varepsilon\right)^{R/2+o(1)}\right)\). The single-qubit unitaries \(G_j\) are obtained after constructing a complementary polynomial and applying the recursive GQSP factorization; the cumulative unitaries \(C_j\), pulse directions \(\omega_j\), and boundary states then follow directly from their definitions. No variational optimization over the \(2d\) Rabi-pulse directions is required. This proves the theorem.
\end{proof}

\subsection{Complete resource bounds}
\label{app:complete-resource-bounds}

This subsection combines the approximation and Rabi-realization results above to derive the pulse-count, total-interaction-time, and postselection bounds stated in Theorem~1 of the main text.

\begin{corollary}[Polylogarithmic Rabi synthesis of polynomial phases]
\label{cor:complete-resource-bounds}
Fix an integer \(n\ge0\), a real polynomial \(p(x)=\sum_{r=0}^{R}a_rx^r\) of degree \(R\ge2\) with \(a_R\ne0\), and an evolution time \(T>0\).
For all sufficiently small \(0<\varepsilon<1\), one can choose
\(L=\Theta\!\left(\sqrt{n+1+\log(1/\varepsilon)}\right)\) and
\(\tau=\pi/(4L)\) such that an even-length \(X\)-Rabi sequence with postselected block \(P_N\) satisfies
\begin{equation}
    \left\|
    \left(P_N-\e^{iTp(\hat X)}\right)\ket\psi
    \right\|
    <\varepsilon
    \label{eq:complete-resource-error}
\end{equation}
for every normalized \(\ket\psi\in\mathcal H_n\).
The required number of Rabi pulses obeys
\begin{equation}
    N
    =
    O\!\left[
    \left(
    T\sum_{r=0}^{R}|a_r|L^r
    +\log\frac1\varepsilon
    \right)^{1+o(1)}
    \right].
    \label{eq:complete-resource-pulse}
\end{equation}
\end{corollary}
\begin{proof}
We apply the preceding results to the polynomial
\(Tp(x)=\sum_{r=0}^{R}(Ta_r)x^r\).
For a quadrature window \([-L,L]\), the corresponding quantity \(S_L\) in Theorem~\ref{thm:polynomial-laurent-approx} is
\(T\sum_{r=0}^{R}|a_r|L^r\).

By Theorem~\ref{thm:polynomial-laurent-approx}, for window accuracy \(\varepsilon/2\) and \(\tau=\pi/(4L)\), there exists a Laurent polynomial \(P_d(z)=\sum_{|m|\le d}p_mz^m\) satisfying \(\sup_{|x|\le L}\left|P_d(e^{i\tau x})-e^{iTp(x)}\right|<\varepsilon/2\) and \(\sup_{|z|=1}|P_d(z)|\le1\). The same theorem gives
\begin{equation}
    d = O\!\left[\left(T\sum_{r=0}^{R}|a_r|L^r+\log\frac1\varepsilon\right)^{1+o(1)}\right].
    \label{eq:complete-bandwidth}
\end{equation}
The window size is chosen according to Theorem~\ref{thm:polynomial-operator-approx}. Its proof shows that
\(L=\Theta\!\left(\sqrt{n+1+\log(1/\varepsilon)}\right)\)
makes the contribution outside the window sufficiently small for every normalized input in \(\mathcal H_n\).
With this choice, the windowed Laurent approximation above gives
\begin{equation}
    \left\|
    \left(
    P_d(e^{i\tau\hat X})-e^{iTp(\hat X)}
    \right)\ket\psi
    \right\|
    <\varepsilon
    \label{eq:complete-operator-error}
\end{equation}
for every normalized \(\ket\psi\in\mathcal H_n\).

It remains to realize this Laurent polynomial by Rabi control.
Since \(P_d\) satisfies the unit-circle constraint, Lemma~\ref{lem:laurent-gqsp} gives a GQSP realization using \(2d\) signal operations.
Equation~\eqref{gqsp_rabi} and Theorem~\ref{thm:polynomial-rabi-realization} rewrite this realization as an equal-duration \(X\)-Rabi sequence with suitable pulse directions and normalized ancilla input and postselection states.
Its postselected oscillator block is exactly \(P_d(e^{i\tau\hat X})\).
Denoting this block by \(P_N\) and setting \(N=2d\), the state-norm estimate in Eq.~\eqref{eq:complete-operator-error} is precisely the error bound in Eq.~\eqref{eq:complete-resource-error}.
Combining \(N=2d\) with Eq.~\eqref{eq:complete-bandwidth} gives Eq.~\eqref{eq:complete-resource-pulse}.
This proves the corollary.
\end{proof}

For fixed \(p\), \(n\), and \(T\), the window choice gives
\(L=\Theta\!\left(\sqrt{\log(1/\varepsilon)}\right)\) as \(\varepsilon\to0\).
Since \(p\) is fixed and \(R\ge2\), one has
\(T\sum_{r=0}^{R}|a_r|L^r+\log(1/\varepsilon)
=O\!\left((\log(1/\varepsilon))^{R/2}\right)\).
Equation~\eqref{eq:complete-resource-pulse} therefore gives
\begin{equation}
    N
    =
    O\!\left[
    \left(
    \log\frac1\varepsilon
    \right)^{R/2+o(1)}
    \right].
    \label{eq:complete-pulse-fixed}
\end{equation}

The same approximation bound also determines the postselection performance.
For every normalized \(\ket\psi\in\mathcal H_n\), Eq.~\eqref{eq:complete-operator-error} and the unitarity of \(e^{iTp(\hat X)}\) give
\begin{equation}
    \left\|P_N\ket\psi\right\|
    \ge
    \left\|e^{iTp(\hat X)}\ket\psi\right\|
    -
    \left\|
    \left(
    P_N-e^{iTp(\hat X)}
    \right)\ket\psi
    \right\|
    >
    1-\varepsilon .
\end{equation}
Hence the postselection success probability satisfies $p_{\rm succ}(\psi) = \left\|P_N\ket\psi\right\|^2 > (1-\varepsilon)^2.$

The normalized postselected output is also close to the ideal state.
Equation~\eqref{eq:complete-operator-error} gives
\(\left\|P_N\ket\psi-e^{iTp(\hat X)}\ket\psi\right\|<\varepsilon\).
Since \(\left\|e^{iTp(\hat X)}\ket\psi\right\|=1\), the reverse triangle inequality also gives
\(\left|\,\left\|P_N\ket\psi\right\|-1\,\right|<\varepsilon\).
It follows that
\begin{equation}
\begin{aligned}
    \left\|
    \frac{P_N\ket\psi}
    {\left\|P_N\ket\psi\right\|}
    -
    e^{iTp(\hat X)}\ket\psi
    \right\|
    &\le
    \left\|
    \frac{P_N\ket\psi}
    {\left\|P_N\ket\psi\right\|}
    -
    P_N\ket\psi
    \right\|
    +
    \left\|
    P_N\ket\psi-e^{iTp(\hat X)}\ket\psi
    \right\| \\
    &=
    \left|
    1-\left\|P_N\ket\psi\right\|
    \right|
    +
    \left\|
    P_N\ket\psi-e^{iTp(\hat X)}\ket\psi
    \right\|
    <2\varepsilon .
\end{aligned}
\label{eq:complete-normalized-error}
\end{equation}

Finally, all \(N\) Rabi pulses have the common duration
\(\tau=\pi/(4L)\), so the total Rabi interaction time is
\(T_{\rm tot}=N\tau=\pi N/(4L)\).
Combining this relation with Eq.~\eqref{eq:complete-resource-pulse} gives
\begin{equation}
    T_{\rm tot}
    =
    O\!\left[
    \frac1L
    \left(
    T\sum_{r=0}^{R}|a_r|L^r
    +
    \log\frac1\varepsilon
    \right)^{1+o(1)}
    \right].
    \label{eq:complete-time-general}
\end{equation}
For fixed \(p\), \(n\), and \(T\), using
\(L=\Theta\!\left(\sqrt{\log(1/\varepsilon)}\right)\) and
Eq.~\eqref{eq:complete-pulse-fixed} yields
\begin{equation}
    T_{\rm tot}
    =
    O\!\left[
    \left(
    \log\frac1\varepsilon
    \right)^{(R-1)/2+o(1)}
    \right].
    \label{eq:complete-time-fixed}
\end{equation}

The pulse directions and ancilla boundary states are obtained constructively from the Laurent coefficients through the complementary-polynomial construction and recursive GQSP factorization described in Lemma~\ref{lem:laurent-gqsp} and Theorem~\ref{thm:polynomial-rabi-realization}.

\section{Lower bound on the total interaction time of arbitrary same-quadrature Rabi sequences}
\label{app:lower-bound}

This section proves the lower bound stated in Theorem~2 of the main text for arbitrary finite same-quadrature Rabi sequences with unequal pulse durations.
Using the notation of the main text, let
\(\mathcal U_N=\mathcal X_{\omega_N}(t_N)\cdots\mathcal X_{\omega_1}(t_1)\),
with \(T_{\rm tot}=\sum_{j=1}^{N}t_j\), and let \(P_N\) denote the corresponding postselected oscillator block.
The proof first relates \(T_{\rm tot}\) to the variation of the scalar function \(P_N(x)\), and then uses the approximation to \(e^{iTp(x)}\) to lower-bound this variation.

To lower-bound the variation of \(P_N(x)\), we use the alternating behavior of the real part of the target phase. For each integer \(k\), define
\(J_k=[k\pi-\pi/6,\,k\pi+\pi/6]\).
On adjacent intervals \(J_k\) and \(J_{k+1}\), the cosine has opposite signs and remains bounded away from zero. The following two lemmas use this property to relate the approximation error to the variation of \(P_N(x)\).

\begin{lemma}[Function-value separation on adjacent phase intervals]
\label{lem:unequal-time-value-gap}
Let $L>0$ and $\ell>0$, and suppose that a real-valued function $\Phi$ and a complex-valued function $P$ satisfy
\begin{equation}
  \int_{\mathbb R}\frac{\mathrm e^{-x^2}}{\sqrt\pi}
  \left\lvert P(x)-\mathrm e^{\mathrm i\Phi(x)}\right\rvert^2\,\mathrm dx
  \le \varepsilon^2.
  \label{eq:unequal-time-lemma-one-error}
\end{equation}
If, for some integer $j$, there exist intervals $I_j,I_{j+1}\subset[L/2,L]$ such that $\Phi(I_k)\subset J_k$ and $\lvert I_k\rvert\ge\ell$ for $k=j,j+1$, and if $\frac{\ell\,\mathrm e^{-L^2}}{16\sqrt\pi}>\varepsilon^2,$ then there exist $x_j\in I_j$ and $x_{j+1}\in I_{j+1}$ such that
\begin{equation}
  \left\lvert
  \operatorname{Re}P(x_{j+1})-\operatorname{Re}P(x_j)
  \right\rvert
  \ge \sqrt3-\frac12.
  \label{eq:unequal-time-lemma-one-conclusion}
\end{equation}
\end{lemma}

\begin{proof}
Fix $k\in\{j,j+1\}$. We first show that $I_k$ contains at least one point $x_k$ satisfying $\lvert P(x_k)-\mathrm e^{\mathrm i\Phi(x_k)}\rvert\le1/4$. Suppose, to the contrary, that no such point exists. Then, for every $x\in I_k$, one has $\lvert P(x)-\mathrm e^{\mathrm i\Phi(x)}\rvert>1/4$. Since $I_k\subset[L/2,L]$, one also has $x^2\le L^2$ and hence $\mathrm e^{-x^2}\ge\mathrm e^{-L^2}$. It follows that
\begin{align}
  \int_{\mathbb R}\frac{\mathrm e^{-x^2}}{\sqrt\pi}
  \left\lvert P(x)-\mathrm e^{\mathrm i\Phi(x)}\right\rvert^2\,\mathrm dx
  &\ge
  \int_{I_k}\frac{\mathrm e^{-x^2}}{\sqrt\pi}
  \left\lvert P(x)-\mathrm e^{\mathrm i\Phi(x)}\right\rvert^2\,\mathrm dx
  \notag\\
  &>
  \int_{I_k}\frac{\mathrm e^{-L^2}}{16\sqrt\pi}\,\mathrm dx =
  \frac{\lvert I_k\rvert\,\mathrm e^{-L^2}}{16\sqrt\pi}
  \notag\\
  &\ge
  \frac{\ell\,\mathrm e^{-L^2}}{16\sqrt\pi} > \varepsilon^2.
  \label{eq:unequal-time-lemma-one-contradiction}
\end{align}
Equation~\eqref{eq:unequal-time-lemma-one-contradiction} contradicts Eq.~\eqref{eq:unequal-time-lemma-one-error}, so the required point $x_k$ must exist. Applying this conclusion separately to $k=j$ and $k=j+1$, we may choose $x_j\in I_j$ and $x_{j+1}\in I_{j+1}$ such that
\begin{equation}
  \left\lvert P(x_k)-\mathrm e^{\mathrm i\Phi(x_k)}\right\rvert
  \le\frac14.
  \label{eq:unequal-time-selected-point-errors}
\end{equation}
Equation~\eqref{eq:unequal-time-selected-point-errors} holds separately for $k=j$ and $k=j+1$.

Because $\Phi(x_k)\in J_k$, there exists $u_k\in[-\pi/6,\pi/6]$ such that $\Phi(x_k)=k\pi+u_k$. Therefore,
\begin{align}
  (-1)^k\cos\Phi(x_k)=(-1)^k\cos(k\pi+u_k)=\cos u_k\ge\frac{\sqrt3}{2}.
  \label{eq:unequal-time-target-real-margin}
\end{align}
Moreover, $\cos\Phi(x_k)=\operatorname{Re}\mathrm e^{\mathrm i\Phi(x_k)}$, and the absolute value of the real part of any complex number is no greater than its modulus. Equation~\eqref{eq:unequal-time-selected-point-errors} thus gives
\begin{align}
  (-1)^k\operatorname{Re}P(x_k)
  &=(-1)^k\cos\Phi(x_k)
  +(-1)^k\operatorname{Re}\!\left[P(x_k)-\mathrm e^{\mathrm i\Phi(x_k)}\right]
  \notag\\
  &\ge
  \frac{\sqrt3}{2}
  -\left\lvert\operatorname{Re}\!\left[P(x_k)-\mathrm e^{\mathrm i\Phi(x_k)}\right]\right\rvert
  \notag\\
  &\ge
  \frac{\sqrt3}{2}
  -\left\lvert P(x_k)-\mathrm e^{\mathrm i\Phi(x_k)}\right\rvert
  \notag\\
  &\ge
  \frac{\sqrt3}{2}-\frac14.
  \label{eq:unequal-time-real-part-margin}
\end{align}
For $k=j$, Eq.~\eqref{eq:unequal-time-real-part-margin} gives $(-1)^j\operatorname{Re}P(x_j)\ge\sqrt3/2-1/4$. For $k=j+1$, using $(-1)^{j+1}=-(-1)^j$, the same inequality gives $(-1)^j\operatorname{Re}P(x_{j+1})\le-(\sqrt3/2-1/4)$. Hence,
\begin{align}
  \left\lvert\operatorname{Re}P(x_{j+1})-\operatorname{Re}P(x_j)  \right\rvert
  &=\left\lvert(-1)^j\operatorname{Re}P(x_{j+1})-(-1)^j\operatorname{Re}P(x_j)\right\rvert\notag\\
  &\ge2\left(\frac{\sqrt3}{2}-\frac14\right) =\sqrt3-\frac12.
\end{align}
This proves the lemma.
\end{proof}

\begin{lemma}[Separation between points in adjacent phase intervals]
\label{lem:unequal-time-point-distance}
Let $\Phi$ be a real-valued function that is continuous on $[L/2,L]$ and differentiable in its interior, and suppose that $\lvert\Phi'(x)\rvert\ge\mu>0$ for every $x\in[L/2,L]$. If $x_j,x_{j+1}\in[L/2,L]$ satisfy $\Phi(x_j)\in J_j$ and $\Phi(x_{j+1})\in J_{j+1}$, then
\begin{equation}
  \left\lvert x_{j+1}-x_j\right\rvert
  \le\frac{4\pi}{3\mu}.
  \label{eq:unequal-time-lemma-two-conclusion}
\end{equation}
\end{lemma}

\begin{proof}
Because $\Phi(x_j)\in J_j$ and $\Phi(x_{j+1})\in J_{j+1}$, their absolute phase difference is no greater than the difference between the left endpoint of $J_j$ and the right endpoint of $J_{j+1}$. Thus,
\begin{align}
  \left\lvert\Phi(x_{j+1})-\Phi(x_j)\right\rvert \le\left((j+1)\pi+\frac{\pi}{6}\right)-\left(j\pi-\frac{\pi}{6}\right) = \frac{4\pi}{3}.
  \label{eq:unequal-time-phase-difference-upper}
\end{align}
Since $J_j$ and $J_{j+1}$ are disjoint, $x_j$ and $x_{j+1}$ cannot coincide. The function $\Phi$ is continuous on the closed interval determined by these two points and differentiable in its interior. The mean-value theorem therefore guarantees a point $\xi$ between them such that
\begin{align}
  \left\lvert\Phi(x_{j+1})-\Phi(x_j)\right\rvert =\left\lvert\Phi'(\xi)\right\rvert \left\lvert x_{j+1}-x_j\right\rvert \ge
  \mu\left\lvert x_{j+1}-x_j\right\rvert.
  \label{eq:unequal-time-distance-mvt}
\end{align}
Combining Eqs.~\eqref{eq:unequal-time-phase-difference-upper} and~\eqref{eq:unequal-time-distance-mvt} yields Eq.~\eqref{eq:unequal-time-lemma-two-conclusion}.
\end{proof}

\begin{theorem}[Lower bound on the total interaction time of arbitrary unequal-duration same-quadrature Rabi sequences]
\label{thm:unequal-rabi-total-time-lower-bound}
In the unequal-duration Rabi control model defined above, suppose that every normalized state $\lvert\psi\rangle\in\mathcal H_n$ satisfies
\begin{equation}
  \left\|
  \left(P_N-U_p(T)\right)
  \lvert\psi\rangle
  \right\|
  \le\varepsilon,
  \label{eq:unequal-time-state-error}
\end{equation}
Then, for the fixed polynomial $p$, there exists a constant $c_p>0$, and for every fixed $T>0$ there exists $\varepsilon_{p,T}>0$, such that whenever $0<\varepsilon<\varepsilon_{p,T}$, one has $T_{\rm tot} \ge c_p T\left(\log\frac1\varepsilon\right)^{\frac{R-1}{2}}$. Equivalently, $T_{\rm tot}=\Omega\!\left(T(\log(1/\varepsilon))^{(R-1)/2}\right)$.
\end{theorem}

\begin{proof}
The proof has two parts. We first show that, for every $x\in\mathbb R$, one has $T_{\rm tot}\ge2\lvert P_N'(x)\rvert$. We then set $L=\sqrt{\log(1/\varepsilon)}$ and prove that $\sup_{x\in\mathbb R}\lvert P_N'(x)\rvert$ is bounded below by a constant multiple of $T L^{R-1}$. Combining these two estimates gives the lower bound $c_p T\left(\log\frac1\varepsilon\right)^{\frac{R-1}{2}}$.

In the $X$ representation, denote the eigenvalue of the position quadrature by $x$. The $j$th Rabi pulse at fixed $x$ reduces to the ancilla-qubit matrix $V_j(x)=\exp(\mathrm i t_jx\sigma_{\omega_j}/2)$. Accordingly, $\mathcal U_N(x)=V_N(x)\cdots V_1(x)$, and the postselected function is $P_N(x)={}_a\langle\phi_{\rm out}\rvert\mathcal U_N(x)  \lvert\phi_{\rm in}\rangle_a.$ Differentiating a single factor gives $V_j'(x)=\mathrm i (t_j/2)\sigma_{\omega_j}V_j(x)$. Applying the product rule to the finite product $\mathcal U_N(x)=V_N(x)\cdots V_1(x)$ yields
\begin{equation}
  \bigl(\mathcal U_N\bigr)'(x)=
  \sum_{k=1}^{N}
  V_N(x)\cdots V_{k+1}(x)
  V_k'(x)
  V_{k-1}(x)\cdots V_1(x).
  \label{eq:unequal-time-product-derivative}
\end{equation}
The operator norm is defined by $\lVert A\rVert_{\rm op}=\sup_{\lVert v\rVert=1}\lVert Av\rVert$. Each $V_j(x)$ is unitary, as is $\sigma_{\omega_j}$. Therefore,
\begin{align}
  \left\lVert V_j'(x)\right\rVert_{\rm op}
   = \left\lVert\mathrm i \frac{t_j}{2}\sigma_{\omega_j}V_j(x)\right\rVert_{\rm op}=\frac{t_j}{2}\left\lVert\sigma_{\omega_j}V_j(x)\right\rVert_{\rm op} = \frac{t_j}{2}.
  \label{eq:unequal-time-single-derivative-norm}
\end{align}
In the $k$th term of Eq.~\eqref{eq:unequal-time-product-derivative}, every factor other than $V_k'(x)$ is unitary. Left or right multiplication by a unitary leaves the operator norm invariant, so the norm of this term equals $t_k/2$. The triangle inequality for the operator norm then gives
\begin{align}
  \left\lVert\bigl(\mathcal U_N\bigr)'(x)\right\rVert_{\rm op}
  &\le\sum_{k=1}^{N}\left\lVert V_N(x)\cdots V_{k+1}(x)V_k'(x) V_{k-1}(x)\cdots V_1(x) \right\rVert_{\rm op}\notag\\
  &\le \frac12\sum_{k=1}^{N}t_k = \frac{T_{\rm tot}}{2}.
  \label{eq:unequal-time-unitary-derivative-bound}
\end{align}
The ancilla input and postselection states are normalized. By the Cauchy--Schwarz inequality and the definition of the operator norm,
\begin{align}
  \left\lvert P_N'(x)\right\rvert
  =\left\lvert{}_a\langle\phi_{\rm out}\rvert \bigl(\mathcal U_N\bigr)'(x)\lvert\phi_{\rm in}\rangle_a\right\rvert
  \le\left\lVert\phi_{\rm out}\right\rVert\left\lVert\bigl(\mathcal U_N\bigr)'(x)\lvert\phi_{\rm in}\rangle\right\rVert
  \le\left\lVert \bigl(\mathcal U_N\bigr)'(x)  \right\rVert_{\rm op} \left\lVert\phi_{\rm in}\right\rVert
  \le \frac{T_{\rm tot}}{2}.
  \label{eq:unequal-time-pointwise-time-bound}
\end{align}
Since Eq.~\eqref{eq:unequal-time-pointwise-time-bound} holds for every real $x$, it follows that
\begin{equation}
  T_{\rm tot}\ge 2\sup_{x\in\mathbb R}\left\lvert P_N'(x)\right\rvert.
  \label{eq:unequal-time-time-vs-derivative}
\end{equation}

We next derive a lower bound on this derivative supremum. Set $L=\sqrt{\log(1/\varepsilon)}$. It suffices to obtain a lower bound on the smaller interval $[L/2,L]$ for $\lvert\operatorname{Re}P_N'(x)\rvert$, which in turn controls $\lvert P_N'(x)\rvert$ on the full real line, because
\begin{align}
  \sup_{x\in\mathbb R}\left\lvert P_N'(x)\right\rvert
  \ge\sup_{x\in[L/2,L]}\left\lvert P_N'(x)\right\rvert
  \ge\sup_{x\in[L/2,L]}\left\lvert\operatorname{Re}P_N'(x)\right\rvert.
  \label{eq:unequal-time-restrict-window}
\end{align}
We take the real part so that the alternating signs of $\operatorname{Re}\mathrm e^{\mathrm iT p(x)}=\cos[Tp(x)]$ on adjacent phase intervals can be used below. For any two distinct points in $[L/2,L]$, denoted by $x_j$ and $x_{j+1}$, the fundamental theorem of calculus gives
\begin{align}
  \left\lvert
  \operatorname{Re}P_N(x_{j+1})-
  \operatorname{Re}P_N(x_j)
  \right\rvert
  &=\left\lvert
  \int_{x_j}^{x_{j+1}}
  \operatorname{Re}P_N'(x)\,\mathrm dx
  \right\rvert
  \notag\\
  &\le
  \int_{\min\{x_j,x_{j+1}\}}^{\max\{x_j,x_{j+1}\}}
  \left\lvert\operatorname{Re}P_N'(x)\right\rvert\,\mathrm dx
  \notag\\
  &\le
  \left\lvert x_{j+1}-x_j\right\rvert
  \sup_{x\in[L/2,L]}
  \left\lvert\operatorname{Re}P_N'(x)\right\rvert.
  \label{eq:unequal-time-quotient-origin}
\end{align}
Hence, for any such pair,
\begin{equation}
  \sup_{x\in[L/2,L]}
  \left\lvert\operatorname{Re}P_N'(x)\right\rvert
  \ge
  \frac{
  \left\lvert
  \operatorname{Re}P_N(x_{j+1})-
  \operatorname{Re}P_N(x_j)
  \right\rvert
  }{
  \left\lvert x_{j+1}-x_j\right\rvert
  }.
  \label{eq:unequal-time-quotient-bound}
\end{equation}
Equation~\eqref{eq:unequal-time-quotient-bound} holds for every pair of distinct points in the interval. Thus, to lower-bound its left-hand side, it is enough to exhibit one pair for which the numerator has a positive lower bound independent of $\varepsilon$, while the denominator is at most a constant multiple of $L^{1-R}/T$. The resulting quotient is then at least a constant multiple of $T L^{R-1}$. We now construct such a pair.

Termwise differentiation of $p(x)=\sum_{r=0}^{R}a_r x^r$ from the main text gives
\begin{align}
  p'(x)=\sum_{r=1}^{R}r a_r x^{r-1}=R a_Rx^{R-1}+\sum_{r=1}^{R-1}r a_r x^{r-1}.
  \label{eq:unequal-time-polynomial-derivative}
\end{align}
Define $x_p=\max\!\left\{ 1, \frac{2\sum_{r=1}^{R-1}r\lvert a_r\rvert}{R\lvert a_R\rvert} \right\}.$ If $x\ge x_p$, then in particular $x\ge1$. For every $1\le r\le R-1$, one has $r-1\le R-2$, and therefore $x^{r-1}\le x^{R-2}$. The triangle inequality gives
\begin{align}
  \left\lvert\sum_{r=1}^{R-1}r a_r x^{r-1}\right\rvert
  \le\sum_{r=1}^{R-1}r\lvert a_r\rvert x^{r-1}
  \le\left(\sum_{r=1}^{R-1}r\lvert a_r\rvert\right)x^{R-2}.
  \label{eq:unequal-time-lower-order-prebound}
\end{align}
The condition $x\ge x_p$ implies $\sum_{r=1}^{R-1}r\lvert a_r\rvert\le R\lvert a_R\rvert x/2$. Multiplying this inequality by the nonnegative factor $x^{R-2}$ and substituting the result into Eq.~\eqref{eq:unequal-time-lower-order-prebound} yields
\begin{equation}
  \left\lvert
  \sum_{r=1}^{R-1}r a_r x^{r-1}
  \right\rvert
  \le
  \frac{R\lvert a_R\rvert}{2}x^{R-1}.
  \label{eq:unequal-time-lower-order-half}
\end{equation}

Applying the reverse triangle inequality $\lvert A+B\rvert\ge\lvert A\rvert-\lvert B\rvert$ to Eq.~\eqref{eq:unequal-time-polynomial-derivative} and then using Eq.~\eqref{eq:unequal-time-lower-order-half}, we obtain
\begin{align}
  \left\lvert p'(x)\right\rvert
  \ge \left\lvert R a_Rx^{R-1}\right\rvert-\left\lvert \sum_{r=1}^{R-1}r a_r x^{r-1}\right\rvert
  \ge R\lvert a_R\rvert x^{R-1} -\frac{R\lvert a_R\rvert}{2}x^{R-1}
  =\frac{R\lvert a_R\rvert}{2}x^{R-1}.
  \label{eq:unequal-time-derivative-lower}
\end{align}
Similarly, using $\lvert A+B\rvert\le\lvert A\rvert+\lvert B\rvert$ gives
\begin{align}
  \left\lvert p'(x)\right\rvert
  \le\left\lvert R a_Rx^{R-1}\right\rvert+\left\lvert\sum_{r=1}^{R-1}r a_r x^{r-1}\right\rvert
  \le R\lvert a_R\rvert x^{R-1}+\frac{R\lvert a_R\rvert}{2}x^{R-1}
  =\frac{3R\lvert a_R\rvert}{2}x^{R-1}.
  \label{eq:unequal-time-derivative-upper}
\end{align}
Equation~\eqref{eq:unequal-time-lower-order-half} shows that the absolute value of the lower-order contribution is at most half that of the leading term. Hence $p'(x)$, for $x\ge x_p$, has the same sign as $a_R$. If $a_R<0$, replace $p$ by $-p$ and $P_N$ by $\overline{P_N}$ in the scalar-function estimates below. The weighted error and $\lvert P_N'(x)\rvert$ are unchanged. It therefore suffices to consider the case $a_R>0$.

Since $L=\sqrt{\log(1/\varepsilon)}$, the condition $L/2\ge x_p$ is equivalent to $\log(1/\varepsilon)\ge4x_p^2$. Thus, whenever $\varepsilon\le\mathrm e^{-4x_p^2}$, one has $L/2\ge x_p$. Under this condition, Eqs.~\eqref{eq:unequal-time-derivative-lower} and~\eqref{eq:unequal-time-derivative-upper} imply, respectively, that every $x\in[L/2,L]$ satisfies
\begin{equation}
  \frac{R\lvert a_R\rvert}{2^R}L^{R-1}
  \le p'(x)
  \le\frac{3R\lvert a_R\rvert}{2}L^{R-1}.
  \label{eq:unequal-time-window-derivative}
\end{equation}
The lower bound uses $x\ge L/2$, whereas the upper bound uses $x\le L$. Define $\Phi(x)=T p(x)$. Since $T>0$, the function $\Phi$ is also strictly increasing on $[L/2,L]$, and
\begin{equation}
  \frac{TR\lvert a_R\rvert}{2^R}L^{R-1}
  \le \Phi'(x)
  \le\frac{3TR\lvert a_R\rvert}{2}L^{R-1}.
  \label{eq:unequal-time-phase-derivative-window}
\end{equation}

The fundamental theorem of calculus and the lower bound in Eq.~\eqref{eq:unequal-time-phase-derivative-window} give
\begin{align}
  \Phi(L)-\Phi(L/2)=\int_{L/2}^{L}\Phi'(x)\,\mathrm dx
 \ge\int_{L/2}^{L}\frac{TR\lvert a_R\rvert}{2^R}L^{R-1}\,\mathrm dx
 =\frac{TR\lvert a_R\rvert}{2^{R+1}}L^R.
  \label{eq:unequal-time-phase-span}
\end{align}
If $L^R \ge \frac{3\pi\,2^{R+1}}{TR\lvert a_R\rvert},$ then Eq.~\eqref{eq:unequal-time-phase-span} immediately gives $\Phi(L)-\Phi(L/2)\ge3\pi$. Because $L^R=(\log(1/\varepsilon))^{R/2}$ diverges as $\varepsilon\to0$, this condition holds for every fixed $T>0$ once $\varepsilon$ is sufficiently small. More explicitly, it is sufficient that $\varepsilon\le\exp\!\left[-\left(\frac{3\pi\,2^{R+1}}{TR\lvert a_R\rvert}  \right)^{2/R}\right].$

Choose $j=\left\lceil(\Phi(L/2)+\pi/6)/\pi\right\rceil$. The ceiling function satisfies
\begin{equation}
  \Phi(L/2)+\frac{\pi}{6}
  \le j\pi
  <\Phi(L/2)+\frac{7\pi}{6}.
  \label{eq:unequal-time-ceiling-bound}
\end{equation}
The left inequality implies $j\pi-\pi/6\ge\Phi(L/2)$. From the right inequality,
\begin{align}
  (j+1)\pi+\frac{\pi}{6}
  =j\pi+\frac{7\pi}{6}
  <\Phi(L/2)+\frac{7\pi}{3}
  <\Phi(L/2)+3\pi
  \le\Phi(L).
  \label{eq:unequal-time-platform-containment-upper}
\end{align}
Thus, both $J_j$ and $J_{j+1}$ are contained in $\Phi([L/2,L])=[\Phi(L/2),\Phi(L)]$.

Because $\Phi$ is continuous and strictly increasing on $[L/2,L]$, it has a single-valued inverse on this interval. For $k=j,j+1$, denote the unique preimage of $k\pi-\pi/6$ on this interval by $\ell_k$, namely $\ell_k:=\Phi^{-1}(k\pi-\pi/6)$. Denote the unique preimage of $k\pi+\pi/6$ by $r_k$, namely $r_k:=\Phi^{-1}(k\pi+\pi/6)$. Define $I_k:=[\ell_k,r_k]$. Strict monotonicity gives $\ell_k<r_k$ and $\Phi(I_k)=J_k$. The endpoint definitions imply $\Phi(\ell_k)=k\pi-\pi/6$ and $\Phi(r_k)=k\pi+\pi/6$, and therefore
\begin{align}
  \frac{\pi}{3}
  =\Phi(r_k)-\Phi(\ell_k)
  =T\bigl[p(r_k)-p(\ell_k)\bigr].
  \label{eq:unequal-time-platform-phase-width}
\end{align}
The first equality is the difference between the endpoints of the phase interval $J_k$, and the second uses $\Phi(x)=Tp(x)$.

The function $\Phi$ is continuous on $[\ell_k,r_k]$ and differentiable in its interior. The mean-value theorem therefore guarantees a point $\xi_k\in(\ell_k,r_k)$ such that
\begin{align}
  \frac{\pi}{3}=\Phi'(\xi_k)(r_k-\ell_k)\le\frac{3TR\lvert a_R\rvert}{2}L^{R-1}\lvert I_k\rvert.
  \label{eq:unequal-time-platform-length-mvt}
\end{align}
The final inequality uses the upper bound in Eq.~\eqref{eq:unequal-time-phase-derivative-window} and the fact that $\xi_k\in I_k\subset[L/2,L]$. Hence,
\begin{equation}
  \lvert I_k\rvert
  \ge
  \frac{2\pi}{9TR\lvert a_R\rvert}L^{1-R}.
  \label{eq:unequal-time-platform-length-lower}
\end{equation}
Equation~\eqref{eq:unequal-time-platform-length-lower} holds for both $k=j$ and $k=j+1$.

The hypothesis~\eqref{eq:unequal-time-state-error} holds for every normalized state in $\mathcal H_n$. Since the vacuum $\lvert0\rangle$ is a normalized state in $\mathcal H_n$, every sequence satisfying the theorem hypothesis must also obey
\begin{equation}
  \left\|
  \left(P_N-U_p(T)\right)
  \lvert0\rangle
  \right\|
  \le\varepsilon.
  \label{eq:unequal-time-vacuum-error}
\end{equation}
Consequently, any lower bound derived from this necessary vacuum-state condition also applies to sequences required to approximate the target uniformly over the entire subspace $\mathcal H_n$.

The vacuum wave function in the $X$ representation is $\langle x\vert0\rangle=\pi^{-1/4}\mathrm e^{-x^2/2}$. In the $X$ representation, $U_p(T)$ acts by multiplication with the scalar function $\mathrm e^{\mathrm iT p(x)}=\mathrm e^{\mathrm i\Phi(x)}$, whereas $P_N$ acts by multiplication with $P_N(x)$. Squaring Eq.~\eqref{eq:unequal-time-vacuum-error} therefore gives
\begin{align}
  \left\|
  \left(P_N-U_p(T)\right)
  \lvert0\rangle
  \right\|^2
  &=\int_{\mathbb R}
  \left\lvert P_N(x)-\mathrm e^{\mathrm iT p(x)}\right\rvert^2
  \left\lvert\langle x\vert0\rangle\right\rvert^2\,\mathrm dx
  \notag\\
  &=\int_{\mathbb R}\frac{\mathrm e^{-x^2}}{\sqrt\pi}
  \left\lvert P_N(x)-\mathrm e^{\mathrm i\Phi(x)}\right\rvert^2\,\mathrm dx
  \notag\\
  &\le\varepsilon^2.
  \label{eq:unequal-time-vacuum-weighted-error}
\end{align}

Define $\ell= \frac{2\pi}{9TR\lvert a_R\rvert}L^{1-R}.$ Equation~\eqref{eq:unequal-time-platform-length-lower} shows that $\lvert I_k\rvert\ge\ell$ for $k=j,j+1$. The quantity $\ell\mathrm e^{-L^2}/(16\sqrt\pi)$ in Lemma~\ref{lem:unequal-time-value-gap} is precisely the lower bound on the weighted squared error contributed by a single phase interval under the assumption that $\lvert P_N(x)-\mathrm e^{\mathrm i\Phi(x)}\rvert>1/4$ everywhere on that interval. Dividing it by the total admissible error $\varepsilon^2$ measures whether this local lower bound alone exceeds the entire error budget.

Since $L^2=\log(1/\varepsilon)$, one has $\varepsilon=\mathrm e^{-L^2}$. Therefore,
\begin{align}
  \frac{\ell\,\mathrm e^{-L^2}}{16\sqrt\pi\,\varepsilon^2}
  =\frac{\sqrt\pi}{72TR\lvert a_R\rvert}\frac{L^{1-R}}{\varepsilon}
  =\frac{\sqrt\pi}{72TR\lvert a_R\rvert}\frac{\mathrm e^{L^2}}{L^{R-1}}.
  \label{eq:unequal-time-detectability-ratio}
\end{align}
To evaluate the limiting behavior of the final expression directly, use the power-series expansion of the exponential. For $L>0$,
\begin{align}
  \frac{\mathrm e^{L^2}}{L^{R-1}}
  =\frac{1}{L^{R-1}}
  \sum_{m=0}^{\infty}\frac{L^{2m}}{m!}
  \ge
  \frac{1}{L^{R-1}}\frac{L^{2R}}{R!}
  =\frac{L^{R+1}}{R!}
  \longrightarrow+\infty
  \qquad(L\to\infty).
  \label{eq:unequal-time-exponential-dominates}
\end{align}
As $\varepsilon\to0$, one has $L=\sqrt{\log(1/\varepsilon)}\to\infty$. The ratio in Eq.~\eqref{eq:unequal-time-detectability-ratio} therefore eventually exceeds $1$. Equivalently, for all sufficiently small $\varepsilon$, one has $\ell\,\mathrm e^{-L^2}/(16\sqrt\pi)>\varepsilon^2$.

The interval-length estimate and the Gaussian-weighted vacuum error together rule out the possibility that the approximation error exceeds \(1/4\) everywhere on either of the two adjacent phase intervals. Lemma~\ref{lem:unequal-time-value-gap} therefore allows us to select \(x_j\in I_j\) and \(x_{j+1}\in I_{j+1}\) such that
\begin{equation}
  \left\lvert
  \operatorname{Re}P_N(x_{j+1})-
  \operatorname{Re}P_N(x_j)
  \right\rvert
  \ge\sqrt3-\frac12.
  \label{eq:unequal-time-selected-value-gap}
\end{equation}

For the same pair of points, Eq.~\eqref{eq:unequal-time-phase-derivative-window} gives $\lvert\Phi'(x)\rvert\ge TR\lvert a_R\rvert L^{R-1}/2^R$ for every $x\in[L/2,L]$, while the interval construction gives $\Phi(x_j)\in J_j$ and $\Phi(x_{j+1})\in J_{j+1}$. Taking $\mu=TR\lvert a_R\rvert L^{R-1}/2^R$ in Lemma~\ref{lem:unequal-time-point-distance}, we obtain
\begin{align}
  \left\lvert x_{j+1}-x_j\right\rvert
  \le \frac{4\pi}{3}\frac{2^R}{TR\lvert a_R\rvert L^{R-1}}
  =\frac{2^{R+2}\pi}{3TR\lvert a_R\rvert}L^{1-R}.
  \label{eq:unequal-time-selected-distance}
\end{align}

Substituting Eqs.~\eqref{eq:unequal-time-selected-value-gap} and~\eqref{eq:unequal-time-selected-distance} into Eq.~\eqref{eq:unequal-time-quotient-bound} gives
\begin{align}
  \sup_{x\in[L/2,L]}\left\lvert\operatorname{Re}P_N'(x)\right\rvert
  \ge \frac{3TR\lvert a_R\rvert}{2^{R+2}\pi}\left(\sqrt3-\frac12\right)L^{R-1}.
  \label{eq:unequal-time-real-derivative-lower}
\end{align}
Finally, combining Eqs.~\eqref{eq:unequal-time-time-vs-derivative} and~\eqref{eq:unequal-time-restrict-window} with $L=\sqrt{\log(1/\varepsilon)}$ yields
\begin{align}
  T_{\rm tot}
  \ge 2\sup_{x\in\mathbb R}\left\lvert P_N'(x)\right\rvert
  \ge \frac{3R\lvert a_R\rvert}{2^{R+1}\pi} \left(\sqrt3-\frac12\right)
  T\left(\log\frac1\varepsilon\right)^{\frac{R-1}{2}}.
  \label{eq:unequal-time-final-chain}
\end{align}
Thus, one may take $c_p=3R\lvert a_R\rvert(\sqrt3-1/2)/(2^{R+1}\pi)$. The derivation requires $L/2\ge x_p$, $L^R \ge \frac{3\pi\,2^{R+1}}{TR\lvert a_R\rvert}$, and the ratio in Eq.~\eqref{eq:unequal-time-detectability-ratio} to exceed $1$. For every fixed $p$ and $T>0$, these three conditions hold simultaneously once $\varepsilon$ is sufficiently small, which establishes the existence of the corresponding $\varepsilon_{p,T}>0$. This proves the theorem.
\end{proof}

\section{Efficient realization of phase-block products and Gaussian-gate decompositions}

This section extends the single-phase construction to finite products of polynomial phase blocks and applies the resulting composition bound to the Gaussian gates used in the main text.
Section~IV.A establishes the approximation and resource bounds for finite phase-block products, while Sec.~IV.B gives the explicit phase-block decompositions of the rotation and controlled-phase gates.

\subsection{Efficient realization of phase-block products}
\label{app:product-stability-proof}

\begin{lemma}[Tail estimate for fixed-degree polynomial intermediate observables]
\label{lem:polynomial-intermediate-tail}
Let $\mathcal H_n=\operatorname{span}\{|0\rangle,\ldots,|n\rangle\}.$ Let \(B=B^\dagger\) be a self-adjoint polynomial of fixed degree \(\mu\) in the single-mode operators \(X,P\).
For any \(L>0\), let \(E_B\) denote the projection-valued spectral measure of \(B\), and define the spectral projection
\(\Pi_L^B:=E_B([-L,L])\).
Then there exist constants \(c_*>0\) and \(L_*>0\), depending only on \(B\) and \(n\), such that for all \(L\ge L_*\),
\begin{equation}
    \sup_{\substack{|\psi\rangle\in\mathcal H_n\\ \|\psi\|=1}}
    \left\|(I-\Pi_L^B)|\psi\rangle\right\|^2
    \le\exp\!\left[-c_*L^{2/\mu}\right].
    \label{eq:polynomial-tail-final}
\end{equation}
\end{lemma}

\begin{proof}
If \(B=0\), then \(\Pi_L^B=I\) for every \(L>0\), and hence \((I-\Pi_L^B)|\psi\rangle=0.\) The claim is then immediate. We therefore assume \(B\neq0\).
Take an arbitrary unit vector \( |\psi\rangle\in\mathcal H_n\), and denote by
\(d\mu_\psi^B(b):=\langle\psi|\,dE_B(b)\,|\psi\rangle\)
the spectral measure of \(B\) associated with \(|\psi\rangle\).
Let \(k\ge1\) be an integer. On the domain \(|b|>L\), one has
\(1\le |b|^{2k}/L^{2k}\).
Using the spectral theorem,
\begin{align}
    \left\|(I-\Pi_L^B)|\psi\rangle\right\|^2
    &= \int_{|b|>L} d\mu_\psi^B(b)
    \le \int_{|b|>L}\frac{|b|^{2k}}{L^{2k}}\,d\mu_\psi^B(b) \notag\\
    &\le L^{-2k}\int_{\mathbb R}|b|^{2k}\,d\mu_\psi^B(b).
    \label{eq:outside-probability-controlled-by-moment-integral}
\end{align}
The same spectral theorem gives
\(\|B^k|\psi\rangle\|^2=\int_{\mathbb R}|b|^{2k}\,d\mu_\psi^B(b)\).
Substituting this identity into Eq.~\eqref{eq:outside-probability-controlled-by-moment-integral}, we obtain
\begin{equation}
    \left\|
        (I-\Pi_L^B)|\psi\rangle
    \right\|^2
    \le
    L^{-2k}
    \|B^k|\psi\rangle\|^2.
    \label{eq:markov-bound-for-B-window}
\end{equation}
Thus the problem reduces to estimating the high-order moment \(\|B^k|\psi\rangle\|^2\).

Since \(X=\frac{a+a^\dagger}{\sqrt2}\) and \(P=\frac{a-a^\dagger}{i\sqrt2}\), any self-adjoint polynomial of fixed degree \(\mu\) in \(X\) and \(P\) can be written as a finite-degree polynomial in \(a\) and \(a^\dagger\). Using \(aa^\dagger=a^\dagger a+I\), each product of \(a\) and \(a^\dagger\) can be put into normal-ordered form, with all \(a^\dagger\)'s to the left and all \(a\)'s to the right. Each interchange of adjacent \(a\) and \(a^\dagger\) produces only lower-degree terms, so this reordering does not increase the highest degree. Hence \(B\) admits a finite expansion
\begin{equation}
    B=\sum_{r+s\le\mu}c_{r,s}(a^\dagger)^r a^s.
    \label{eq:B-normal-ordered}
\end{equation}
Here \(c_{r,s}\in\mathbb C\), and \((a^\dagger)^0a^0=I\).

We next estimate the growth of \(B\) on finite Fock subspaces. Let
\(|\varphi\rangle = \sum_{\ell=0}^{n}d_\ell|\ell\rangle\in\mathcal H_n.\)
We first bound \(a^s|\varphi\rangle\):
\begin{align}
    \|a^s|\varphi\rangle\|^2
    &=\sum_{\ell=s}^{n}|d_\ell|^2\ell(\ell-1)\cdots(\ell-s+1)
    \le (n+1)^s\sum_{\ell=0}^{n}|d_\ell|^2 
    =(n+1)^s\|\varphi\|^2.
    \label{eq:as-bound}
\end{align}
We then bound \((a^\dagger)^r\). If
\(|\chi\rangle=\sum_{\ell=0}^{n}e_\ell|\ell\rangle   \in\mathcal H_n,\)
then
\begin{align}
    \|(a^\dagger)^r|\chi\rangle\|^2
    =\sum_{\ell=0}^{n}|e_\ell|^2(\ell+1)(\ell+2)\cdots(\ell+r) 
    \le (n+r+1)^r\sum_{\ell=0}^{n}|e_\ell|^2 
    \le (n+\mu+1)^r\|\chi\|^2.
    \label{eq:adag-r-bound}
\end{align}
The last step uses \(r\le\mu\).
Taking \(|\chi\rangle=a^s|\varphi\rangle\), and noting that \(a^s\) only lowers the Fock number, we have
\(a^s|\varphi\rangle\in\mathcal H_n.\)
Equations~\eqref{eq:as-bound} and \eqref{eq:adag-r-bound} then give
\begin{align}
    \|(a^\dagger)^r a^s|\varphi\rangle\|
    &\le (n+\mu+1)^{r/2}\|a^s|\varphi\rangle\| 
    \le (n+\mu+1)^{r/2}(n+1)^{s/2}\|\varphi\| \notag\\
    &\le (n+\mu+1)^{(r+s)/2} \|\varphi\| 
    \le (n+\mu+1)^{\mu/2} \|\varphi\|.
    \label{eq:normal-monomial-bound}
\end{align}
The last step uses \(r+s\le\mu\). Substituting Eq.~\eqref{eq:normal-monomial-bound} into Eq.~\eqref{eq:B-normal-ordered}, we obtain
\begin{align}
    \|B|\varphi\rangle\|
    &\le \sum_{r+s\le\mu}|c_{r,s}|\,\|(a^\dagger)^r a^s|\varphi\rangle\|
    \le \sum_{r+s\le\mu}|c_{r,s}|\,(n+\mu+1)^{\mu/2} \|\varphi\| \notag\\
    &:= \Gamma_B(n+\mu+1)^{\mu/2}\|\varphi\|,
    \label{eq:B-one-step-growth}
\end{align}
where \(\Gamma_B:= \sum_{r+s\le\mu}|c_{r,s}|\). Since this is a finite sum, \(\Gamma_B<\infty\).
Moreover, because \(r+s\le\mu\), each monomial \((a^\dagger)^r a^s\) maps \(\mathcal H_n\) into \(\mathcal H_{n+\mu}\). A finite linear combination of such monomials has the same property, and therefore
\(B\mathcal H_n \subseteq \mathcal H_{n+\mu}.\)

We now estimate \(\|B^k|\psi\rangle\|\). For every integer \(\ell\ge0\), one has
\(B^\ell|\psi\rangle\in  \mathcal H_{n+\ell\mu}.\)
Before the \(\ell\)-th application of \(B\), the state \(B^{\ell-1}|\psi\rangle\) belongs to
\(\mathcal H_{n+(\ell-1)\mu}.\)
Thus, applying Eq.~\eqref{eq:B-one-step-growth} with \(n\) replaced by \(n+(\ell-1)\mu\), we get
\begin{align}
    \|B^\ell|\psi\rangle\|
    &=\|B(B^{\ell-1}|\psi\rangle)\| 
    \le \Gamma_B\bigl(n+(\ell-1)\mu+\mu+1\bigr)^{\mu/2}\|B^{\ell-1}|\psi\rangle\| \notag\\
    &= \Gamma_B(n+\ell\mu+1)^{\mu/2}\|B^{\ell-1}|\psi\rangle\|.
    \label{eq:Bell-recursion}
\end{align}
Multiplying these estimates for \(\ell=1,2,\ldots,k\), and using \(\|\psi\|=1\), gives
\begin{align}
    \|B^k|\psi\rangle\|
    \le \prod_{\ell=1}^{k}\Gamma_B(n+\ell\mu+1)^{\mu/2} 
    =\Gamma_B^k\left[\prod_{\ell=1}^{k}(n+\ell\mu+1)\right]^{\mu/2}.
\end{align}
We now simplify the product. For \(1\le\ell\le k\), we have
\(n+\ell\mu+1 \le n+k\mu+1\), and
\( k(n+\mu+1)-(n+k\mu+1)=(k-1)(n+1)\ge0\).
Hence
\begin{align}
    \|B^k|\psi\rangle\|^2
    \le \Gamma_B^{2k}\left[\bigl(k(n+\mu+1)\bigr)^k\right]^\mu 
    =\left[\Gamma_B^2(n+\mu+1)^\mu k^\mu\right]^k :=\left(A_{B,n}k^\mu\right)^k,
    \label{eq:moment-bound-final}
\end{align}
where \(A_{B,n}:=\Gamma_B^2(n+\mu+1)^\mu.\)

Substituting Eq.~\eqref{eq:moment-bound-final} into Eq.~\eqref{eq:markov-bound-for-B-window}, we find that, for every integer \(k\ge1\),
\begin{align}
    \left\|(I-\Pi_L^B)|\psi\rangle\right\|^2
    \le L^{-2k}\|B^k|\psi\rangle\|^2 \le L^{-2k}\left(A_{B,n}k^\mu\right)^k 
    = \left(\frac{A_{B,n}k^\mu}{L^2}\right)^k.
    \label{eq:tail-before-k-choice}
\end{align}
Choose
\(k=\left\lfloor(4A_{B,n})^{-1/\mu}L^{2/\mu}\right\rfloor .\)
Define \( L_*:=\left[2(4A_{B,n})^{1/\mu}\right]^{\mu/2}.\)
For \(L\ge L_*\), we have
\((4A_{B,n})^{-1/\mu}L^{2/\mu}\ge 2\), and hence \(k\ge 1\).
We first use the upper bound on \(k\) to control the factor in parentheses. By the definition of the floor function,
\(k\le (4A_{B,n})^{-1/\mu}L^{2/\mu}.\)
Since \(\mu>0\), raising both sides to the power \(\mu\) gives
\(k^\mu \le (4A_{B,n})^{-1}L^2 .\)
Therefore
\begin{align}
    \frac{A_{B,n}k^\mu}{L^2}\le\frac{A_{B,n}(4A_{B,n})^{-1}L^2}{L^2} = \frac14 .
    \label{eq:factor-one-fourth-direct}
\end{align}
Substituting this estimate into Eq.~\eqref{eq:tail-before-k-choice}, we obtain
\begin{align}
    \left\|(I-\Pi_L^B)|\psi\rangle\right\|^2
    \le \left(\frac{A_{B,n}k^\mu}{L^2}\right)^k \le\left(\frac14\right)^k = e^{-k\log 4}.
    \label{eq:tail-after-k-choice-direct}
\end{align}

It remains to use the lower bound on \(k\) to control the exponent. Again by the definition of the floor function,
\(k \ge (4A_{B,n})^{-1/\mu}L^{2/\mu}-1.\)
Using \(x-1\ge \frac12 x\) for \(x\ge2\), we get
\(k \ge \frac12(4A_{B,n})^{-1/\mu}L^{2/\mu}.\)
Substituting this lower bound into the exponent \(e^{-k\log4}\), we obtain
\begin{equation}
    -k\log 4\le -\frac{\log4}{2}(4A_{B,n})^{-1/\mu}L^{2/\mu}.
\end{equation}
Equation~\eqref{eq:tail-after-k-choice-direct} therefore gives
\begin{equation}
    \left\|(I-\Pi_L^B)|\psi\rangle\right\|^2
    \le\exp\!\left[-\frac{\log4}{2}(4A_{B,n})^{-1/\mu}L^{2/\mu}\right]:=e^{-c_*L^{2/\mu}},
\end{equation}
where \(c_*:=\frac{\log4}{2}(4A_{B,n})^{-1/\mu}.\)

\end{proof}

\begin{theorem}[Efficient realization of finite products of phase blocks]
\label{thm:efficient-product-phase-app}
Let \(\mathcal H_n=\operatorname{span}\{|0\rangle,\ldots,|n\rangle\}\), and let
\(U_{\rm tot}=U_MU_{M-1}\cdots U_1\), with
\(U_j=e^{ip_j(Q_j)}\).
Here \(p_j(q)=\sum_{r=0}^{R_j}a_{j,r}q^r\) is a real polynomial of fixed degree \(R_j\), and \(Q_j\) is a real linear quadrature.
The fixed evolution time \(T_j\) used in the main text is absorbed into the coefficients \(a_{j,r}\).

Define \(W_j:=U_{j-1}U_{j-2}\cdots U_1\), with \(W_1:=I\), and
\(B_j:=W_j^\dagger Q_jW_j\).
Suppose each \(B_j\) is a self-adjoint polynomial in \(X,P\) of fixed degree \(\mu_j\).
Then, for any \(0<\varepsilon<1\), there exist Laurent-polynomial approximants
\(K_j=P_{j,d_j}(e^{i\tau_jQ_j})\) such that, for every normalized state
\(|\psi\rangle\in\mathcal H_n\),
\begin{equation}
    \left\|
    \left(
    K_MK_{M-1}\cdots K_1
    -
    U_MU_{M-1}\cdots U_1
    \right)
    |\psi\rangle
    \right\|
    <\varepsilon .
    \label{target_M}
\end{equation}
If \(M\), \(p_j\), \(B_j\), and \(n\) are held fixed as
\(\varepsilon\to0\), and \(R_j\mu_j/2\ge1\), the Laurent bandwidths can be chosen as
\(d_j=O[(\log(M/\varepsilon))^{R_j\mu_j/2+o(1)}]\).
The corresponding total Rabi interaction time satisfies
\begin{equation}
    T_{\rm tot} = O\!\left[\sum_{j=1}^{M} \left( \log\frac{M}{\varepsilon} \right)^{(R_j-1)\mu_j/2+o(1)} \right].
    \label{eq:phase-product-total-time}
\end{equation}
The total number of Rabi pulses is \(2\sum_{j=1}^{M}d_j\), and the joint postselection success probability satisfies
\(p_{\rm succ}(\psi)>(1-\varepsilon)^2\).
\end{theorem}

\begin{proof}
Let \( |\psi\rangle\in\mathcal H_n \) be an arbitrary normalized state. By the telescoping identity for products, we have
\begin{align}
    &\left\|\left[K_MK_{M-1}\cdots K_1-U_MU_{M-1}\cdots U_1\right]|\psi\rangle\right\| \notag\\
    &= \left\| \left[ \sum_{j=1}^{M} K_MK_{M-1}\cdots K_{j+1} (K_j-U_j)U_{j-1}U_{j-2}\cdots U_1 \right]|\psi\rangle\right\| \notag\\
    &\le\sum_{j=1}^{M}\left\|K_MK_{M-1}\cdots K_{j+1}(K_j-U_j)W_j|\psi\rangle\right\|.
    \label{eq:telescoping-norm-proof}
\end{align}
Here, when \(j=M\), the product \(K_MK_{M-1}\cdots K_{j+1}\) is an empty product and is defined to be the identity operator \(I\). Similarly, when \(j=1\), the product \(U_{j-1}U_{j-2}\cdots U_1\) is also an empty product and is defined to be \(I\).

We now fix \(j\in\{1,\ldots,M\}\) and estimate
\(\|(K_j-U_j)W_j|\psi\rangle\|\).
Let \( |\phi_j\rangle:=W_j|\psi\rangle.\) Choose a window \(L_j>0\), to be specified below. Decompose \(|\phi_j\rangle\) according to the \(Q_j\)-window:
\begin{align}
    \|(K_j-U_j)|\phi_j\rangle\|
    &= \left\|(K_j-U_j)\left[\Pi_{L_j}^{Q_j} + (I-\Pi_{L_j}^{Q_j})\right]    |\phi_j\rangle \right\|\notag \\
    &\le \|(K_j-U_j)\Pi_{L_j}^{Q_j}|\phi_j\rangle\| + \|(K_j-U_j)(I-\Pi_{L_j}^{Q_j})|\phi_j\rangle\|.
    \label{eq:single-step-window-decomposition-proof}
\end{align}

We first estimate the contribution inside the window. Here we use the scalar windowed Laurent approximation, rather than the single-gate state-error theorem on \(\mathcal H_n\), so we do not need to assume \(|\phi_j\rangle\in\mathcal H_n\). It suffices to approximate the scalar function \(q\mapsto e^{ip_j(q)}\) on the real interval \([-L_j,L_j]\).
Write \(p_j(q)=\sum_{r=0}^{R_j}a_{j,r}q^r\) as a real polynomial. By the scalar windowed Laurent approximation result, for the window \([-L_j,L_j]\) and accuracy \(\varepsilon/(2M)\), with \(\tau_j:=\pi/(4L_j)\), one can choose a Laurent polynomial
\(P_{j,d_j}(z)=\sum_{m=-d_j}^{d_j}p_{j,m}z^m\) such that
\begin{equation}
    \sup_{|z|=1}|P_{j,d_j}(z)|\le1,\qquad \sup_{|q|\le L_j}\left|P_{j,d_j}(e^{i\tau_jq})-e^{ip_j(q)}\right|
    <\frac{\varepsilon}{2M}.
    \label{eq:bounded-laurent-j-proof}
\end{equation}
Therefore, in the \(Q_j\)-representation,
\begin{align}
    &\|(K_j-U_j)\Pi_{L_j}^{Q_j}|\phi_j\rangle\|^2 \notag\\
    &=\int_{|q|\le L_j}\left|P_{j,d_j}(e^{i\tau_jq}) - e^{ip_j(q)}\right|^2|\phi_{j,Q}(q)|^2\,dq \notag\\
    &\le\left[\sup_{|q|\le L_j}\left|P_{j,d_j}(e^{i\tau_jq})-e^{ip_j(q)}\right|\right]^2 \int_{|q|\le L_j}|\phi_{j,Q}(q)|^2\,dq <\left(\frac{\varepsilon}{2M}\right)^2.
    \label{eq:inside-window-final-proof}
\end{align}
Here \(\phi_{j,Q}(q)\) is the wave function of \(|\phi_j\rangle\) in the \(Q_j\)-representation,
\(\phi_{j,Q}(q):={}_{Q_j}\langle q|\phi_j\rangle\), and in the last step we used
\(\int_{|q|\le L_j}|\phi_{j,Q}(q)|^2\,dq \le\|\phi_j\|^2=1\).

We next estimate the outside-window contribution. First, we bound \(\|K_j\|\). For an arbitrary state \(|\chi\rangle\), let
\(\chi_{Q_j}(q):={}_{Q_j}\langle q|\chi\rangle\) be its wave function in the \(Q_j\)-representation. Then
\begin{align}
    \|K_j|\chi\rangle\|^2
    =\int_{\mathbb R}\left|P_{j,d_j}(e^{i\tau_jq})\right|^2|\chi_{Q_j}(q)|^2\,dq \le\int_{\mathbb R}|\chi_{Q_j}(q)|^2\,dq =\|\chi\|^2.
\end{align}
Here we used \(|P_{j,d_j}(e^{i\tau_jq})|\le1.\) Hence \(\|K_j\|\le1.\) It follows that
\begin{align}
    \|&(K_j-U_j)(I-\Pi_{L_j}^{Q_j})|\phi_j\rangle\| \notag\\
    &\le\bigl(\|K_j\|+\|U_j\|\bigr)\|(I-\Pi_{L_j}^{Q_j})|\phi_j\rangle\| \le2\|(I-\Pi_{L_j}^{Q_j})W_j|\psi\rangle\| \notag \\
    &=2 \|W_j^\dagger(I-\Pi_{L_j}^{Q_j})W_j|\psi\rangle\| =2\|(I-W_j^\dagger\Pi_{L_j}^{Q_j}W_j)|\psi\rangle\|.
    \label{eq:outside-window-before-pullback-proof}
\end{align}
The penultimate step uses the fact that the unitary \(W_j^\dagger\) preserves the norm.
Since \(B_j=W_j^\dagger Q_jW_j\), the operators \(B_j\) and \(Q_j\) are unitarily equivalent. Their spectral projections onto the same interval are therefore related by
\begin{align}
    \Pi_{L_j}^{B_j}
    = W_j^\dagger\Pi_{L_j}^{Q_j}W_j.
    \label{eq:projection-pullback-proof}
\end{align}
Thus Eq.~\eqref{eq:outside-window-before-pullback-proof} becomes
\begin{equation}
    \|(K_j-U_j)(I-\Pi_{L_j}^{Q_j})|\phi_j\rangle\| \le 2\|(I-\Pi_{L_j}^{B_j})|\psi\rangle\|.
    \label{eq:tail-pullback-proof}
\end{equation}

By Lemma~\ref{lem:polynomial-intermediate-tail}, since \(B_j\) is a self-adjoint polynomial in \(X,P\) of fixed degree \(\mu_j\), there exist constants \(c_{j,*},L_{j,*}>0\), depending only on \(B_j\) and \(n\), such that for \(L_j\ge L_{j,*}\),
\begin{equation}
    \left\|(I-\Pi_{L_j}^{B_j})|\psi\rangle\right\|^2
    \le
    \exp[-c_{j,*}L_j^{2/\mu_j}]
    \label{eq:Bj-tail-lemma-proof}
\end{equation}
for every normalized state \(|\psi\rangle\in\mathcal H_n\).
To make the outside-window contribution smaller than \(\varepsilon/(2M)\), Eq.~\eqref{eq:tail-pullback-proof} shows that it is enough to require
\(2\|(I-\Pi_{L_j}^{B_j})|\psi\rangle\| < \frac{\varepsilon}{2M}.\)
By Eq.~\eqref{eq:Bj-tail-lemma-proof}, a sufficient condition is
\(\exp[-c_{j,*}L_j^{2/\mu_j}] < \frac{\varepsilon^2}{16M^2}.\)
Taking logarithms, it suffices to impose
\(c_{j,*}L_j^{2/\mu_j}>\log\frac{16M^2}{\varepsilon^2} = 2\log\frac{4M}{\varepsilon}.\)
Thus we may choose
\begin{equation}
    L_j \ge \max\left\{L_{j,*},\left(\frac{2\log(4M/\varepsilon)}{c_{j,*}}  \right)^{\mu_j/2}\right\}.
    \label{eq:Lj-choice-proof}
\end{equation}
With this choice,
\(\|(K_j-U_j)(I-\Pi_{L_j}^{Q_j})|\phi_j\rangle\|<\frac{\varepsilon}{2M}.\)
Combining the inside- and outside-window estimates, and using Eqs.~\eqref{eq:single-step-window-decomposition-proof} and \eqref{eq:inside-window-final-proof}, we obtain
\begin{align}
    \|(K_j-U_j)W_j|\psi\rangle\|
    =\|(K_j-U_j)|\phi_j\rangle\| 
    <\frac{\varepsilon}{2M}+\frac{\varepsilon}{2M}
    =\frac{\varepsilon}{M}.
    \label{eq:single-step-final-proof}
\end{align}

The preceding construction and estimates hold for each \(j=1,\ldots,M\). We now return to the telescoping error in Eq.~\eqref{eq:telescoping-norm-proof}. Since every \(K_\ell\) satisfies \(\|K_\ell\|\le1\), each term obeys
\begin{align}
    &\left\|K_MK_{M-1}\cdots K_{j+1}(K_j-U_j)W_j|\psi\rangle\right\| \notag\\
    &\le\|K_MK_{M-1}\cdots K_{j+1}\|\cdot\|(K_j-U_j)W_j|\psi\rangle\| \notag\\
    &\le\|(K_j-U_j)W_j|\psi\rangle\| <
    \frac{\varepsilon}{M}.
    \label{eq:telescoping-term-final-proof}
\end{align}
Substituting Eq.~\eqref{eq:telescoping-term-final-proof} into Eq.~\eqref{eq:telescoping-norm-proof}, we find
\begin{align}
    &\left\|\left[K_MK_{M-1}\cdots K_1-U_MU_{M-1}\cdots U_1\right]|\psi\rangle\right\| \notag\\
    &\le
    \sum_{j=1}^{M}
    \left\|K_MK_{M-1}\cdots K_{j+1}(K_j-U_j)W_j|\psi\rangle\right\|<\sum_{j=1}^{M}\frac{\varepsilon}{M} = \varepsilon.
\end{align}
Since \(|\psi\rangle\in\mathcal H_n\) was an arbitrary normalized state, Eq.~\eqref{target_M} follows.

It remains to estimate the Laurent degree. From Eq.~\eqref{eq:Lj-choice-proof} and the explicit expressions for \(c_{j,*}\) and \(L_{j,*}\) obtained in the proof of Lemma~\ref{lem:polynomial-intermediate-tail}, for fixed \(B_j\) and \(n\) one has
\begin{equation}
    L_j = O\!\left[\left(\log\frac{M}{\varepsilon}\right)^{\mu_j/2}\right].
    \label{eq:Lj-scale-proof}
\end{equation}
On the other hand, by the scalar windowed Laurent approximation result, if
\(p_j(q)=\sum_{r=0}^{R_j}a_{j,r}q^r,\)
then one can choose
\begin{equation}
    d_j=O\!\left[\left(\sum_{r=0}^{R_j}|a_{j,r}|L_j^r + \log\frac{M}{\varepsilon}\right)^{1+o(1)}\right].
    \label{eq:dj-window-proof}
\end{equation}
Here the factor \(\log(M/\varepsilon)\) comes from the single-step scalar approximation accuracy \(\varepsilon/(2M)\) inside the window.
Substituting Eq.~\eqref{eq:Lj-scale-proof} into Eq.~\eqref{eq:dj-window-proof} gives
\begin{align}
    d_j=O\!\left[\left(\sum_{r=0}^{R_j}|a_{j,r}|\left(\log\frac{M}{\varepsilon}\right)^{r\mu_j/2}+\log\frac{M}{\varepsilon}\right)^{1+o(1)}\right].
    \label{eq:dj-general-proof}
\end{align}
When \(M\), \(p_j\), \(B_j\), and \(n\) are held fixed as \(\varepsilon\to0\), the coefficients \(a_{j,r}\) and the constants implicit in Eq.~\eqref{eq:Lj-scale-proof} are independent of \(\varepsilon\). If \(R_j\mu_j/2\ge1\), the term with \(r=R_j\) determines the asymptotic logarithmic power, and hence
\begin{equation}
    d_j=O\!\left[\left(\log\frac{M}{\varepsilon}\right)^{R_j\mu_j/2+o(1)}\right].
\end{equation}

The total number of Rabi pulses is therefore \(2\sum_{j=1}^{M}d_j\).
We next estimate the total Rabi interaction time. With the explicit choice in Eq.~\eqref{eq:Lj-choice-proof}, for fixed \(B_j\) and \(n\) one may take
\(L_j=\Theta[(\log(M/\varepsilon))^{\mu_j/2}]\), and hence
\(\tau_j=\pi/(4L_j)=\Theta[(\log(M/\varepsilon))^{-\mu_j/2}]\).
Since the \(j\)th Laurent polynomial is realized by \(2d_j\) equal-duration Rabi pulses, its contribution to the total Rabi interaction time is
\(T_{{\rm tot},j}=2d_j\tau_j
=O[(\log(M/\varepsilon))^{(R_j-1)\mu_j/2+o(1)}]\).
Summing over all phase blocks gives
\begin{equation}
    T_{\rm tot}
    =
    O\!\left[
    \sum_{j=1}^{M}
    \left(
    \log\frac{M}{\varepsilon}
    \right)^{(R_j-1)\mu_j/2+o(1)}
    \right].
\end{equation}

Finally, we prove the success-probability bound. If each \(K_j\) is implemented by a postselected Rabi sequence, then, conditioned on all measurements succeeding, the unnormalized oscillator output is
\(K_{\rm tot}|\psi\rangle\), where \(K_{\rm tot}:=K_MK_{M-1}\cdots K_1\). Thus the total success probability is
\(p_{\rm succ}(\psi)=\left\|K_{\rm tot}|\psi\rangle\right\|^2.\)
Let \(U_{\rm tot}:=U_MU_{M-1}\cdots U_1\). By Eq.~\eqref{target_M},
\(\|(K_{\rm tot}-U_{\rm tot})|\psi\rangle\|<\varepsilon.\)
Since \(U_{\rm tot}\) is unitary, \(\|U_{\rm tot}|\psi\rangle\|=1\). The reverse triangle inequality then gives
\begin{equation}
        \|K_{\rm tot}|\psi\rangle\|\ge\|U_{\rm tot}|\psi\rangle\|-\|(K_{\rm tot}-U_{\rm tot})|\psi\rangle\|>1-\varepsilon.
\end{equation}
Therefore \(p_{\rm succ}(\psi)=\|K_{\rm tot}|\psi\rangle\|^2>(1-\varepsilon)^2.\) This proves the theorem.
\end{proof}

\begin{corollary}[Multimode extension]
\label{cor:multimode-phase-products}
Under the notation of Theorem~\ref{thm:efficient-product-phase-app}, let \(\mathcal H_n^{(m)}\) denote the span of the \(m\)-mode Fock states \(|\nu_1,\ldots,\nu_m\rangle\) with total excitation number \(\nu_1+\cdots+\nu_m\le n\). Let each \(Q_j\) be a real linear quadrature of the \(m\) modes, and suppose that each \(B_j\) is a self-adjoint polynomial of fixed total degree \(\mu_j\) in the multimode quadratures. Then all conclusions of Theorem~\ref{thm:efficient-product-phase-app} remain valid for every normalized state \(|\psi\rangle\in\mathcal H_n^{(m)}\).
\end{corollary}

\begin{proof}
The only part of the proof of Theorem~\ref{thm:efficient-product-phase-app} that uses the single-mode assumption is the spectral-tail estimate in Lemma~\ref{lem:polynomial-intermediate-tail}. Let \(B\) be a self-adjoint polynomial of total degree \(\mu\) in the \(m\) modes. After normal ordering, \(B\) is a finite sum of monomials of the form \(\prod_{\ell=1}^{m}(a_\ell^\dagger)^{r_\ell}a_\ell^{s_\ell}\), with \(\sum_{\ell=1}^{m}(r_\ell+s_\ell)\le\mu\). For every \(|\varphi\rangle\in\mathcal H_N^{(m)}\), each such monomial satisfies \(\left\|\prod_{\ell=1}^{m}(a_\ell^\dagger)^{r_\ell}a_\ell^{s_\ell}|\varphi\rangle\right\|\le(N+\mu+1)^{\frac12\sum_{\ell=1}^{m}(r_\ell+s_\ell)}\|\varphi\|\le(N+\mu+1)^{\mu/2}\|\varphi\|\), and maps \(\mathcal H_N^{(m)}\) into \(\mathcal H_{N+\mu}^{(m)}\). Since only finitely many monomials occur, the one-step estimate and the moment bound used in Lemma~\ref{lem:polynomial-intermediate-tail} remain valid, with constants depending on the fixed polynomial \(B\). The same choice of the moment order therefore gives \(\|(I-\Pi_L^B)|\psi\rangle\|^2\le\exp[-c_*L^{2/\mu}]\) for sufficiently large \(L\).

The remaining proof of Theorem~\ref{thm:efficient-product-phase-app} uses only this tail estimate, the identity \(\Pi_{L_j}^{B_j}=W_j^\dagger\Pi_{L_j}^{Q_j}W_j\), the bound \(\|K_j\|\le1\), and the telescoping identity for the product error. These steps are unchanged in the multimode setting, and the result follows.
\end{proof}

\subsection{Phase-block decompositions of rotation and controlled-phase gates}
\label{app:bounded-gaussian-decomposition}

This subsection derives the rotation- and controlled-phase-gate decompositions used in the main text. For a single-mode Gaussian unitary \(V\), let \(\mathsf M[V]\) denote its Heisenberg-picture symplectic matrix:
\begin{equation}
    V^\dagger\begin{pmatrix}\hat X\\ \hat P\end{pmatrix}V
    =\mathsf M[V]\begin{pmatrix}\hat X\\ \hat P\end{pmatrix}.
\end{equation}
The quadratic phase gates satisfy
\begin{equation}
    \mathsf M\!\left[e^{ib\hat P^2}\right]
    =U(b):=\begin{pmatrix}1&-2b\\0&1\end{pmatrix},
    \qquad
    \mathsf M\!\left[e^{ia\hat X^2}\right]
    =L(a):=\begin{pmatrix}1&0\\2a&1\end{pmatrix}.
\end{equation}

We first consider the rotation gate
\(R(\varphi):=e^{i\varphi(\hat X^2+\hat P^2)}\).
Since \(\hat X^2+\hat P^2=2\hat N+I\), one has
\(R(\varphi+\pi)=-R(\varphi)\), and hence the angle may be reduced to
\(\varphi\in[-\pi/2,\pi/2]\) up to a global phase.
Its Heisenberg action is
\(R(\varphi)^\dagger\hat X R(\varphi)
=\hat X\cos(2\varphi)-\hat P\sin(2\varphi)\)
and
\(R(\varphi)^\dagger\hat P R(\varphi)
=\hat X\sin(2\varphi)+\hat P\cos(2\varphi)\).
Therefore,
\begin{equation}
    \mathsf M[R(\varphi)]
    =\begin{pmatrix}
    \cos(2\varphi)&-\sin(2\varphi)\\
    \sin(2\varphi)&\cos(2\varphi)
    \end{pmatrix}.
\end{equation}
A direct multiplication gives
\begin{equation}
    L(a)U(b)L(a)
    =\begin{pmatrix}
    1-4ab&-2b\\
    4a-8a^2b&1-4ab
    \end{pmatrix}.
\end{equation}
For an angle \(\nu\), define
\(a_\nu:=\frac12\tan\nu\) and
\(b_\nu:=\frac12\sin(2\nu)\).
Then
\(1-4a_\nu b_\nu=\cos(2\nu)\),
\(-2b_\nu=-\sin(2\nu)\), and
\(4a_\nu-8a_\nu^2b_\nu=\sin(2\nu)\).
Hence
\begin{equation}
    R(\nu)
    =e^{ia_\nu\hat X^2}e^{ib_\nu\hat P^2}e^{ia_\nu\hat X^2},
\end{equation}
up to an irrelevant global phase.

To keep the strengths of all quadratic phase blocks uniformly bounded for
\(\varphi\in[-\pi/2,\pi/2]\), we apply this identity to four rotations of angle
\(\varphi/4\).
Defining
\(a_\varphi:=\frac12\tan(\varphi/4)\) and
\(b_\varphi:=\frac12\sin(\varphi/2)\), we obtain
\begin{equation}
    R(\varphi)
    =\left[e^{ia_\varphi\hat X^2}e^{ib_\varphi\hat P^2}e^{ia_\varphi\hat X^2}\right]^4.
\end{equation}
After expansion, the adjacent \(\hat X^2\) phase blocks at the boundaries of neighboring factors can be merged.
The two end blocks have strength \(a_\varphi\), the three internal \(\hat X^2\) blocks have strength \(2a_\varphi\), and the four \(\hat P^2\) blocks have strength \(b_\varphi\).
Since \(\varphi\in[-\pi/2,\pi/2]\), one has
\(2|a_\varphi|\le\tan(\pi/8)=\sqrt2-1<1\) and
\(|b_\varphi|\le1/(2\sqrt2)<1\).
Thus the decomposition contains nine quadratic phase blocks, all with strengths bounded by one.

We next consider the controlled-phase gate
\(CZ_{12}(\zeta)=e^{i\zeta\hat X_1\hat X_2}\).
Define
\(Q_\pm:=(\hat X_1\pm\hat X_2)/\sqrt2\).
Since \([\hat X_1,\hat X_2]=0\), one has
\(\hat X_1\hat X_2=(Q_+^2-Q_-^2)/2\), and the two quadratic terms commute.
Therefore,
\begin{equation}
    CZ_{12}(\zeta)
    =e^{i\zeta Q_+^2/2}e^{-i\zeta Q_-^2/2}.
\end{equation}
Thus the rotation and controlled-phase gates reduce, respectively, to nine and two quadratic phase blocks.

\section{Finite-resource applications}
\label{app:cvhhl-decomposition-and-simulation}

\subsection{Numerical implementation of representative CV evolutions}
\label{app:finite-resource-cv-dynamics}

This subsection describes the finite-resource numerical procedure used for Fig.~1 of the main text. We consider the cubic phase gate \(e^{i\hat X^3}\) and the double-well evolution \(e^{i(\hat X^4-\hat X^2)}\), both acting on the vacuum state \(\ket0\), together with the rotation \(e^{i(\hat X^2+\hat P^2)}\) acting on \((\ket0+\ket1)/\sqrt2\). The evolution time is \(T=1\) in all cases.

For a single polynomial phase \(e^{ip(\hat X)}\), the asymptotic choices of the window parameter \(L\) and the smoothing parameter \(\delta\) used above are sufficient to establish the resource bounds, but need not give the smallest error at finite Laurent bandwidth. In the numerical calculations, we therefore fix the Laurent bandwidth \(d\) and optimize \(L\) and \(\delta\). For each pair \((L,\delta)\), we construct the periodic function \(F_\delta\) and its truncated Fourier series \(\widetilde P_d\) as in Sec.~\ref{app:bounded-laurent-approximation}. The bounded Laurent polynomial is then obtained from the normalization following Eq.~\eqref{eq:Ptilde-bound},
\(P_{d,L,\delta}(z)=\widetilde P_d(z)/(1+\eta_d)\), with \(\tau_L=\pi/(4L)\).

For an input state \(\ket{\psi_{\rm in}}\), we evaluate the state-dependent implementation error
\begin{equation}
    E_d(L,\delta)
    =
    \left\|
    \left[
    P_{d,L,\delta}(e^{i\tau_L\hat X})
    -
    e^{ip(\hat X)}
    \right]
    \ket{\psi_{\rm in}}
    \right\|.
    \label{eq:numerical-single-phase-error}
\end{equation}
For each fixed bandwidth \(d\), we vary \(L\) and \(\delta\) and define the finite-resource error by
\(\varepsilon_d:=\min_{L,\delta}E_d(L,\delta)\).
The optimizing window, denoted by \(L_d\), determines the pulse duration
\(\tau_d=\pi/(4L_d)\), and therefore also fixes the total Rabi interaction time of the corresponding \(2d\)-pulse sequence.
\begin{equation}
    T_{\rm tot}
    =
    2d\tau_d
    =
    \frac{\pi d}{2L_d}.
    \label{eq:numerical-single-phase-time}
\end{equation}
This procedure is applied directly to the cubic phase gate with \(p(x)=x^3\) and to the double-well evolution with \(p(x)=x^4-x^2\).

The rotation involves both \(\hat X\) and \(\hat P\), and is implemented using the exact quadratic-phase decomposition derived in Sec.~\ref{app:bounded-gaussian-decomposition}.
After expanding the four three-block factors and merging adjacent \(\hat X^2\) phase blocks, the rotation contains nine quadratic phase blocks.
We write the resulting decomposition as \(U_{\rm rot}=U_M\cdots U_1\), with \(M=9\), where each block has the form \(U_j=e^{ic_j\hat Q_j^2}\) with \(\hat Q_j=\hat X\) or \(\hat P\).
For the \(j\)th block, let \(P_{j,d_j,L_j,\delta_j}\) denote the corresponding bounded Laurent polynomial and set \(\tau_j=\pi/(4L_j)\).  The complete postselected approximation is $K_{\rm rot}=K_M\cdots K_1,$ where $K_j=P_{j,d_j,L_j,\delta_j}(e^{i\tau_j\hat Q_j}).$ 
Its state-dependent implementation error is evaluated directly on the input state,
\begin{equation}
    \varepsilon_{\rm rot}
    =
    \left\|
    \left(
    K_{\rm rot}-U_{\rm rot}
    \right)
    \frac{\ket0+\ket1}{\sqrt2}
    \right\|.
    \label{eq:numerical-rotation-error}
\end{equation}
The total Rabi interaction time of the complete rotation sequence is the sum over the nine phase blocks,
\begin{equation}
    T_{\rm tot}
    =
    \sum_{j=1}^{M}2d_j\tau_j
    =
    \sum_{j=1}^{M}\frac{\pi d_j}{2L_j}.
    \label{eq:numerical-rotation-time}
\end{equation}

The errors obtained in this way are plotted against the corresponding \(T_{\rm tot}\) in Fig.~1 of the main text. The decrease of the numerical error with increasing interaction time illustrates the finite-resource convergence of the Rabi synthesis for both single polynomial phases and finite products of phase blocks.

\subsection{Polynomial-phase decomposition of the core unitary}
\label{app:pde-phase-decomposition}

We consider the regularized biharmonic equation
\begin{equation}
    \left(
    \mu-\alpha\Delta_D+\beta\Delta_D^2
    \right)
    u(\mathbf x)
    =
    f(\mathbf x),
    \qquad
    \mu,\alpha,\beta>0.
    \label{eq:pde-app}
\end{equation}
The source and solution are encoded as continuous-variable wave functions, and the goal is to prepare a state proportional to
\(\hat A_D^{-1}\ket f\), where
\(\hat A_D=\mu-\alpha\Delta_D+\beta\Delta_D^2\).
The first two modes serve as resource modes and the remaining \(D\) modes as the data register.
Taking \(\hat P_j=-i\partial_{x_j}\), one has
\(-\Delta_D=\sum_{j=1}^{D}\hat P_{j+2}^{\,2}\), and hence
\begin{equation}
    \hat A_D
    =
    \mu
    +
    \alpha\sum_{j=1}^{D}\hat P_{j+2}^{\,2}
    +
    \beta
    \left(
    \sum_{j=1}^{D}\hat P_{j+2}^{\,2}
    \right)^2 .
    \label{eq:AD-app}
\end{equation}
The core unitary is
\(U_A=\exp[-i\hat X_1\hat X_2\hat A_D]\).
Since the data-mode momenta commute with one another and with \(\hat X_1\) and \(\hat X_2\), its generator can be expanded as
\begin{equation}
\begin{aligned}
    \hat X_1\hat X_2\hat A_D
    ={}&
    \mu\hat X_1\hat X_2
    +
    \alpha\sum_{j=1}^{D}
    \hat X_1\hat X_2\hat P_{j+2}^{\,2}
    +
    \beta\sum_{j=1}^{D}
    \hat X_1\hat X_2\hat P_{j+2}^{\,4}
    \\
    &+
    2\beta\sum_{1\le j<k\le D}
    \hat X_1\hat X_2
    \hat P_{j+2}^{\,2}
    \hat P_{k+2}^{\,2}.
\end{aligned}
\label{eq:UA-generator-expansion-app}
\end{equation}
Thus the generator contains only commuting monomials of degrees \(2\), \(4\), and \(6\).

We use the polarization identity to reduce the higher-order monomials to powers of linear quadratures.
For pairwise commuting operators \(A_1,\ldots,A_m\),
\begin{equation}
    A_1A_2\cdots A_m
    =
    \frac{1}{2^m m!}
    \sum_{\boldsymbol\nu\in\{\pm1\}^m}
    \left(
    \prod_{\ell=1}^{m}\nu_\ell
    \right)
    \left(
    \sum_{\ell=1}^{m}\nu_\ell A_\ell
    \right)^m .
    \label{eq:polarization-app}
\end{equation}
Define
\(\eta_{\boldsymbol\nu}^{(m)}:=\prod_{\ell=1}^{m}\nu_\ell\), where
\(\boldsymbol\nu=(\nu_1,\ldots,\nu_m)\in\{\pm1\}^m\).

For each data mode \(j=1,\ldots,D\), define
\begin{equation}
    Q_{\boldsymbol\nu,j}^{(4)}
    :=
    \frac{1}{2}
    \left[
    \nu_1\hat X_1
    +
    \nu_2\hat X_2
    +
    (\nu_3+\nu_4)\hat P_{j+2}
    \right],
    \qquad
    \boldsymbol\nu\in\{\pm1\}^4 .
    \label{eq:Q4-app}
\end{equation}
Taking
\((A_1,A_2,A_3,A_4)
=(\hat X_1,\hat X_2,\hat P_{j+2},\hat P_{j+2})\)
in Eq.~\eqref{eq:polarization-app} gives
\(\hat X_1\hat X_2\hat P_{j+2}^{\,2}
=\frac{1}{24}
\sum_{\boldsymbol\nu\in\{\pm1\}^4}
\eta_{\boldsymbol\nu}^{(4)}
(Q_{\boldsymbol\nu,j}^{(4)})^4\).
Since all linear quadratures appearing here commute, the corresponding phase gate can be written exactly as
\begin{equation}
    \exp\!\left(
    -i\alpha\hat X_1\hat X_2\hat P_{j+2}^{\,2}
    \right)
    =
    \prod_{\boldsymbol\nu\in\{\pm1\}^4}
    \exp\!\left[
    -i\frac{\alpha}{24}
    \eta_{\boldsymbol\nu}^{(4)}
    \left(
    Q_{\boldsymbol\nu,j}^{(4)}
    \right)^4
    \right].
    \label{eq:X1X2P2-phase-Q4-app}
\end{equation}

For the sixth-order monomial on a single data mode, define
\begin{equation}
    Q_{\boldsymbol\nu,j}^{(6)}
    :=
    \frac{1}{\sqrt6}
    \left[
    \nu_1\hat X_1
    +
    \nu_2\hat X_2
    +
    (\nu_3+\nu_4+\nu_5+\nu_6)\hat P_{j+2}
    \right],
    \qquad
    \boldsymbol\nu\in\{\pm1\}^6 .
    \label{eq:Q6-single-app}
\end{equation}
Taking
\((A_1,\ldots,A_6)
=(\hat X_1,\hat X_2,\hat P_{j+2},\hat P_{j+2},
\hat P_{j+2},\hat P_{j+2})\)
in Eq.~\eqref{eq:polarization-app} gives
\(\hat X_1\hat X_2\hat P_{j+2}^{\,4}
=\frac{3}{640}
\sum_{\boldsymbol\nu\in\{\pm1\}^6}
\eta_{\boldsymbol\nu}^{(6)}
(Q_{\boldsymbol\nu,j}^{(6)})^6\).
Therefore,
\begin{equation}
    \exp\!\left(
    -i\beta\hat X_1\hat X_2\hat P_{j+2}^{\,4}
    \right)
    =
    \prod_{\boldsymbol\nu\in\{\pm1\}^6}
    \exp\!\left[
    -i\frac{3\beta}{640}
    \eta_{\boldsymbol\nu}^{(6)}
    \left(
    Q_{\boldsymbol\nu,j}^{(6)}
    \right)^6
    \right].
    \label{eq:X1X2P4-phase-Q6-app}
\end{equation}

For the mixed sixth-order monomial involving distinct data modes, with \(1\le j<k\le D\), define
\begin{equation}
    Q_{\boldsymbol\nu,jk}^{(6)}
    :=
    \frac{1}{\sqrt6}
    \left[
    \nu_1\hat X_1
    +
    \nu_2\hat X_2
    +
    (\nu_3+\nu_4)\hat P_{j+2}
    +
    (\nu_5+\nu_6)\hat P_{k+2}
    \right],
    \qquad
    \boldsymbol\nu\in\{\pm1\}^6 .
    \label{eq:Q6-mixed-app}
\end{equation}
Taking
\((A_1,\ldots,A_6)
=(\hat X_1,\hat X_2,\hat P_{j+2},\hat P_{j+2},
\hat P_{k+2},\hat P_{k+2})\)
in Eq.~\eqref{eq:polarization-app} gives
\(\hat X_1\hat X_2\hat P_{j+2}^{\,2}\hat P_{k+2}^{\,2}
=\frac{3}{640}
\sum_{\boldsymbol\nu\in\{\pm1\}^6}
\eta_{\boldsymbol\nu}^{(6)}
(Q_{\boldsymbol\nu,jk}^{(6)})^6\).
Accounting for the coefficient \(2\beta\) in Eq.~\eqref{eq:UA-generator-expansion-app}, the corresponding phase gate is
\begin{equation}
    \exp\!\left(
    -i2\beta
    \hat X_1\hat X_2
    \hat P_{j+2}^{\,2}
    \hat P_{k+2}^{\,2}
    \right)
    =
    \prod_{\boldsymbol\nu\in\{\pm1\}^6}
    \exp\!\left[
    -i\frac{3\beta}{320}
    \eta_{\boldsymbol\nu}^{(6)}
    \left(
    Q_{\boldsymbol\nu,jk}^{(6)}
    \right)^6
    \right].
    \label{eq:X1X2P2P2-phase-Q6-app}
\end{equation}

For the quadratic term, define
\(Q_{+}^{(2)}:=(\hat X_1+\hat X_2)/\sqrt2\) and
\(Q_{-}^{(2)}:=(\hat X_1-\hat X_2)/\sqrt2\).
Then
\(\hat X_1\hat X_2=
[(Q_{+}^{(2)})^2-(Q_{-}^{(2)})^2]/2\).
Substituting the above decompositions into
\(U_A=\exp(-i\hat X_1\hat X_2\hat A_D)\) gives
\begin{equation}
\begin{aligned}
    U_A
    ={}&
    \exp\!\left[
    -i\frac{\mu}{2}
    \left(
    Q_{+}^{(2)}
    \right)^2
    \right]
    \exp\!\left[
    i\frac{\mu}{2}
    \left(
    Q_{-}^{(2)}
    \right)^2
    \right]
    \\
    &\times
    \prod_{j=1}^{D}
    \prod_{\boldsymbol\nu\in\{\pm1\}^4}
    \exp\!\left[
    -i\frac{\alpha}{24}
    \eta_{\boldsymbol\nu}^{(4)}
    \left(
    Q_{\boldsymbol\nu,j}^{(4)}
    \right)^4
    \right]
    \\
    &\times
    \prod_{j=1}^{D}
    \prod_{\boldsymbol\nu\in\{\pm1\}^6}
    \exp\!\left[
    -i\frac{3\beta}{640}
    \eta_{\boldsymbol\nu}^{(6)}
    \left(
    Q_{\boldsymbol\nu,j}^{(6)}
    \right)^6
    \right]
    \\
    &\times
    \prod_{1\le j<k\le D}
    \prod_{\boldsymbol\nu\in\{\pm1\}^6}
    \exp\!\left[
    -i\frac{3\beta}{320}
    \eta_{\boldsymbol\nu}^{(6)}
    \left(
    Q_{\boldsymbol\nu,jk}^{(6)}
    \right)^6
    \right].
\end{aligned}
\label{eq:UA-complete-polynomial-phase-decomposition-app}
\end{equation}

Every exponential on the right-hand side of Eq.~\eqref{eq:UA-complete-polynomial-phase-decomposition-app} is a polynomial phase block of the form \(e^{itQ^r}\), where \(r\in\{2,4,6\}\) and \(Q\) is a real linear combination of \(\hat X_1\), \(\hat X_2\), and \(\hat P_3,\ldots,\hat P_{D+2}\).
These basic quadratures commute pairwise, so all linear quadratures appearing in the decomposition also commute with one another.
The ordering of the phase blocks therefore does not affect the resulting core unitary.

We now apply Corollary~\ref{cor:multimode-phase-products}.
Write Eq.~\eqref{eq:UA-complete-polynomial-phase-decomposition-app} as
\(U_A=U_M\cdots U_1\), where
\(U_\ell=e^{it_\ell Q_\ell^{r_\ell}}\) and
\(r_\ell\in\{2,4,6\}\).
Define \(W_\ell:=U_{\ell-1}\cdots U_1\), with \(W_1:=I\).
Since all \(Q_\ell\) commute with one another, one has
\([U_k,Q_\ell]=0\) for every \(k,\ell\), and hence
\(B_\ell:=W_\ell^\dagger Q_\ell W_\ell=Q_\ell\).
Thus every effective observable \(B_\ell\) remains a linear multimode quadrature and has total degree \(\mu_\ell=1\).

Moreover, each \(Q_\ell\) contains at most four nonzero quadrature coefficients, and their magnitudes are bounded by constants independent of \(D\) and of the position \(\ell\) in the product.
For a fixed total-Fock cutoff, the constants entering the multimode tail estimate can therefore be chosen uniformly over the phase blocks.
Accordingly, the spectral windows may be chosen with
\(L_\ell=\Theta\!\left(\sqrt{\log(M/\varepsilon)}\right)\), with constants independent of \(D\).

Without merging repeated quadratures or removing trivial factors, the direct polarization decomposition contains at most
\begin{equation}
\begin{aligned}
    M_{\rm raw}
    =2+2^4D+2^6D+2^6\binom{D}{2}
    =32D^2+48D+2 .
\end{aligned}
\label{eq:PDE-raw-phase-count-app}
\end{equation}
Therefore, the number of phase blocks satisfies \(M=O(D^2)\).

For the \(\ell\)th phase block, Corollary~\ref{cor:multimode-phase-products} with \(\mu_\ell=1\) gives
\(d_\ell=O[(\log(M/\varepsilon))^{r_\ell/2+o(1)}]\)
for fixed PDE coefficients and a fixed total-Fock cutoff.
The sixth-order blocks give the largest exponent.
Hence the total number of generalized multimode Rabi pulses required for the complete core unitary satisfies
\begin{equation}
    N
    =
    2\sum_{\ell=1}^{M}d_\ell
    =
    O\!\left[
    M
    \left(
    \log\frac{M}{\varepsilon}
    \right)^{3+o(1)}
    \right]
    =
    O\!\left[
    D^2
    \left(
    \log\frac{D}{\varepsilon}
    \right)^{3+o(1)}
    \right].
    \label{eq:PDE-pulse-count-app}
\end{equation}

Finally, the common-duration construction for the \(\ell\)th phase block uses
\(\tau_\ell=\pi/(4L_\ell)\).
With the window choice above,
\(\tau_\ell=\Theta[(\log(M/\varepsilon))^{-1/2}]\).
Since the \(\ell\)th block contains \(2d_\ell\) Rabi pulses, its contribution to the total Rabi interaction time is \(2d_\ell\tau_\ell\).
Therefore,
\begin{equation}
\begin{aligned}
    T_{\rm tot}
    =\sum_{\ell=1}^{M}2d_\ell\tau_\ell
    =O\!\left[D^2\left(\log\frac{D}{\varepsilon}\right)^{5/2+o(1)}\right].
\end{aligned}
\label{eq:PDE-total-time-app}
\end{equation}
This estimate concerns the Rabi synthesis of the core unitary \(U_A\).

\subsection{Finite-resource simulation of the two-dimensional regularized biharmonic equation}
\label{app:pde-numerical-setup}

\begin{figure}[t]
    \centering
    \includegraphics[width=0.5\linewidth]{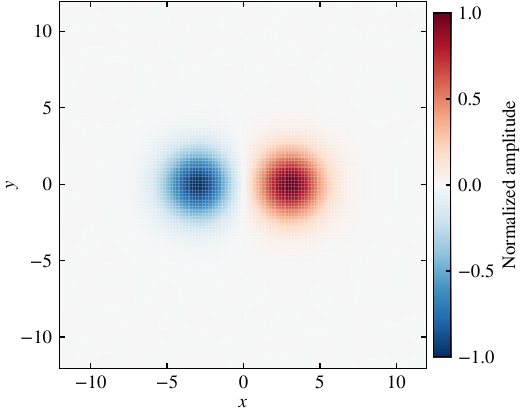}
    \caption{
    Finite-resource output for the two-dimensional regularized biharmonic equation with
    \(\mu=1.0\), \(\alpha=0.8\), and \(\beta=0.15\).
    The normalized amplitude \(u_{\rm Rabi}(x,y)\) is obtained using the Rabi synthesis of the
    \(Q^2\), \(Q^4\), and \(Q^6\) polynomial phase blocks in
    \(U_A=\exp[-i\hat X_1\hat X_2\hat A_D]\).
    For the resource-state cutoff \(n_{\max}=41\) and Gaussian postselection width
    \(\Delta=0.1\), the finite-resource output reproduces the antisymmetric spatial profile of the reference solution with fidelity
    \(F_{\rm sol}=0.9998\).
    }
    \label{fig:pde-inversion-app}
\end{figure}

Following the continuous-variable linear-equation protocols of Ref.~\cite{ArrazolaKalajdzievskiWeedbrookLloyd2019}, the solution state is prepared using the core unitary
\(U_A=\exp[-i\hat X_1\hat X_2\hat A_D]\).
The present work focuses on the Rabi implementation of this core unitary.

The two-dimensional regularized biharmonic equation is simulated using the dimensionless quadrature convention
\([\hat X,\hat P]=i\).
The spatial domain is \([-6,6)\times[-6,6)\), with \(256\) grid points along each direction.
The equation parameters are \(\mu=1.0\), \(\alpha=0.8\), and \(\beta=0.15\).
The source term is chosen as a pair of localized Gaussian profiles with opposite signs,
\begin{equation}
\begin{split}
    f(x,y)=\mathcal N\Big[\exp\!\left(-\frac{(x-x_0)^2+y^2}{2\sigma^2}
    \right)-\exp\!\left(-\frac{(x+x_0)^2+y^2}{2\sigma^2}\right)\Big],
\end{split}
\label{eq:double-gaussian-source-app}
\end{equation}
where \(x_0=3\) and \(\sigma=1.1\).
The normalization constant \(\mathcal N\) is chosen so that the discretized source field has unit spatial \(L^2\) norm.
The resulting state is encoded as the input state \(\ket f\) of the data register.

The first resource mode is prepared in a finite-energy smooth step state with target wavefunction
\begin{equation}
s_{L_s,\sigma_{\rm s}}(x)
=
\frac{1}{2}
\left[
\tanh\!\left(\frac{x}{\sigma_{\rm s}}\right)
-
\tanh\!\left(\frac{x-L_s}{\sigma_{\rm s}}\right)
\right].
\label{eq:smooth-step-state-app}
\end{equation}
We take \(L_s=3.0\) and \(\sigma_{\rm s}=0.08\), normalize this state, and project it onto the finite Fock subspace
\(\operatorname{span}\{\ket0,\ldots,\ket{41}\}\).
The second resource mode is prepared in the single-photon state \(\ket1\).

Finite-precision postselection is implemented using a Gaussian momentum state,
\begin{equation}
\langle p|\Delta\rangle
\propto
\exp\!\left(
-\frac{p^2}{2\Delta^2}
\right),
\qquad
\operatorname{Var}(\hat P)
=
\frac{\Delta^2}{2}.
\label{eq:gaussian-postselection-app}
\end{equation}
We take \(\Delta=0.1\).
Under the present convention, this corresponds to a squeezing level
\(S_{\rm dB}=-20\log_{10}\Delta=20\,{\rm dB}\).
The finite-width Gaussian postselection state is treated as an external Gaussian resource of the algorithm.

For \(D=2\), the polarization decomposition gives \(38\) nonzero monomial phase factors after trivial contributions are removed.
Since all factors commute, terms associated with the same linear quadrature can be combined into a single polynomial phase block.
The numerical implementation therefore uses \(28\) polynomial phase blocks.
Each block is implemented using the Laurent-polynomial construction developed above.
We assign a total approximation-error budget of \(\varepsilon=10^{-2}\) to the complete phase product.

The reference solution is obtained independently using a Fourier spectral method.
Let \(\ket{u_{\rm ref}}\) denote the normalized reference solution state and let \(\ket{u_{\rm Rabi}}\) denote the normalized finite-resource output obtained with the Rabi-synthesized core unitary.
We quantify the agreement between the two states by $F_{\rm sol} =\left|\langle u_{\rm ref}|u_{\rm Rabi}\rangle\right|^2 .$
For the parameters specified above, we obtain
\(F_{\rm sol}=0.9998\).

As shown in Fig.~\ref{fig:pde-inversion-app}, the finite-resource output preserves the antisymmetric spatial structure induced by the double-Gaussian source.
Together with the complexity result in Eq.~\eqref{eq:PDE-total-time-app}, this example demonstrates the use of Rabi-synthesized polynomial phase blocks as the core unitary of a continuous-variable linear-equation algorithm without requiring native fourth- or sixth-order oscillator nonlinearities.

\bibliography{ref}